\documentclass[fleqn,usenatbib]{mnras}
\usepackage{amsmath}
\usepackage{amssymb}
\usepackage{siunitx}
\usepackage{txfonts}
\usepackage{rotating}
\usepackage{float}
\usepackage{epsfig}
\usepackage{epstopdf}
\usepackage{graphicx}
\usepackage{color}
\usepackage{pdflscape}
\usepackage{appendix}
\usepackage{comment}
\usepackage{tabularx}
\usepackage{hyperref}
\renewcommand{\theenumi}{\arabic{enumi}}

\begin{document}

\newcommand{\orcid}[1]{\href{https://orcid.org/#1}{\includegraphics[width=10pt]{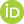}}}
\newcommand{\red}{\textcolor{red}}

\title[Stellar Populations in SNe Ia host galaxies]
{Stellar Populations in type Ia supernova host galaxies at intermediate-high redshift: Star formation and metallicity enrichment histories.}

\author[Millán-Irigoyen et al.]
{I. Millan-Irigoyen$^{1}$\thanks{E-mail:millaniker@gmail.com}\orcid{0000-0003-4115-5140}, 
M. G. del Valle-Espinosa $^{2}$\orcid{0000-0002-0191-4897}, 
R. Fern{\'a}ndez-Aranda$^{3}$\orcid{0000-0002-7714-688X}, 
\newauthor 
L. Galbany$^{4,5}$\orcid{0000-0002-1296-6887}, 
J.M. Gomes$^{6}$\orcid{0000-0002-3861-7482},
M. Moreno-Raya$^{1,7}$\orcid{0000-0003-3437-1339}, 
\'A.R. L{\'o}pez-S{\'a}nchez$^{8,9,10}$\orcid{0000-0001-8083-8046},
\newauthor
and M.Moll{\'a}$^{1}$\orcid{0000-0003-0817-581X}\\
$^{1}$ Dpto. de Investigaci\'{o}n B\'{a}sica, CIEMAT, Avda. Complutense 40, E-28040 Madrid, Spain\\
$^{2}$ Institute for Astronomy, The Univeristy of Edinburgh, Royal Observatory, Blackford Hill, Edinburgh, EH9 3HJ, UK\\
$^{3}$ Institute of Astrophysics - FORTH \& Dept. of Physics ,University of Crete, 70 013 Heraklion, Greece. \\
$^{4}$ Institute of Space Sciences (ICE, CSIC), Campus UAB, Carrer de Can Magrans, s/n, E-08193 Barcelona, Spain. \\
$^{5}$ Institut d'Estudis Espacials de Catalunya (IEEC), E-08034 Barcelona, Spain.\\
$^{6}$ Instituto de Astrofísica e Ciências do Espaço, Universidade do Porto, CAUP, Rua das Estrelas, 4150-762 Porto, Portugal\\
$^{7}$ Instituto de Astrof{\'\i}sica de Andaluc{\'\i}a, -CSIC, Glorieta de la Astronom{\'\i}a s.n., 18008 Granada,Spain\\
$^{8}$ Australian Astronomical Optics, Macquarie University, 105 Delhi Rd, North Ryde, NSW 2113, Australia\\
$^{9}$ Macquarie University Research Centre for Astronomy, Astrophysics \& Astrophotonics, Sydney, NSW 2109, Australia\\
$^{10}$ Australian Research Council Centre of Excellence for All Sky Astrophysics in 3 Dimensions (ASTRO-3D), Australia\\
}

\date{Accepted Received ; in original form }
\pagerange{\pageref{firstpage}--\pageref{lastpage}} \pubyear{}

\maketitle
\label{firstpage}

\begin{abstract}
We present a summary of our project that studies galaxies hosting type Ia supernova (SN Ia) at different redshifts. We present Gran Telescopio de Canarias (GTC) optical spectroscopy of six SN Ia host galaxies at redshift  $z\sim 0.4-0.5$. They are joined to a set of SN Ia host galaxies at intermediate-high redshift, which include galaxies from surveys SDSS and COSMOS. The final sample, after a selection of galaxy spectra in terms of signal-to-noise and other characteristics, consists of 680 galaxies with redshift in the range $0.04 < z < 1$. We perform an inverse stellar population synthesis with the code {\sc fado} to estimate the star formation and enrichment histories of this set of galaxies, simultaneously obtaining their mean stellar age and metallicity and stellar mass. After analysing the correlations among these characteristics, we look for possible dependencies of the Hubble diagram residuals and supernova features (luminosity, color and strength parameter) on these stellar parameters. We find that the Hubble residuals show a clear dependence on the stellar metallicity weighted by mass with a slope of -0.061\,mag\,dex$^{-1}$, when represented in logarithmic scale, $\log{\langle Z_{M}/Z_{\sun}\rangle}$. This result supports our previous findings obtained from gas oxygen abundances for local and SDSS-survey galaxies. Comparing with other works from the literature that also use the stellar metallicity, we find a similar value, but with more precision and a better significance (2.08 vs $\sim$ 1.1), due to the higher number of objects and wider range of redshift of our sample.

\end{abstract}


\section{Introduction}
\label{Intro}

Type Ia supernova (SN Ia) peak brightness can be standardized, so that distances to their host galaxies can be estimated simply obtaining SN~Ia light-curves and measuring their apparent peak magnitude ($m$). 
Assuming a common and universal absolute peak magnitude ($M$), their distance modulus ($\mu =m-M$) can be represented as a function of the redshift ($z$) of their host galaxy, in which is known as the Hubble diagram (HD). 
Using SNe Ia as standard candles allowed astronomers in the mid 1990s to measure the evolution of the Universe with \mbox{sufficient} precision and at distances large enough to detect an accelerated rate of expansion of the Universe. Two projects, the High-z Supernova project \citep{riess1998} and the Supernova Cosmology Project \citep{perlmutter1999}, revealed that high-$z$ SNe Ia appeared to be fainter than expected in a monotonic expansion. This suggested the presence of an additional component with negative pressure in the content of the Universe, later dubbed dark energy.

In fact, SNe Ia are not natural standard candles but {\sl standardizable} by using a relationship between their peak brightness $m$, and both the width of their light-curves (parametrized by $x1$, $s$, or $\Delta m_{15}$), and their color ($c$ or $E(B-V)$) at the peak \citep{phillips1993,riess1996,tripp1998}. 
Once accounted for, it is possible to study the distance to these objects with high precision, with an uncertainty down to 5\% \citep{2018ApJ...859..101S,2019ApJ...874..150B}. Using a large number of SNe Ia data along a range of redshift, it is possible to fit different cosmological models in the Hubble diagram, and find the most suitable to the observational data.

The standardization technique, however, is primarily trained with SNe Ia located in galaxies of the local Universe, so they typically have metal abundances and stellar ages around Solar values. In fact, a dependence of the SN Ia luminosity on the metallicity of the progenitor binary system has been theoretically predicted: \citet{timmes2003} showed that the excess of neutrons in the explosion of a white dwarf is a direct function of the metallicity of the progenitor star, and that this excess is what controls the ratio between radioactive abundances (as $^{56}$Ni, which defines the brightness of the explosion) to non-radioactive (elements of the group of iron). Thus, by varying an order 3 in the metallicity may induce a change of approximately 25\% in the mass of $^{56}$Ni synthesized during the explosion. Later, \citet{bravo2010} also found from different explosion models, that the mass of ejected $^{56}$Ni, and therefore, the luminosity of the explosion, would depend on the metallicity of the components of the binary system. Since the elemental abundance in the interstellar medium (ISM) changes with redshift, due to the chemical enrichment along the universe evolution, the direct light-magnitude-distance curve relationship may not be entirely valid for all redshifts, if the metallicity of the system is not taken into account.

There have been many studies for the last two decades, studying
the dependence of SNe Ia light-curve parameters on global
characteristics of their hosts galaxies
\citep{2006ApJ...648..868S, Gallagher+2008, 2009ApJ...691..661H, 2009ApJ...700..331H, 2010ApJ...715..743K, 2010MNRAS.406..782S, 2010ApJ...722..566L, 2011ApJ...743..172D, 2011ApJ...740...92G, 2011ApJ...734...42N, 2011ApJ...737..102S, Johansson+2013, Pan+2014}.
Most of them show that SNe~Ia are systematically brighter in the massive host galaxies than in the less massive ones, after LC shape and color corrections. Through the mass-metallicity relation \citep{tremonti2004,laralopez2010}, this would indicate that there exists a correlation between magnitudes of SNe~Ia and metallicities of their host galaxies: the more metal-rich galaxies would have the brightest SNe~Ia after corrections with a difference of $\sim ~0.10$\,mag\,dex$^{-1}$. More recent studies devoted to correlate the characteristics of the SN~Ia with galaxies properties have increased the redshift range and the number of objects involved in their samples. This way, \citet{Uddin2017,Uddin2018} included 1388 galaxies with a range in redshift of $0.01 \le z \le 1.1$; \citet{Jones+2018} has 1360 objects and  $z \le 0.7$;
The large PANTHEON sample from \citet{Scolnic+2018} has a sample with 1023 SNe~Ia and $z \le 2.3$; \citet{Kim+2019} do an analysis with 674 galaxies and redshift  $z< 0.85$;  However, these numbers reduce when they obtain the stellar mass. Thus, the \citet{Scolnic+2018} sample reduces to 270 objects in a range $0.03 \le z \le 0.65$ for which they obtained the stellar mass by applying the Z-Pegase code. Besides that, most of them focus again on the dependence of the Hubble Diagram residuals (HR) of the SNe~Ia on the stellar mass or other characteristics of their galaxies. For example, \citet{roman2018} took a sample of 882 galaxies with $0.01 \le z \le 1$ and by using the SED fitting technique studied the stellar populations characteristics, obtaining again the stellar mass and the color U-V, that is neither metallicity nor age; \citet{Uddin2020} determined host galaxy stellar masses, and then studied the correlation between the HR and those masses in the uBgVriYJH photometric bands.  Also, \citet{Smith+2020} study the DES survey supernova Ia finding a difference of 0.040\,mag between low mass and massive host galaxies. Apart from the stellar mass, other groups have also looked for correlations with the mean stellar age, the metallicity or the sSFR, as \citet{Rigault+2020}, who has found a strong correlation between the sSFR of the vicinity of the SN and the Hubble residual. 

The direct dependence of SNe~Ia luminosities on metallicity was early studied by \citet{Gallagher+2005}, who estimated oxygen elemental abundances by using host galaxies with emission lines, and by \citet{Gallagher+2008}, where they analyzed spectral absorption indices in early-type galaxies, by using theoretical evolutionary synthesis models. In both cases, they found a trend between magnitudes of SNe~Ia
and the stellar population metallicity of their host galaxies, in agreement with theoretical expectations. After it, there were some works \citep{2009ApJ...691..661H,konishi2011,dAndrea2011,Johansson+2013,Childress+2013,Pan+2014,betoule2014,mr-2016a,Campbell+2016,Wolf+2016} that studied the possible impact of the metallicity on the luminosity of SNe~Ia. Most of them have studied the emission lines intensities, estimating the Oxygen abundances as a proxy of the metallicity of galaxies. This technique has two issues: 1) The abundances are obtained only for star forming galaxies, that is, they are biased towards a particular type of galaxy; Moreover, this implies the use of spectroscopic methods, which need a certain observational time and, in consequence, thus reducing the number of possible observed galaxies; 2) The obtained abundances are indicating the final state of galaxies related with the gas associated to the youngest stars, which could not represent the metallicity of the SNe~Ia progenitors, since they need a time of evolution before suffer a supernova explosion.

From the works published in the last five years, only \citet{Kang+2020} study the dependence of HR on stellar ages and metallicities in host galaxies by analyzing spectra from the YONSEI project with high signal-to-noise (S/N $\sim 175$). Their sample is composed of 60 early type galaxies, that is without dust, in the range $0.01 \le z \le 0.08$. By using a technique similar to the one from \citet{Gallagher+2008}, by means of the PPF synthesis model, they measure the Lick spectral indices, finding the best age and metallicity for each galaxy. After they study the dependence of the HR on these stellar parameters, claiming that the age is the only parameter with which they found a correlation.

We started a project to estimate the metallicity in SN Ia host galaxies, and to determine whether or not a dependence of SNe Ia brightness with the host galaxy metallicity there exists. Our objective in this particular work is two fold: trying to increase the redshift range and, furthermore, to increase the number of objects. With this aim, we analyze data at different redshift ranges:

\begin{enumerate}
\item {\it Local Universe:} 
A set of 28 local galaxies was presented in \citet{mr-2016a,mr-2016b}, where it was observationally confirmed that, effectively, the absolute magnitude of SNe~Ia in galaxies with distances estimated independently by other methods (e.g. Tully-Fisher, Fundamental Plane), shows a clear dependence on the oxygen abundance of their galactic environments.
\item {\it Galaxies at intermediate redshift ($0.03 < z < 0.4$):}
\citet{2018MNRAS.476..307M} estimated the Oxygen abundance using spectra of emission-line SNe~Ia host galaxies from the Sloan Digital Sky Survey (SDSS, see Sect.~\ref{data:sdss}
Studying the dependence of HR  of SN~Ia on the Oxygen abundance of their host galaxies, a correlation was found with a slope of $-0.186\pm 0.123$\,mag\,dex$^{-1}$ (1.52$\sigma$), in line with other works and in good agreement with theoretical expectations. 
\item {\it Galaxies at redshift $z\sim 0.4-0.5$ from GTC:}
Spectroscopy of six SN Ia host galaxies was obtained with GTC at the Roque de los Muchachos Observatory in La Palma. This data set is presented in this work (see Section~\ref{data:gtc}).
\item {\it Galaxies at high redshift ($z > 0.5$):}
Since we want to increase the sample with galaxies at redshift higher than 0.5, we have compiled spectroscopy of SNe~Ia host galaxies from different sources. The selection of the sample and the obtained spectra are described in Section~\ref{data:highz}. 
\end{enumerate}

\begin{figure*}
\includegraphics[width=\textwidth]{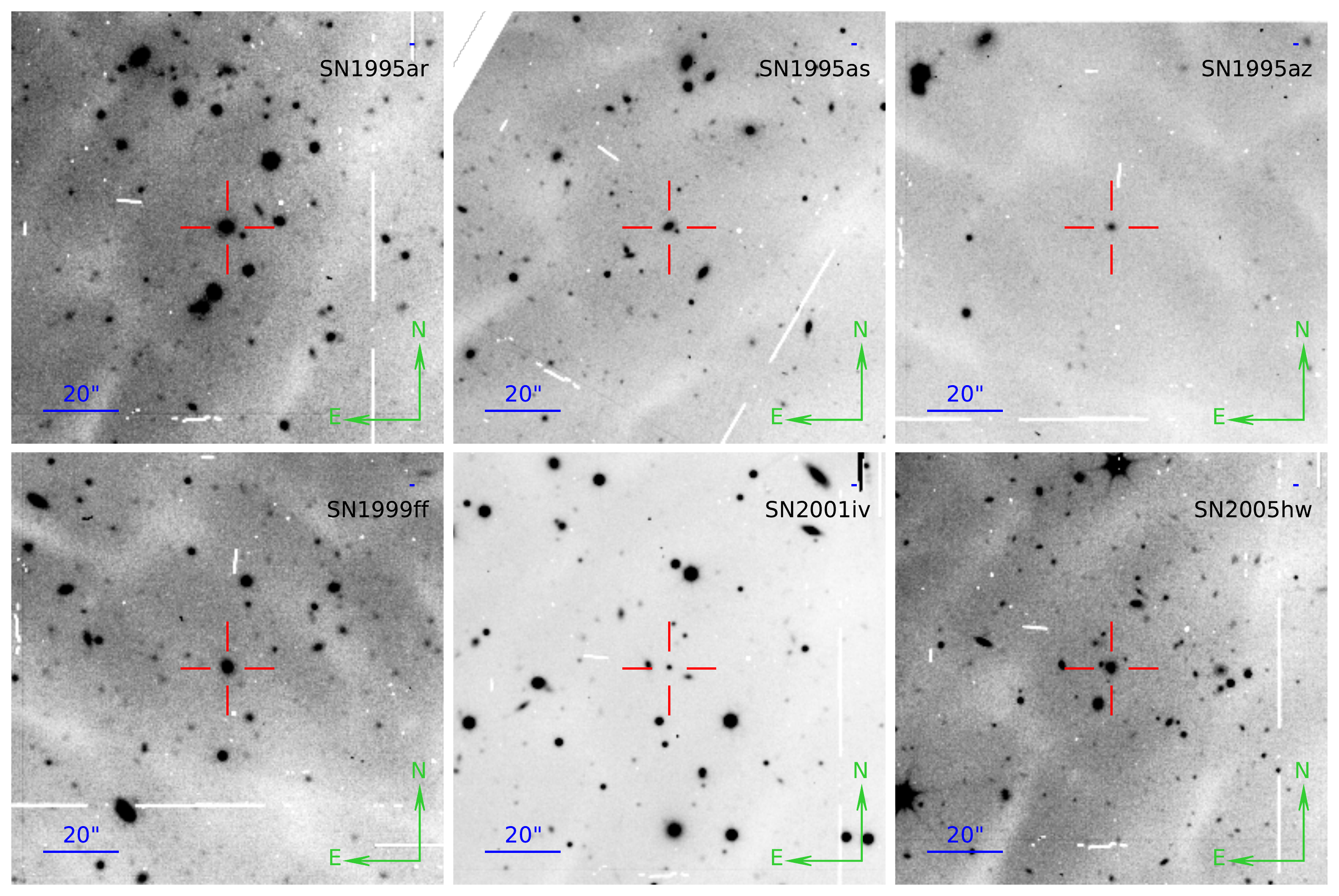}
\caption{\textit{From left to right and top to bottom}, GTC/OSIRIS acquisition images (Sloan-\textit{r}) for the six galaxies at $z\sim 0.45$, chronologically sorted by date of discovery. The field of view of each image is 2x2 arcmin$^2$, (north is up, east is left). Targets are centred in the red cross, and the blue line spans 20 arcsec.}
\label{fig:GTCacq}
\end{figure*}

This paper intends to wrap up previous work at intermediate redshift, and go one step further by extending the sample at higher redshift. On other hand, we have change the technique used in our previous works, studying in this case the stellar populations, instead the emission lines, of the full sample as a whole and the impact of the results on the HD. 
First, we present the data described above in Section ~\ref{sec:data}, 
and the selection criteria applied that resulted in a the final sample used in this work.
Next, we present in Section ~\ref{sec:fado}  the {\sc{fado}} code \citep{fado}, the used technique to extract information of the spectra, based in evolutionary population synthesis, as in \citet{Gallagher+2008} and \citet{Kang+2020}. We obtain information on stellar populations that form each galaxy, such as the star formation and enrichment histories, but also the averaged values of age and metallicity for each one as well the created stellar mass. These results are presented in Section ~\ref{sec:results}. Finally, these results allow to look for any possible dependence of the SNe~Ia brightness on the resulting stellar metallicities and ages. An analysis of the HR is done, by using simultaneously the stellar population information and the 
SNe~Ia data, in Section ~\ref{sec: HR}. Our conclusions are given in Section ~\ref{sec:conclusions}. 

\section{Observational data}
\label{sec:data}

We compiled spectra of SN Ia host galaxies from different sources focusing on different redshift ranges. They are presented in the following subsections. 

\subsection{SDSS galaxies at low and intermediate redshift ($z \leq 0.4$)}
\label{data:sdss}

Our intermediate redshift SN Ia sample is provided by the SDSS-II/SNe in their Data Release (DR, \citealt{Sako+2018}). 
It consists of 1066 SNe Ia confirmed either spectroscopically (352 spec-Ia) or photometrically, identified based on a Bayesian LC fitting using the spectroscopic redshift of their host galaxy (714 photo-Ia, \citealt{2014AJ....147...75O,Sako+2018}), and with publicly available host galaxy spectra in the 16th data release of the SDSS \citep{2020ApJS..249....3A}.
The sample is similar to the one presented in \citet{2018MNRAS.476..307M}, where oxygen abundances of these galaxies were obtained and the dependence of the HD residuals on these abundances was studied, and in \cite{galbany22}, where we performed aperture corrections to several host galaxy parameters.
In the present work, we have excluded those spectra with signal-to-noise ratio S/N $\le 3$, and instead of measuring oxygen abundances, we are going to study their stellar populations.
This means we will obtain the evolution of their star formation rate and stellar metallicities, and also estimate the averaged stellar ages and metallicities, along with the total mass ever created and in current stars. 

\begin{figure*}
\includegraphics[width=\textwidth]{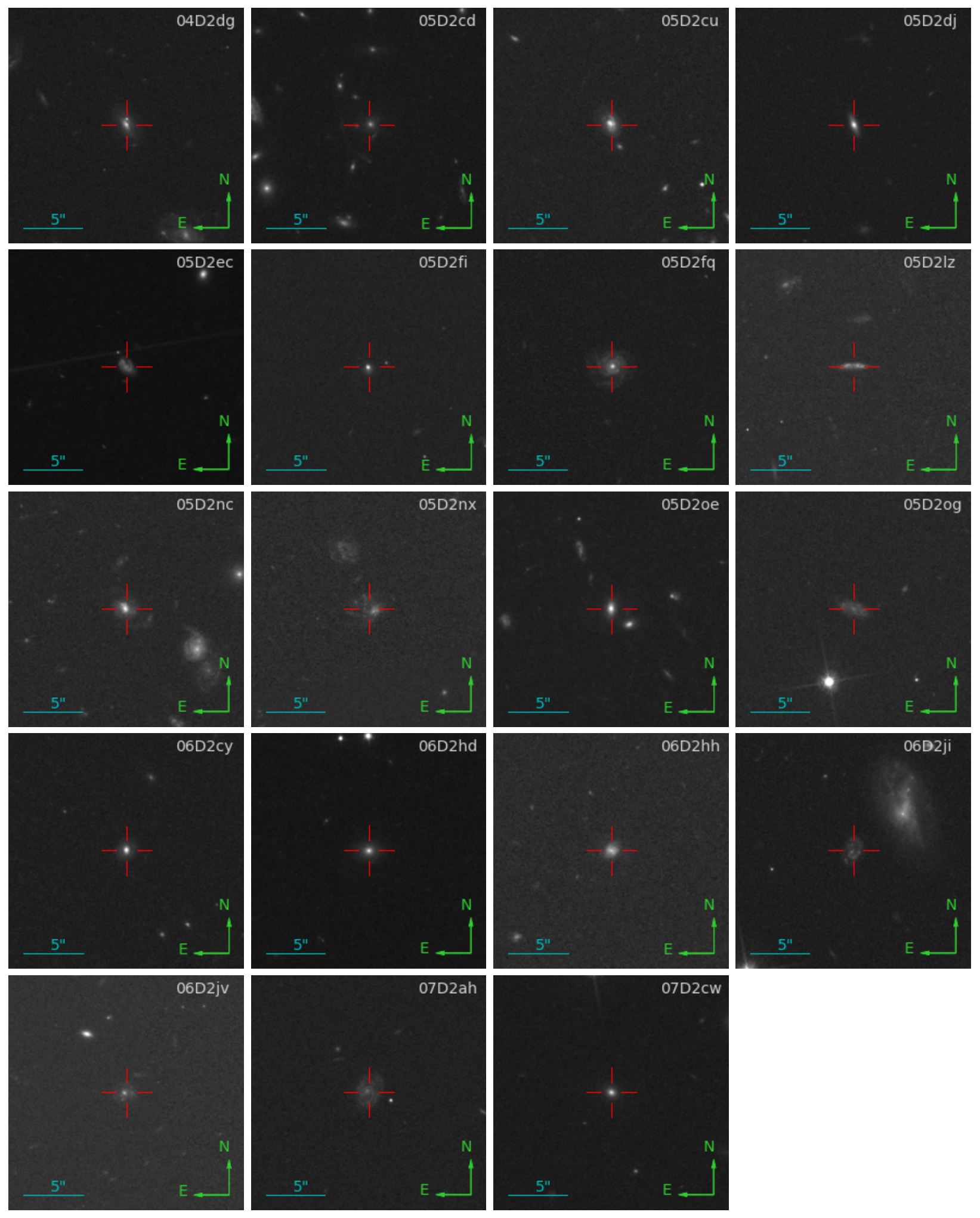}
\caption{\textit{From left to right and top to bottom}, HST-COSMOS acquisition images for the nineteen selected galaxies at  $0.5 < z \le 1$ ordered by SNe name (from SNLS). The field of view of each image is 20$\times$20 arcsec$^2$ (North is up, East is left). SNe are centred in the red cross, and the blue bar spans 5 arcsec.} 
\label{fig:COSMOSacq}
\end{figure*}


\begin{table}
\caption{Data of the observed galaxies chronologically sorted by date of discovery. SN Ia name in (1), equatorial coordinates RA and DEC in (2), and (3) redshift $z$ in (4), the observation date in (5), the exposure time in (6) and the signal-to-noise ratio S/N in (7).}
\resizebox{8cm}{!}{
\centering
\begin{tabular}{lcccccc}
SN-Ia name & RA & DEC & $z$ & Date & Exposure & S/N\\ 
        &        &    &          &      &  (s) & \\
\hline 
(1)  & (2) & (3) & (4) & (5) & (6) & (7) \\
\hline
SN~1995ar & $01^h01^m21^s$ & $\ang{04;18;34}$ & 0.465 & 14.11.2014 & 4x900 &  16.08\\
SN~1995as & $01^h01^m35^s$ & $\ang{04;26;14}$ & 0.498 & 14.11.2014 & 4x1800 &  10.42\\
SN~1995az & $04^h40^m33^s$ & $\ang{-05;26;27}$ & 0.450 & 12.01.2015 & 4x1800 & 7.63\\
SN~1999ff & $02^h33^m54^s$ & $\ang{00;36;24}$ & 0.455 & 16.11.2014 & 2x1800 & 15.28\\
SN~2001iv & $07^h50^m13^s$ & $\ang{10;17;19}$ & 0.397 & 22.01.2015 & 4x1800 & 16.05\\
SN~2005hw & $00^h09^m43^s$ & $\ang{01;09;16}$ & 0.408 & 13.11.2014 & 2x1800 &  38.33\\ 
\hline
\end{tabular}
\label{tab:data}
}
\end{table}
\normalsize


\subsection{Galaxies at intermediate redshift ($z \in 0.4-0.5$)}
\label{data:gtc}

We selected 28 SN Ia host galaxies from the sample of Union2.1 \citep{suzuki} at redshifts around $ z \approx 0.45 $, right above the maximum redshift of the distribution for our SDSS sample in \citet{2018MNRAS.476..307M} and \cite{galbany22},  to measure their oxygen abundances. 
At this redshift range, the main emission lines used in the most common metallicity calibrators, as 
[\ion{O}{iii}]$\lambda\lambda$\,4959,5007\,\AA, [\ion{N}{ii}]$\lambda$6584\,\AA\ and H$\beta$ and H$\alpha$, are
still at wavelengths covered by typical optical spectrographs, with $\lambda\le 9800$\,\AA\ \footnote{ H$\beta$ will be in 7048.45 and 7291.50\,\AA, for $z=0.45$ and 0.5, respectively;  while H$\alpha$ and [N{\sc ii}]$\lambda$\,6584\AA\ will fall in  9516.35\,\AA\ and 9546.8\,\AA\ for $z=0.45$.}. 
From our initial sample, only 6 galaxies were finally observed with the GTC, whose data is presented here.

\begin{figure*}
\hspace{-1.2cm}
\includegraphics[width=0.95\textwidth]{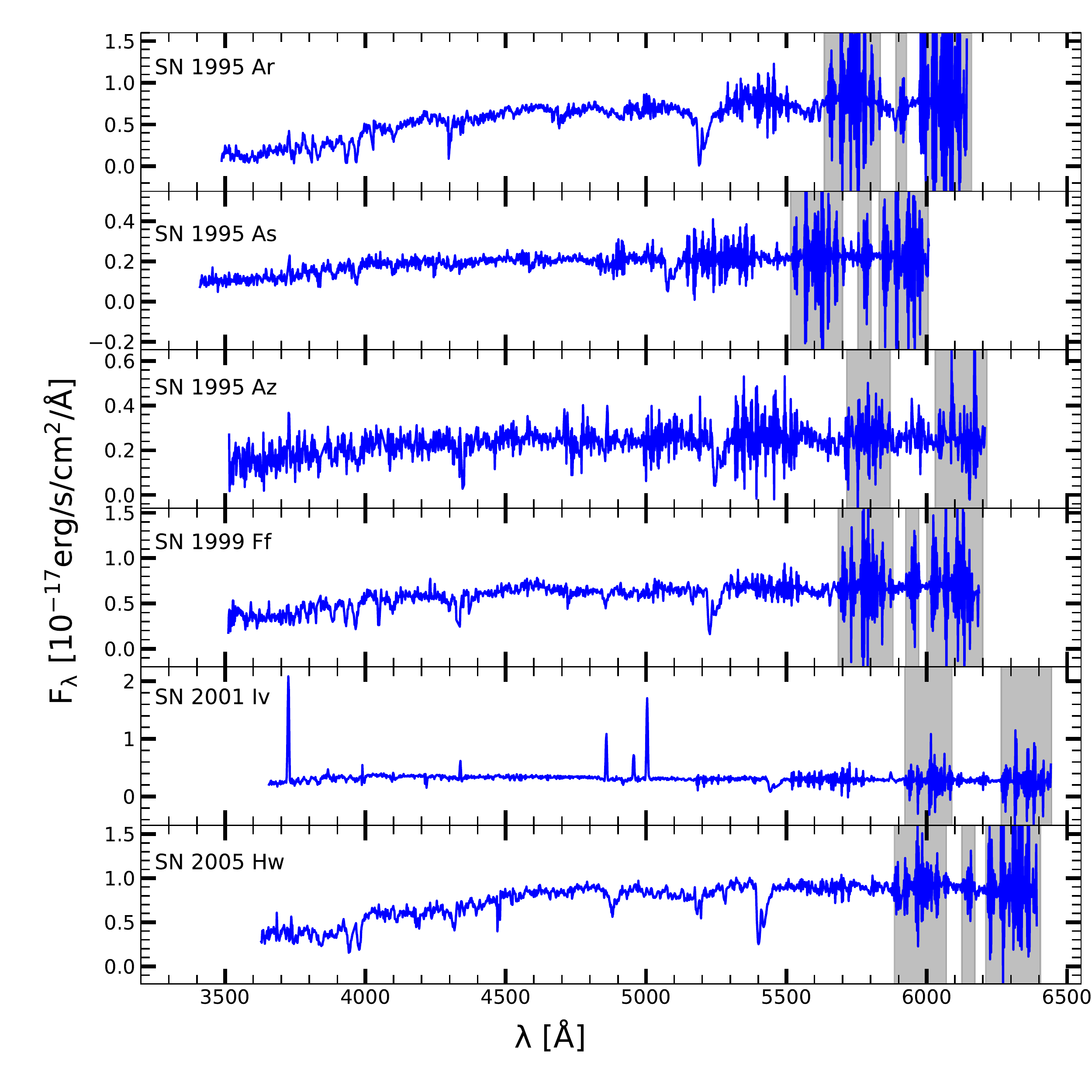}
\vspace{-0.7cm}
\caption{Spectra for the six galaxies observed with GTC-OSIRIS after the reduction processes. In each one, top left, the name of the SN-Ia hosted by the galaxy is indicated. 
} 
\label{fig:spectra-gtc}
\end{figure*}

Spectra of these six galaxies were taken with GTC/OSIRIS \citep[Optical System for Imaging and low-Intermediate-Resolution Integrated Spectroscopy, ][]{Cepa2005,Cepa2010} 
in long-slit mode, between November 2014 and January 2015 under the GTC57-14B programme (PI:Moreno-Raya). We set the width of the slit to 0.8 arcsec and used the R1000R grating, which provides spectral coverage in the range from 5100 to 10000\,\AA~ (although the efficiency drops down drastically beyond 9000 \,\AA), with a resolving power of $R\sim 1100$. The coordinates, redshift, date of observation and exposure time for these six SN-Ia host galaxies are summarised in Table ~\ref{tab:data}. The S/N per pixel unit of the 6 galaxies is also included as column 7 in Table~\ref{tab:data}. The acquisition images of the observed galaxies are shown in Figure~\ref{fig:GTCacq}, with the target of interest in the centre of the red cross.

The data were reduced using the IRAF software \citep{iraf}. 
We used the standard packages available for the bias subtraction, flat field correction and cosmic ray rejection and wavelength and flux calibration using the same procedure as in \cite{mr-2016b}. Since the sky lines presented a curved shape, the construction of an accurate 2D map which transform pixel coordinates into wavelength values is needed. Using the HgXe arc spectra to calibrate in wavelength, we mapped the pixel-to-wavelength distortions across the CCD, and applied the same distortion correction to our science spectra. After the wavelength calibration we extracted the 1D spectrum of every object and removed the sky contribution. Then, we corrected our spectra from the instrument sensibility and performed the flux calibration using standard stars.

Due to the position and distance of the GTC studied galaxies, it is necessary to apply two additional corrections on their spectra before being analyzed.
One of them is the extinction effect, caused by the dust of our galaxy located in the observation line. This correction is performed by the \textsc{deredden} task together with the tabulated values for the galactic extinction map available in \citet{MapaExti}.
The other is the correction by redshift on the wavelengths. With the data in the Table~\ref{tab:data} and by mean of  the \textsc{dopcor} task, this effect is corrected and the spectrum emitted by the galaxy is obtained. 
After the described corrections have been done, we have the final spectra for the six galaxies, as shown in Figure~\ref{fig:spectra-gtc} .
There, we show the calibrated spectra in rest-frame. Although we try to optimise the reduction, there are still present some high residuals of the sky lines, which we mask (shaded grey bands) for the analysis that we will do in next subsection. For our initial goal, to estimate the oxygen abundance from the emission lines, only the SN~2001iv host galaxy were valid, showing emission lines in its spectra. However, to study the stellar populations of these galaxies and to estimate the stellar metallicity (instead the gas abundance of oxygen), as we will present in the following section, all these spectra are useful since they have the required $S/N > 3$ (see last column of Table~\ref{tab:data}). 

\subsection{Galaxies at high redshift ($z\geq 0.5$)}
\label{data:highz}
The sample of SNe Ia for the high-$z$ range ($0.5 \le z \le 1.0$) was taken from the Supernova Legacy Survey\footnote{\url{https://www.cadc-ccda.hia-iha.nrc-cnrc.gc.ca/en/cfht/cfhtls.html}} 
(SNLS, \citealt{Guy+2010,2014A&A...568A..22B}), a project that used the MegaCam instrument at the Canada-France-Hawaii Telescope (CFHT) to discover and follow up SNe, and spectroscopic observations from big ground-based telescopes (Gemini North and South, VLT, and Keck 1 and 2) to classify the SN candidates and estimate their redshift. SNLS targeted four regions of $1^\circ \times 1^\circ$ in the sky, named from D1 to D4. 

The SNLS D2 region overlapped with the COSMOS extragalactic survey \citep{COSMOS} footprint\footnote{\url{https://irsa.ipac.caltech.edu/data/COSMOS/tables/spectra/}}.
COSMOS consisted of a multi-wavelength imaging and spectroscopy (from radio to X-rays) of a region of the sky of $2^\circ \times 2^\circ$ centered at the J2000 coordinates (RA,Dec)=(+150.1191, +2.2058), using a multitude of telescopes. 
In particular, for our study we took spectra from the COSMOS sub-surveys {\it Magellan}
\citep[][using IMACS at the Baade telescope]{magellan} and {\it zCOSMOS}, \citep[][using VIMOS at the ESO-VLT]{zcosmos}. 
Other sub-surveys of COSMOS were not used for different reasons: {\it VUDS} targeted higher redshift objects than those of our interest ($2 < z < 6$), {\it IRS} is composed of mid-infrared spectra (not suitable for the analysis with {\sc fado}), and the spectra of {\it DEIMOS} are not flux-calibrated. Other big surveys were also considered ({\it DEEP2}, that coincides with D3 region of SNLS, and {\it VVDS}, partially in regions D1 and D4 of SNLS), but also discarded for the unavailability of flux-calibrated spectra. 

In the process of looking for pairs galaxy-SN, the coordinates of objects with spectra were crossmatched with the SNLS SNe~Ia coordinates, setting as selection criteria (i) a physical projected distance of 50\,kpc between the SN position and the center of the galaxy, and (ii) a redshift relative error of 10\%. With these first two criteria, from the initial sample of galaxies of 11981 (10643 from zCOSMOS and 1338 from Magellan) we found 44 matches (39 from zCOSMOS and 5 from Magellan), 3 of them being galaxies repeated in both of the COSMOS surveys. A visual inspection with Hubble Space Telescope (HST) $I$-band images was then made for the resulting matches to make sure that the galaxy was the real host of the SN, discarding the galaxies were the SN was obviously located in a different galaxy. A total of 28 galaxy-SN pairs was finally considered for further analysis.

\begin{figure*}
\hspace{-1.5cm}
\includegraphics[width=0.98\textwidth]{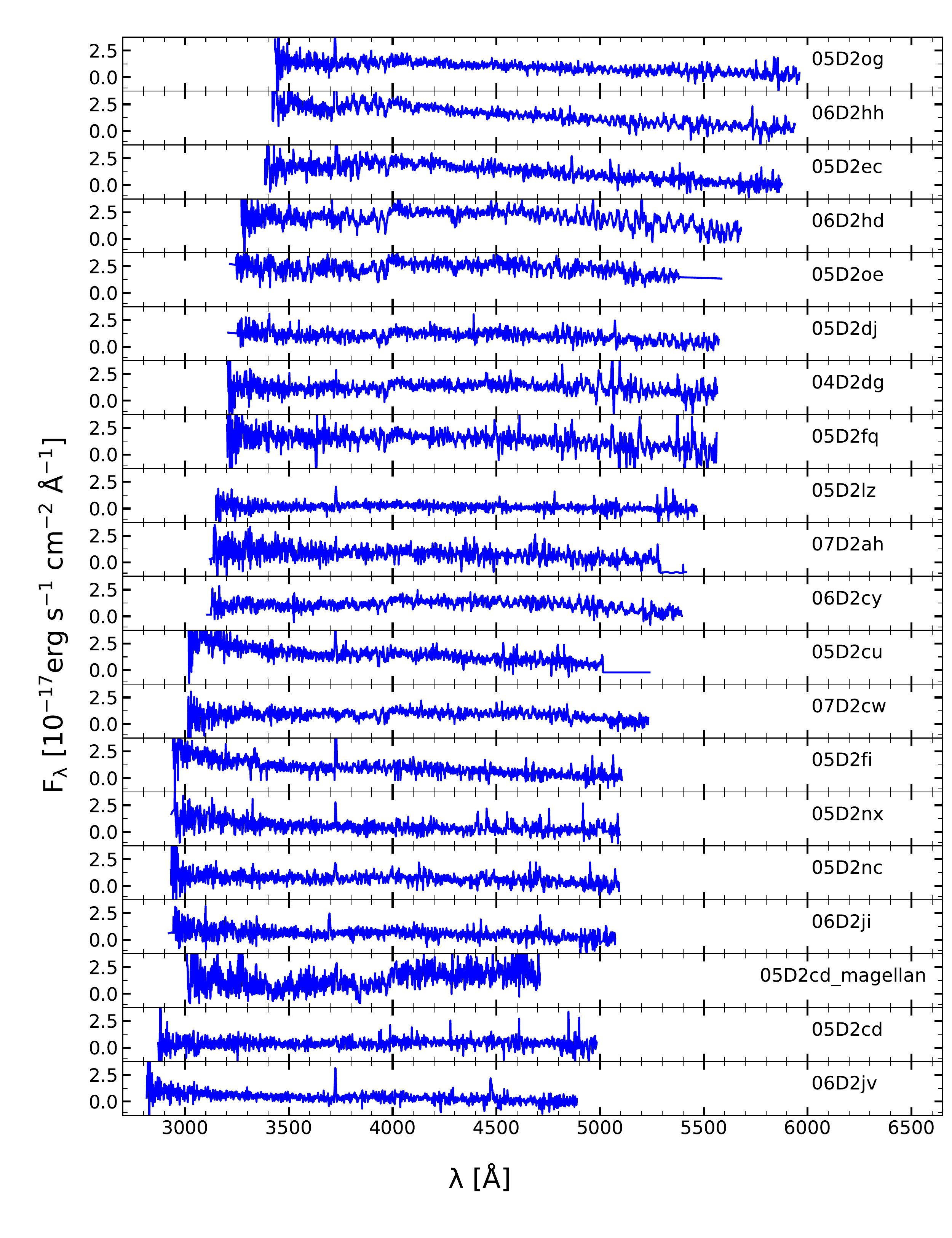}
\vspace{-1cm}
\caption{Spectra for the nineteen galaxies of zCOSMOS and MAGELLAN ordered by the redshift, from the top with $z=0.616$ until the bottom with $z=0.972$. In each one, the name of the SN-Ia hosted by the galaxy is indicated. 
}
\label{fig:spectra-cosmos}
\end{figure*}

The spectra of the 20 galaxies finally selected (see Section~\ref{subsec: subsample}) from these 28 galaxies is shown in Figure~\ref{fig:spectra-cosmos} sorted by the redshift from the top with $z=0.616$ until the bottom with $z=0.972$. In each one, the name of the hosted SN~Ia is indicated.

%
\subsection{Selection of the sub-sample to analyze}
\label{subsec: subsample}

The complete sample consists of 1087 galaxies (1053 from SDSS, 6 from GTC-OSIRIS and 28 from COSMOS) with available spectra. 
After computing the S/N of the spectra, we keep 803 galaxies (769 from SDSS, 6 from GTC and 28 from COSMOS) that have spectra with $S/N>3$.
\begin{figure*}
\centering
\includegraphics[scale=0.3]{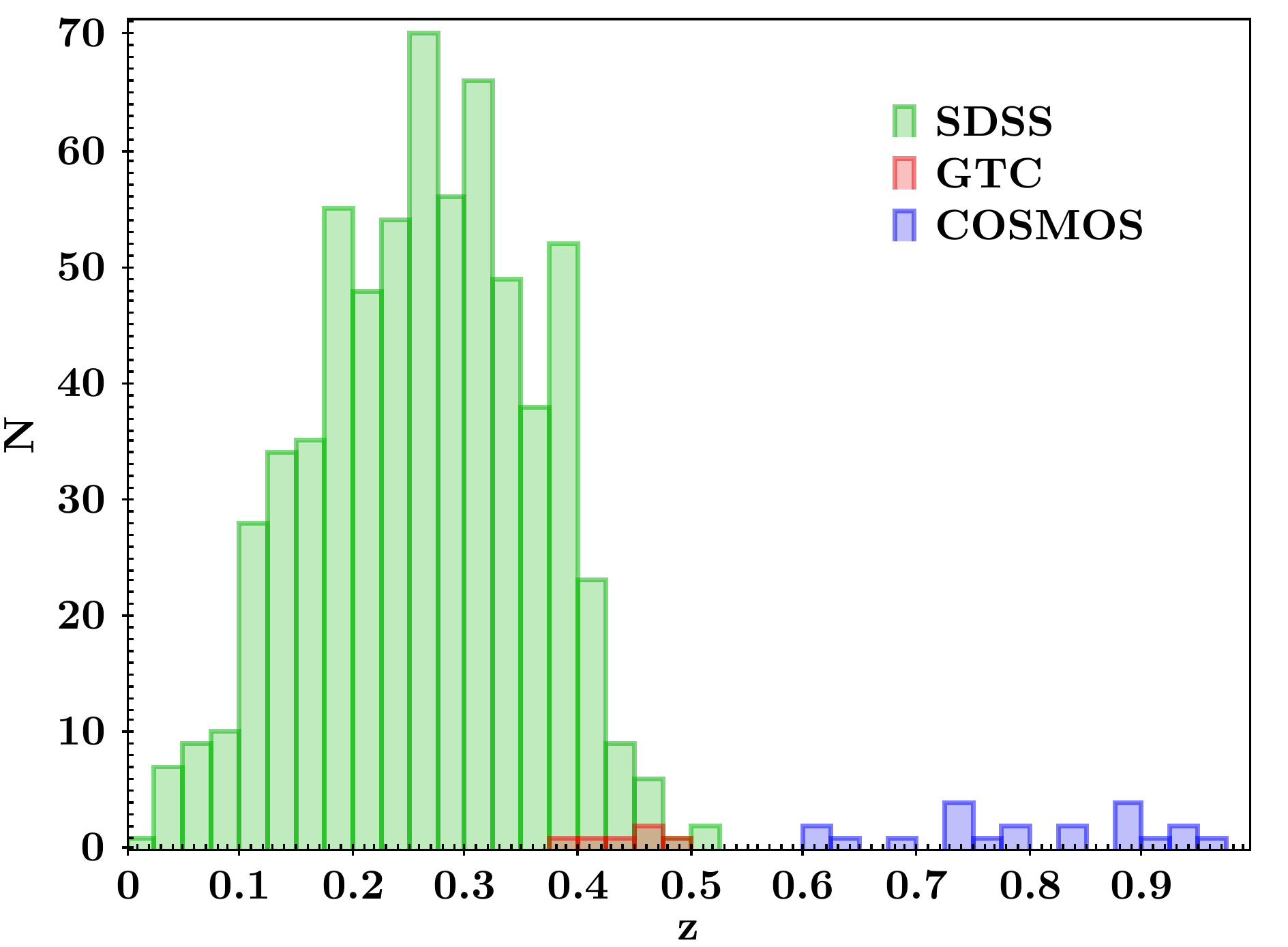}
\includegraphics[scale=0.3]{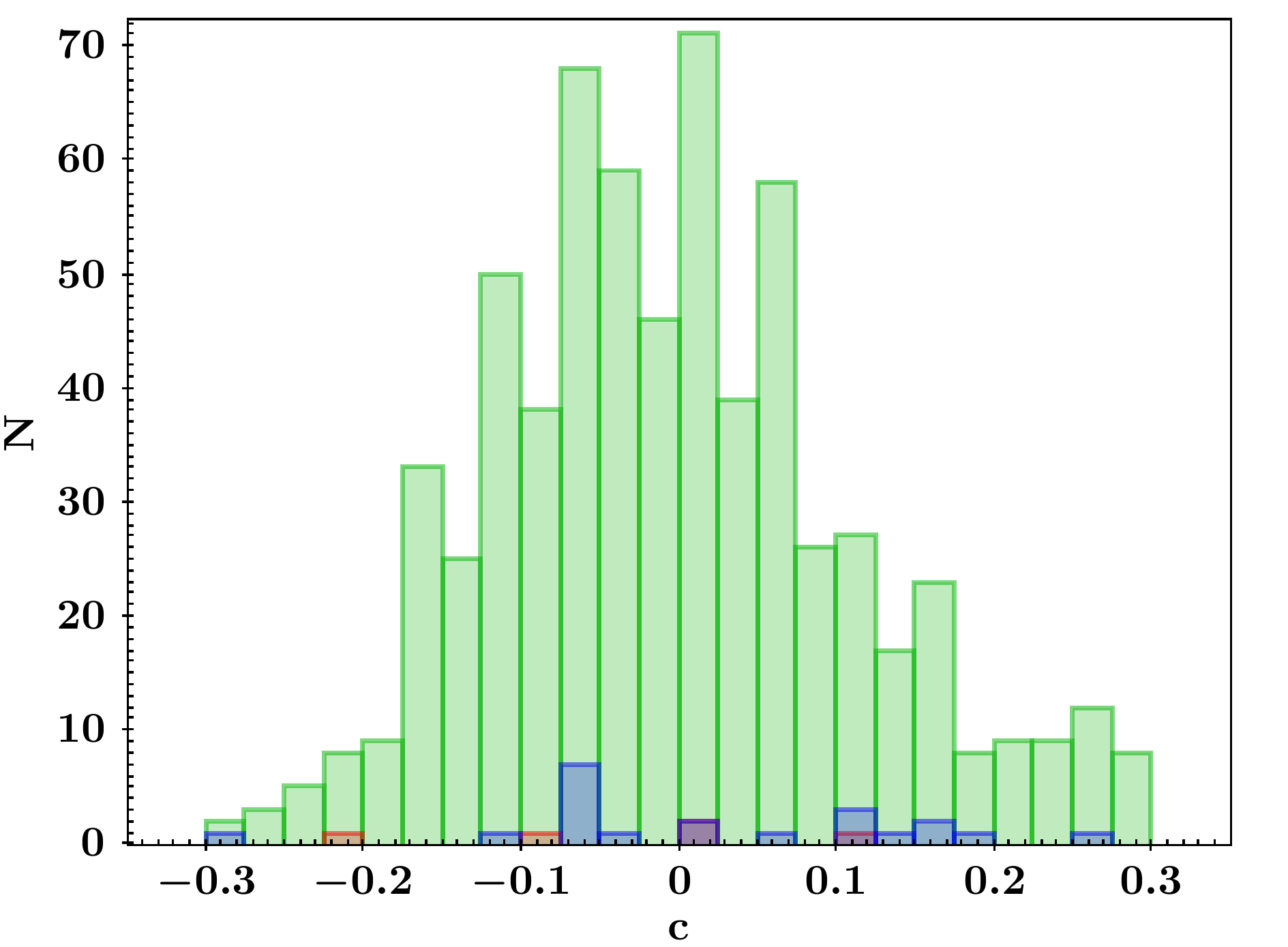}
\includegraphics[scale=0.3]{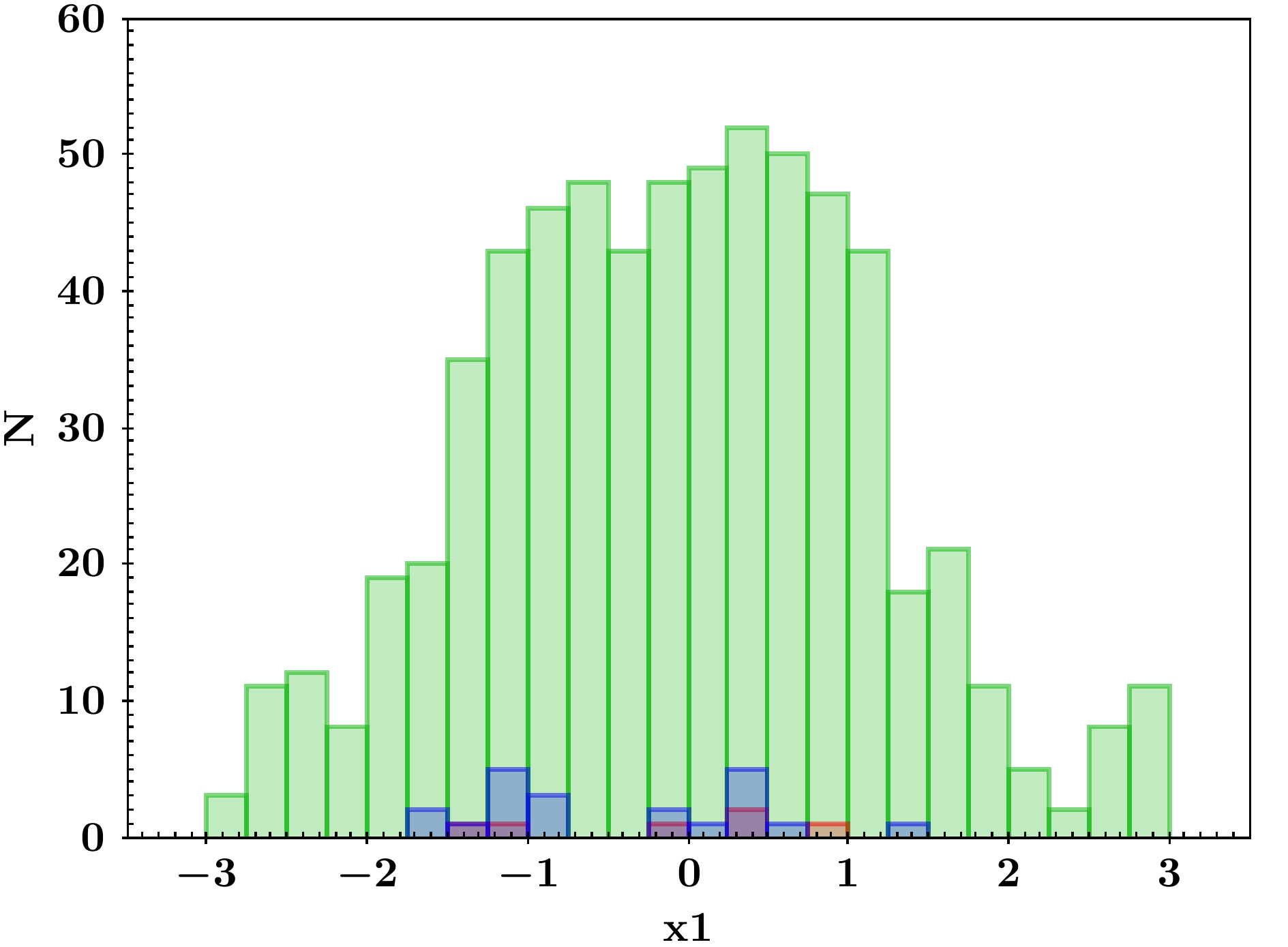}
\caption{The distribution of: left) the redshift of the three subsamples used in this work; middle) The color, $c$; and right) the stretch $x1$; for the complete sample of 680 galaxies used in this work. 
}
\label{fig:SN-information}
\end{figure*}
Additionally, we perform some cuts based on the SN~Ia light-curve properties. 
We take the peak apparent magnitude, as well as the color and the stretch from \cite{Sako+2018} for the SDSS subsample. These same information for the six SNe~Ia in galaxies observed with GTC was taken from Union2.1 \citep{suzuki}, while the information for the higher redshift subsample was taken from the the SNLS. 
We kept galaxies for which their SNe Ia have light-curve stretch in the range $-3 < x1 < 3$ and color in the range $-0.3 < c < 0.3$. These ranges of $x1$ and $c$ have been chosen, because the data of SNLS and SSDS-II/SNe were analysed using the SALT2 \citep{salt2,Guy+2010} empirical SNe Ia model, which, in turn, was trained in those ranges of stretch and color.
In that way, the final sample consists in 680 galaxies: 654 from SDSS, 6 from GTC and 20 from COSMOS, for which we have the complete information necessary for our analysis.
We show in Fig.~\ref{fig:SN-information} the distribution of redshift $z$ for the three subsamples, SDSS, GTC and COSMOS, and also the distribution of color, $c$, and stretch, $x1$, measured for SNe Ia of our sample.

\section{Analyzing galaxy spectra}
\label{sec:fado}

Once the spectra of the 680 galaxies has been selected\footnote{The set of used spectra is in  \url{https://github.com/HOSTFLOWS}}, we now proceed to their analysis. The final aim is to estimate their averaged stellar age and metallicity, but we will obtain as a sub-product their star formation and enrichment histories that we also present here.

To analyze the spectra we used {\sc fado} \citep[see][for further details]{fado}, a code that decomposes the spectrum of a galaxy in the stellar populations that the galaxy contains and estimates its characteristics such as mass, age, and metallicity. 
For that, {\sc fado} tries to find the best combination of stellar spectra that reproduces the spectrum of the galaxy using spectral catalogues of the so-called Single Stellar Populations (SSPs). \footnote{This technique is complementary to the evolutionary synthesis, based on the addition of SSP spectra in the proportion given by a SFH to fit the observed spectrum}. SSPs are sets of stars of the same age and metallicity whose spectra are calculated as the sum of spectra of individual stars that form them. A galaxy is made of several generations of stars, determined by its SFH and its evolution in metallicity, so that the final spectrum is made of various components or SSPs in different proportions. These proportions are just the results given by {\sc fado} from a spectrum using a set of SSP spectra called \emph{Basis}.

There are other codes similar to {\sc fado}, such as {\sc starlight} \citep{starlight,starlight2}, but in general they do not take into account the nebular emission in the spectrum adjustment, which is a problem when observed galaxies have high star formation (SF). However, {\sc fado} has the ability to simultaneously identify the SFH that best reproduces a galaxy and the properties of the observed nebular emission.
Neglecting the nebular emission can introduce a bias in the stellar continuum and in the scale relationships that involve magnitudes such as in Tully-Fisher, or relationships between luminosity and metallicity, diameter, or dispersion velocity. In addition, the nebular emission also affects the calculation of the mass according to the M/L ratio used.
{\sc fado} also includes other improvements such as the determination of the physical conditions of the gas (electronic temperature and density and extinction in the nebular component $A_V$), in order to give derived secondary products of the fit, besides of providing the basic components that fit a given spectrum, and the final spectroscopic classification of the galaxy.

The SSP basis are a very important part of the analysis and it is crucial to choose a set of basis that satisfies the following properties: enough spectral resolution, and best coverage possible in both ages and metallicities. Firstly, the spectra of the basis need to have at least the spectral resolving power (R) of the observed spectra. This is a requisite in order to not degrade the observed spectra and, thus, not lose part of the information of the spectra. Besides that, the age coverage is also a very important factor to consider. If the basis do not contain young populations (few Myr), the code would introduce an error in the determination of stellar characteristics, mainly, averaged age $<age>$, averaged metallicity $<Z>$, and in the properties of the nebular component, as well as the formed stellar mass. The coverage in metallicity is important too, because not having low and/or high metallicity models would affect the results of the analysis given the age-metallicity degeneracy. 

\begin{figure*}
\includegraphics[scale=0.2,angle=0]{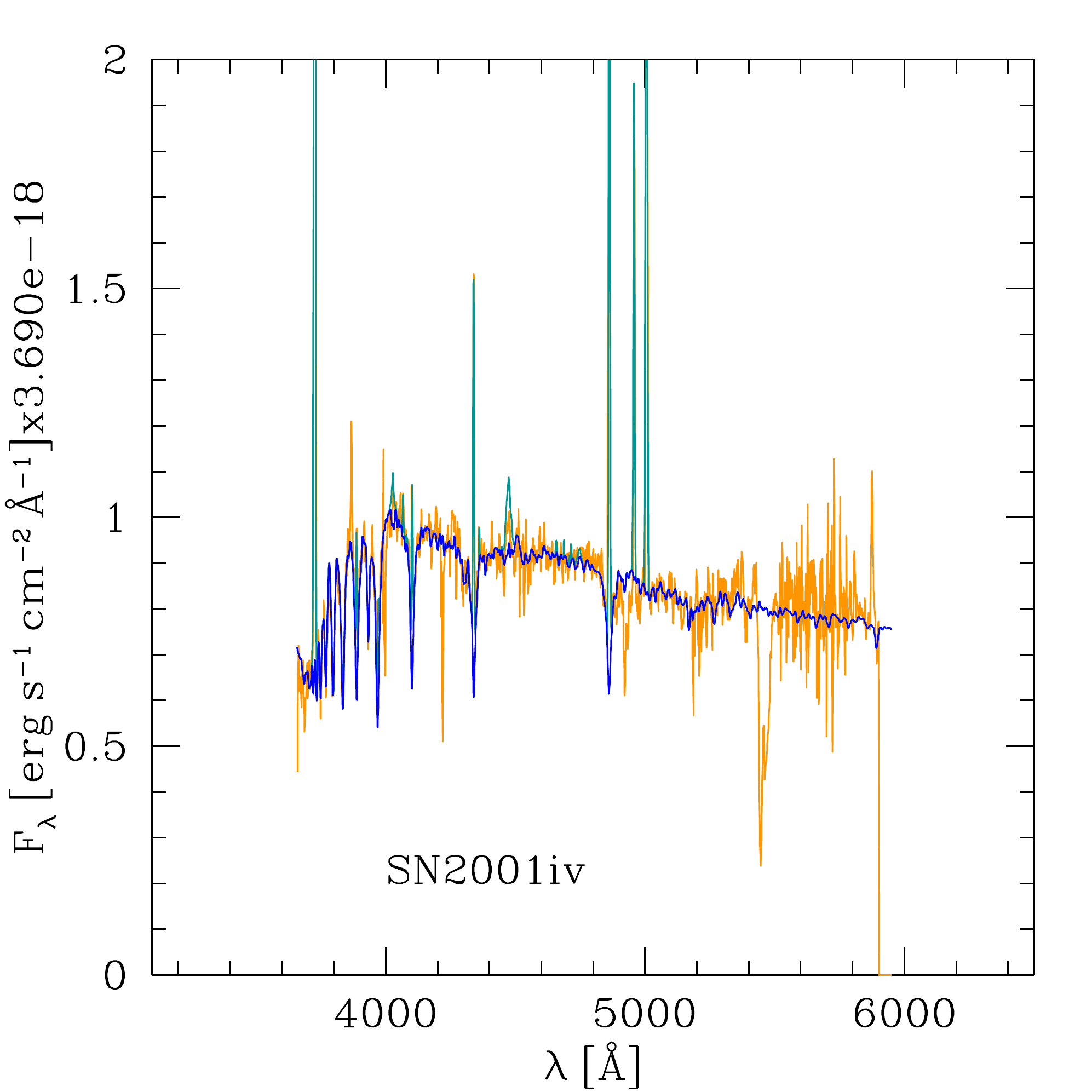}
\includegraphics[scale=0.2,angle=0]{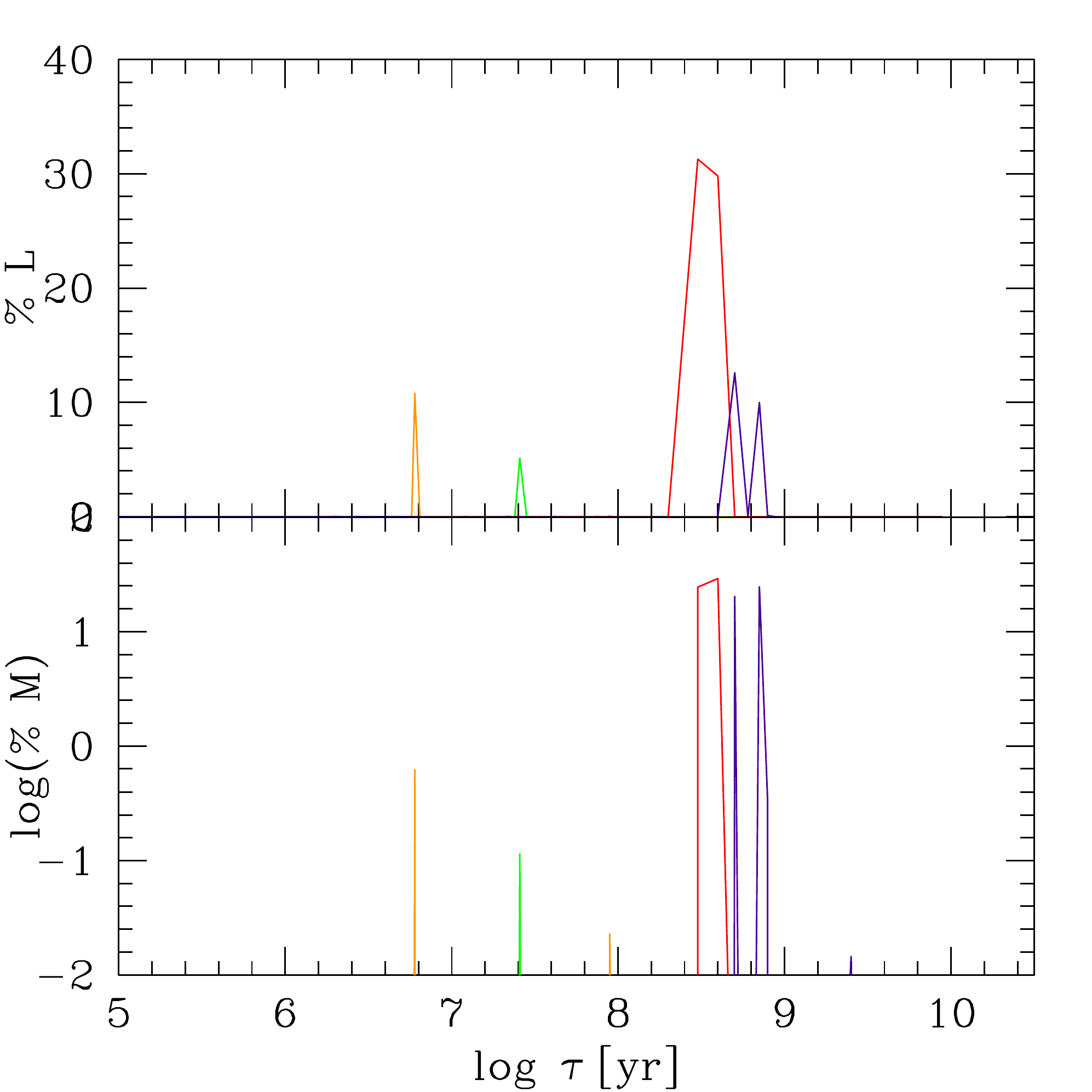}
\includegraphics[scale=0.2,angle=0]{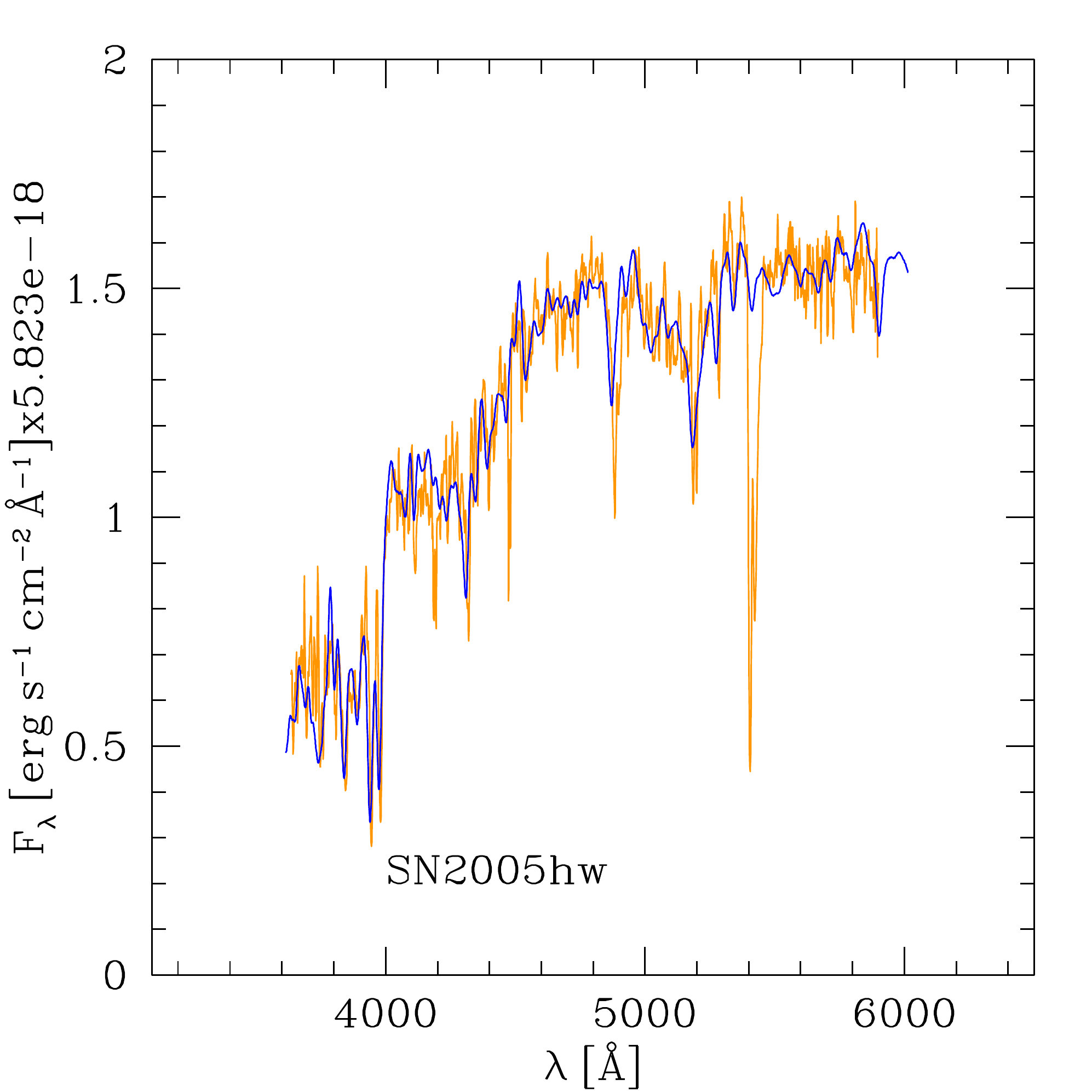}
\includegraphics[scale=0.2,angle=0]{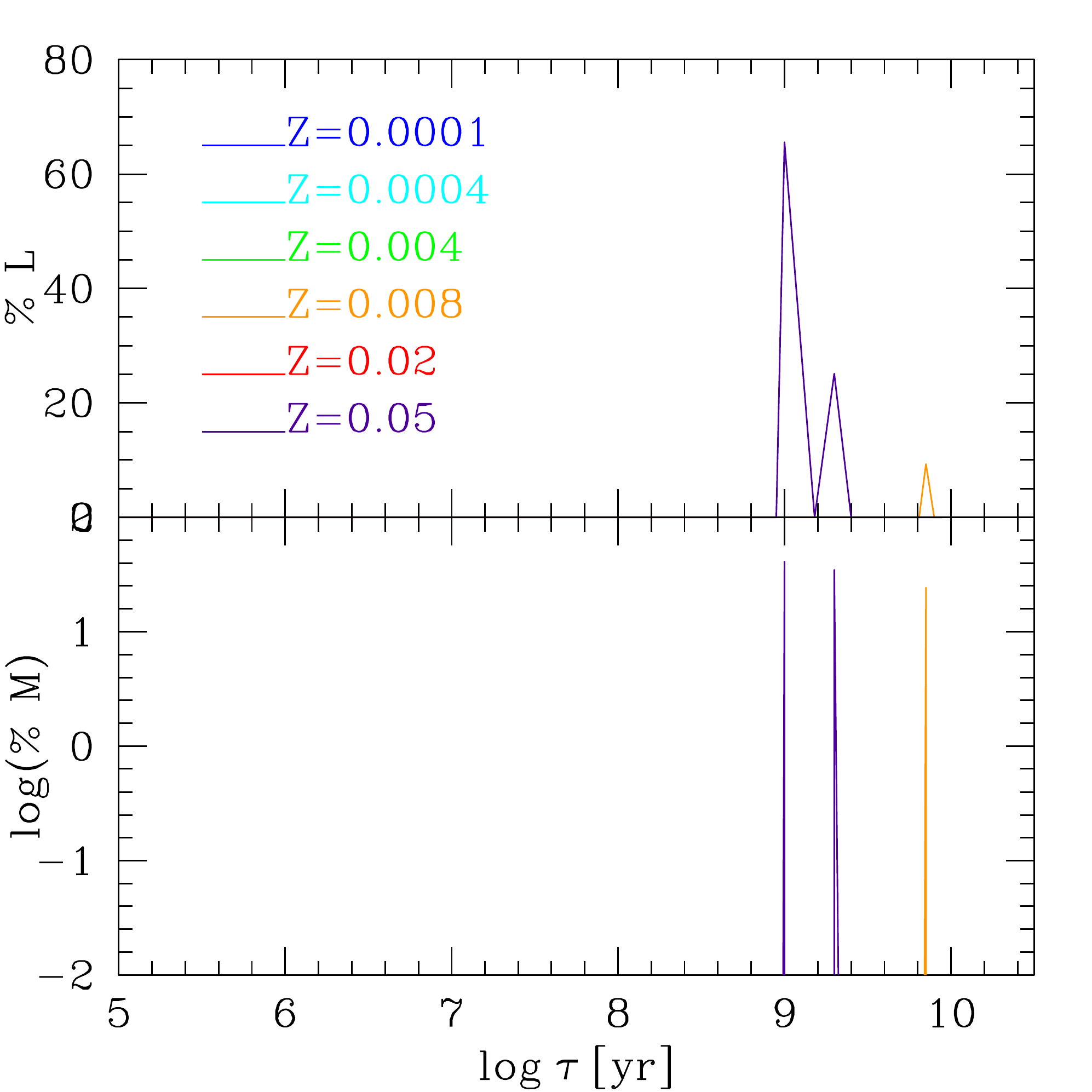}
\includegraphics[scale=0.2,angle=0]{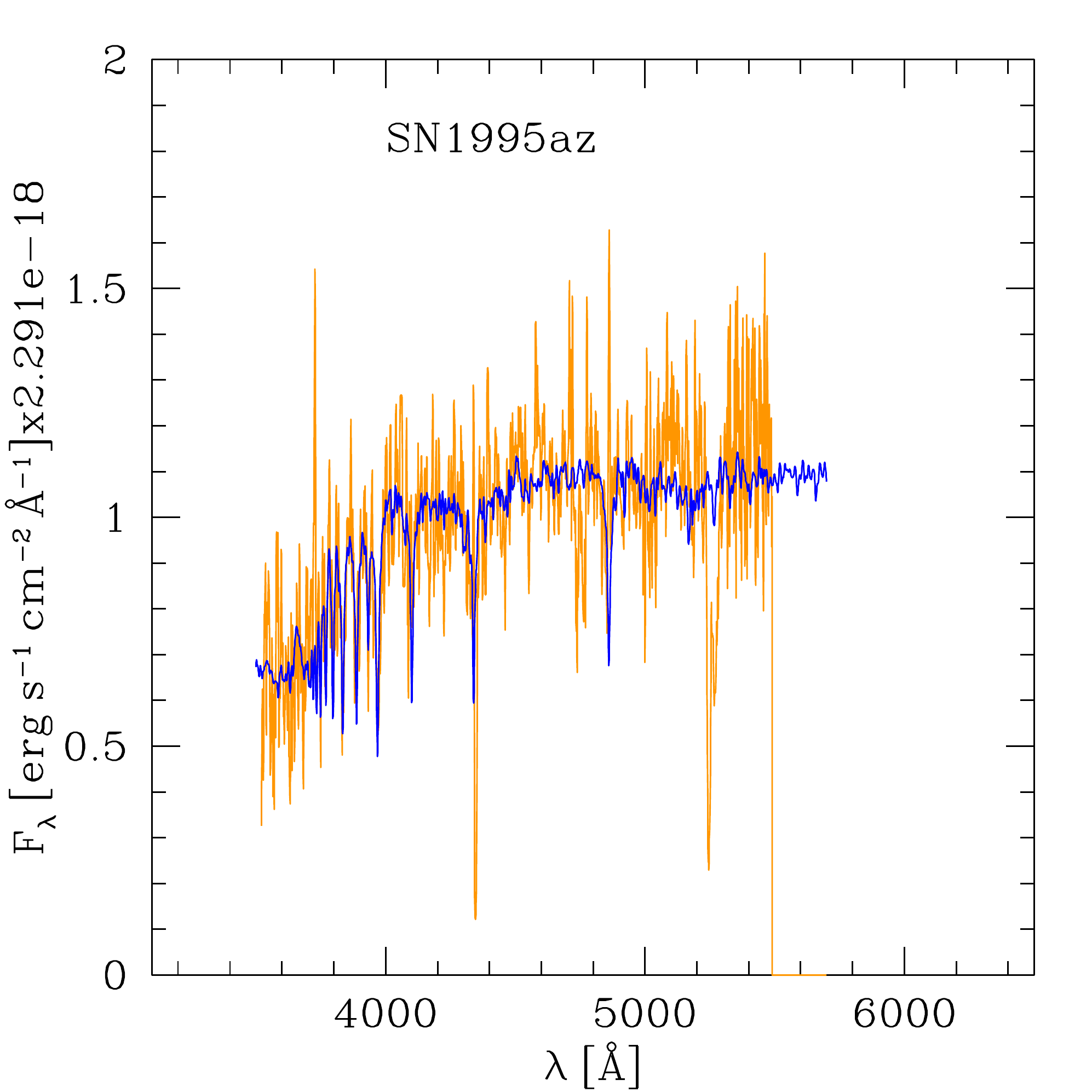}
\includegraphics[scale=0.2,angle=0]{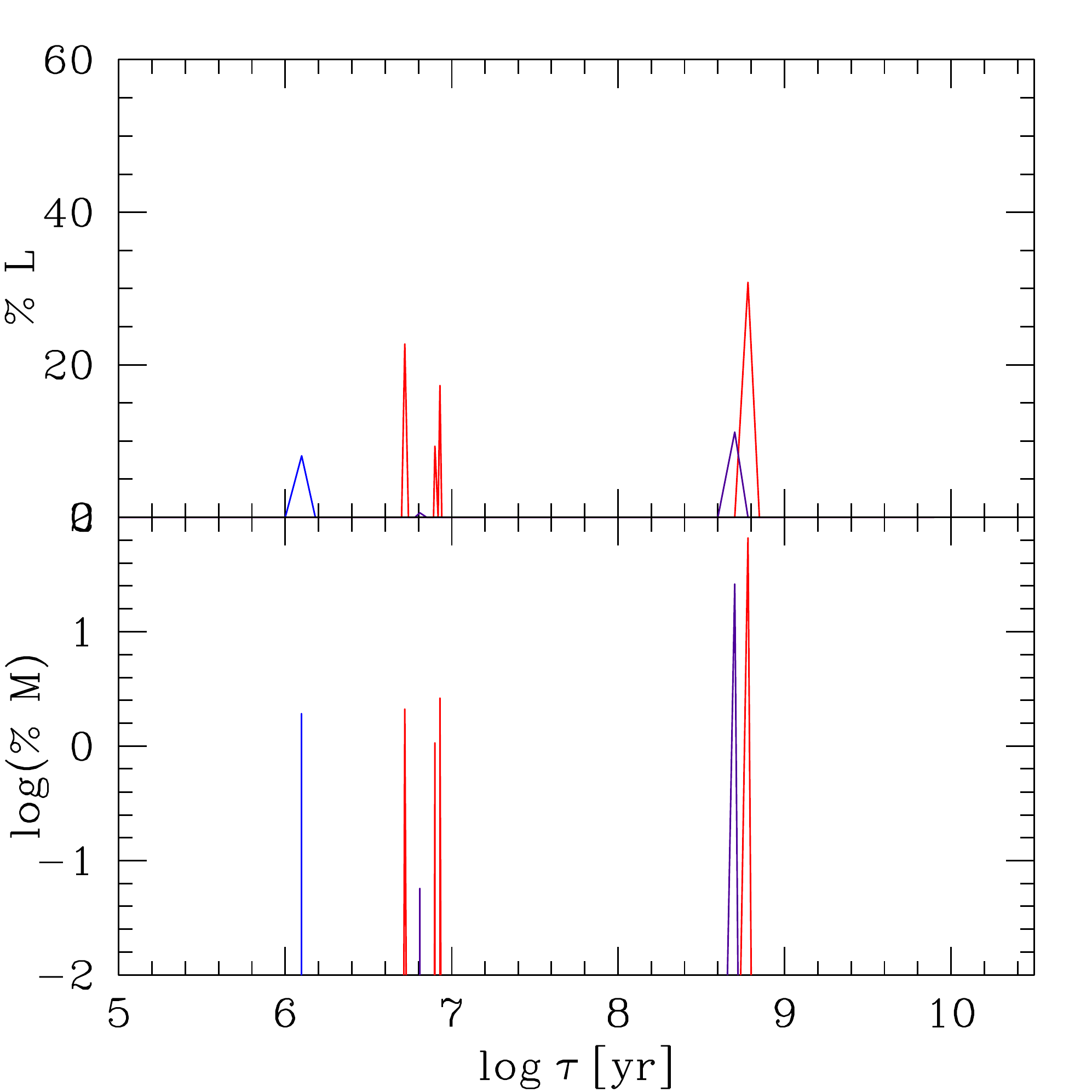}
\includegraphics[scale=0.2,angle=0]{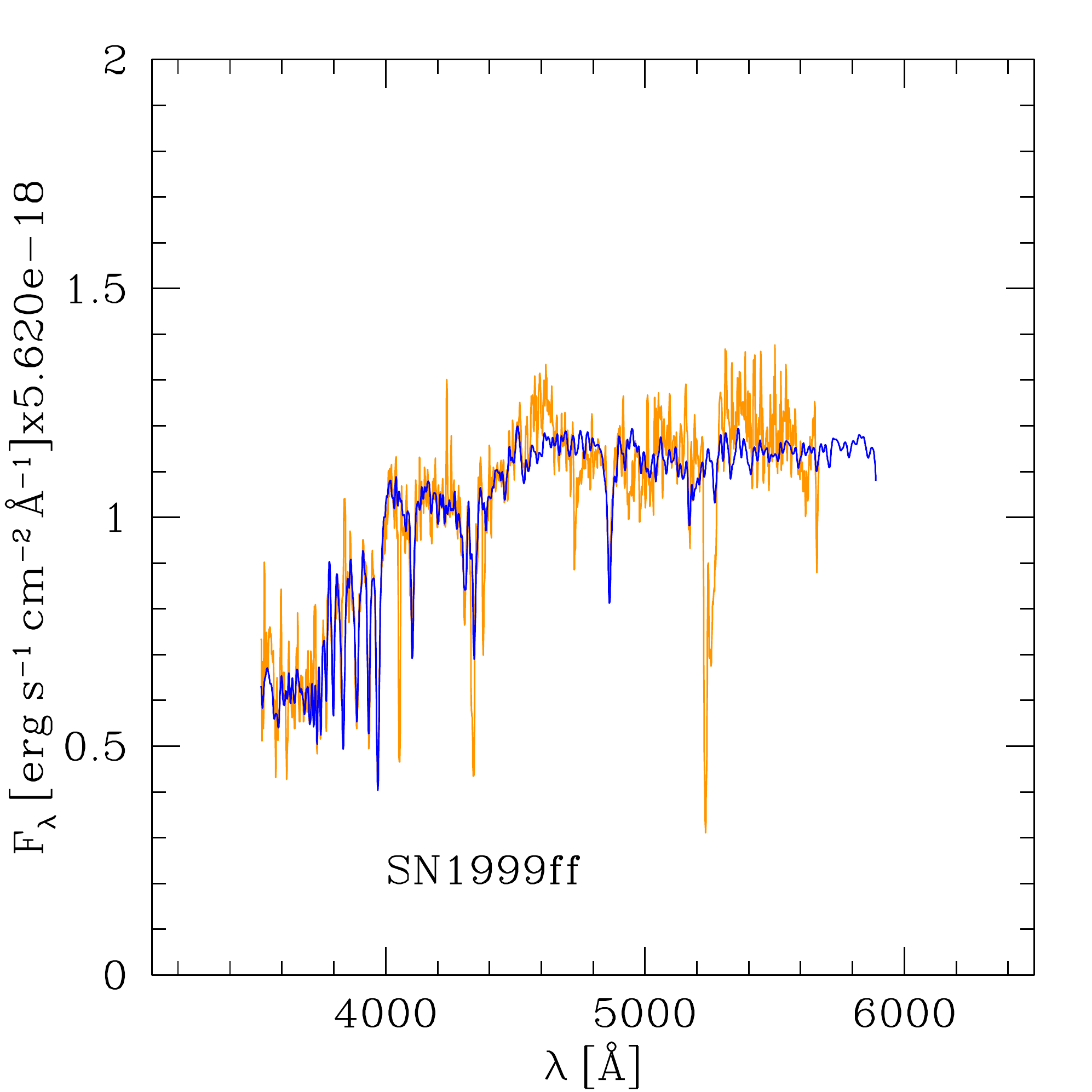}
\includegraphics[scale=0.2,angle=0]{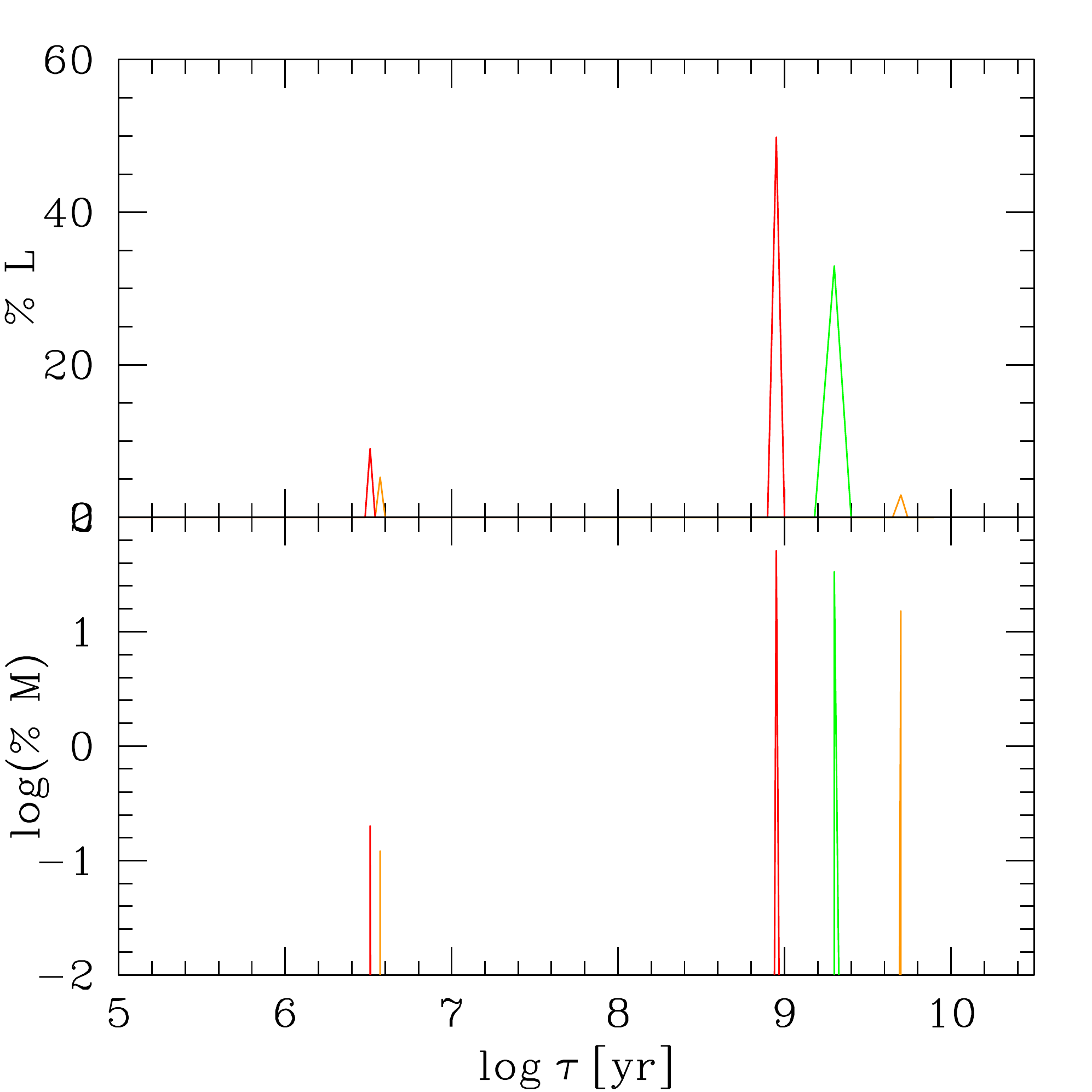}
\includegraphics[scale=0.2,angle=0]{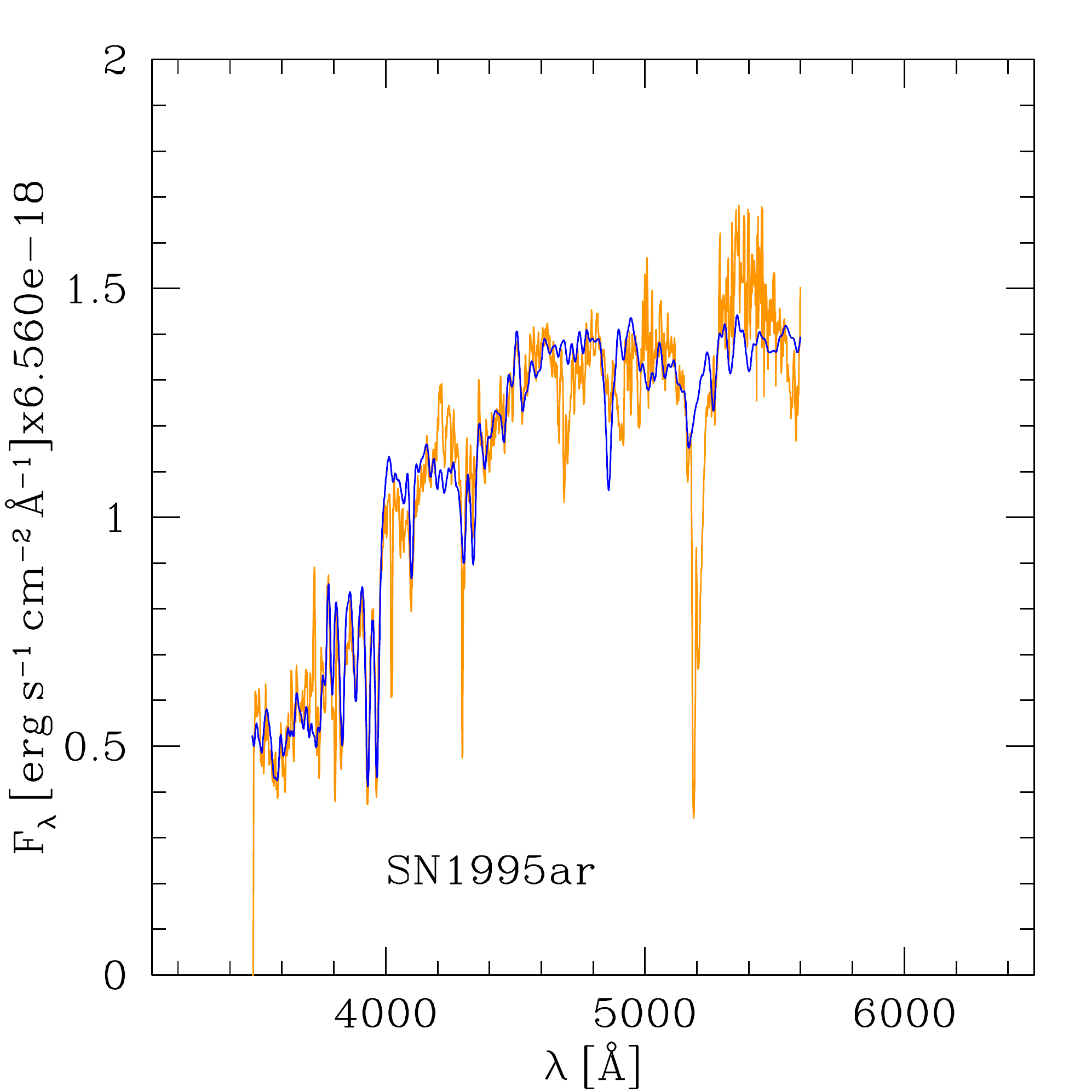}
\includegraphics[scale=0.2,angle=0]{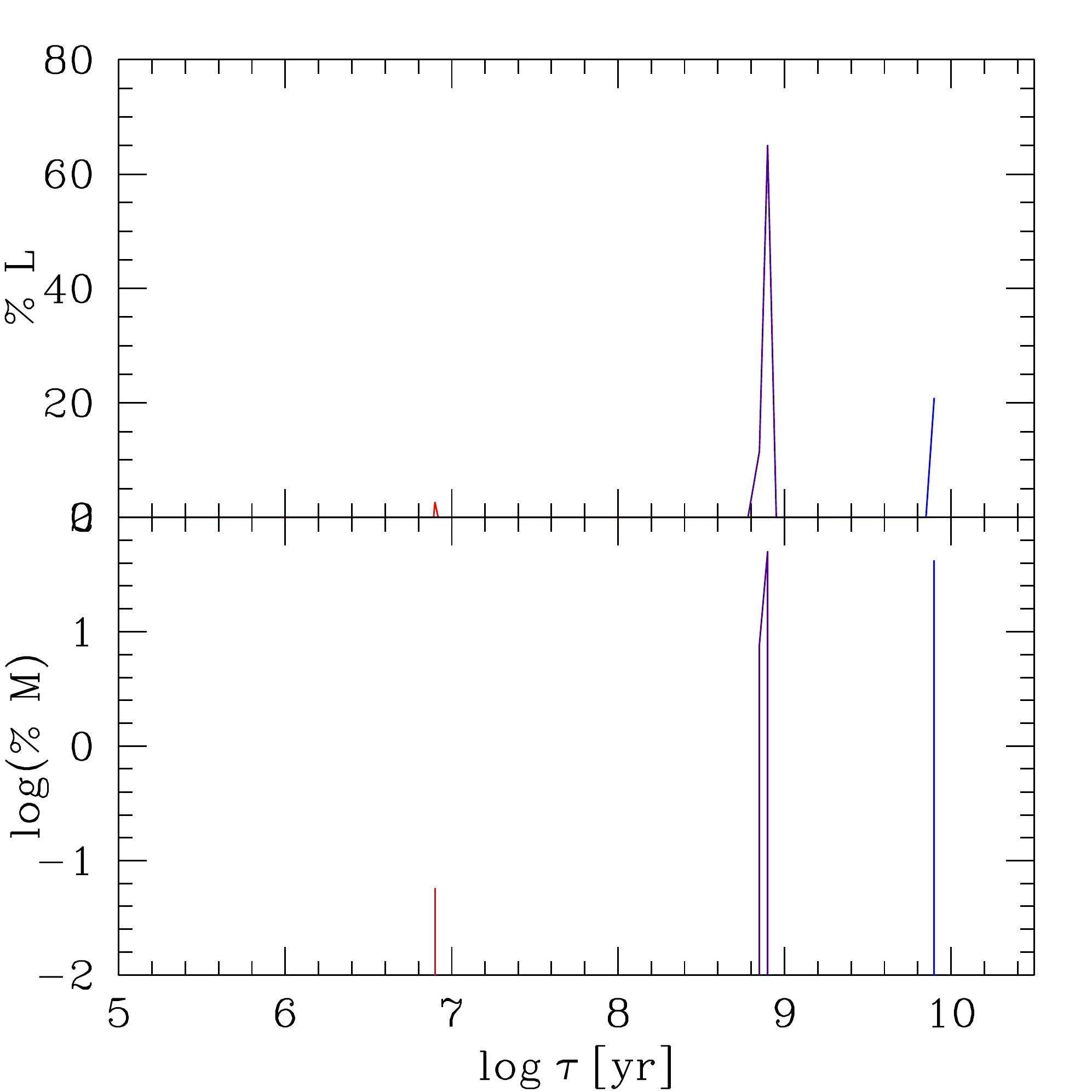}
\includegraphics[scale=0.2,angle=0]{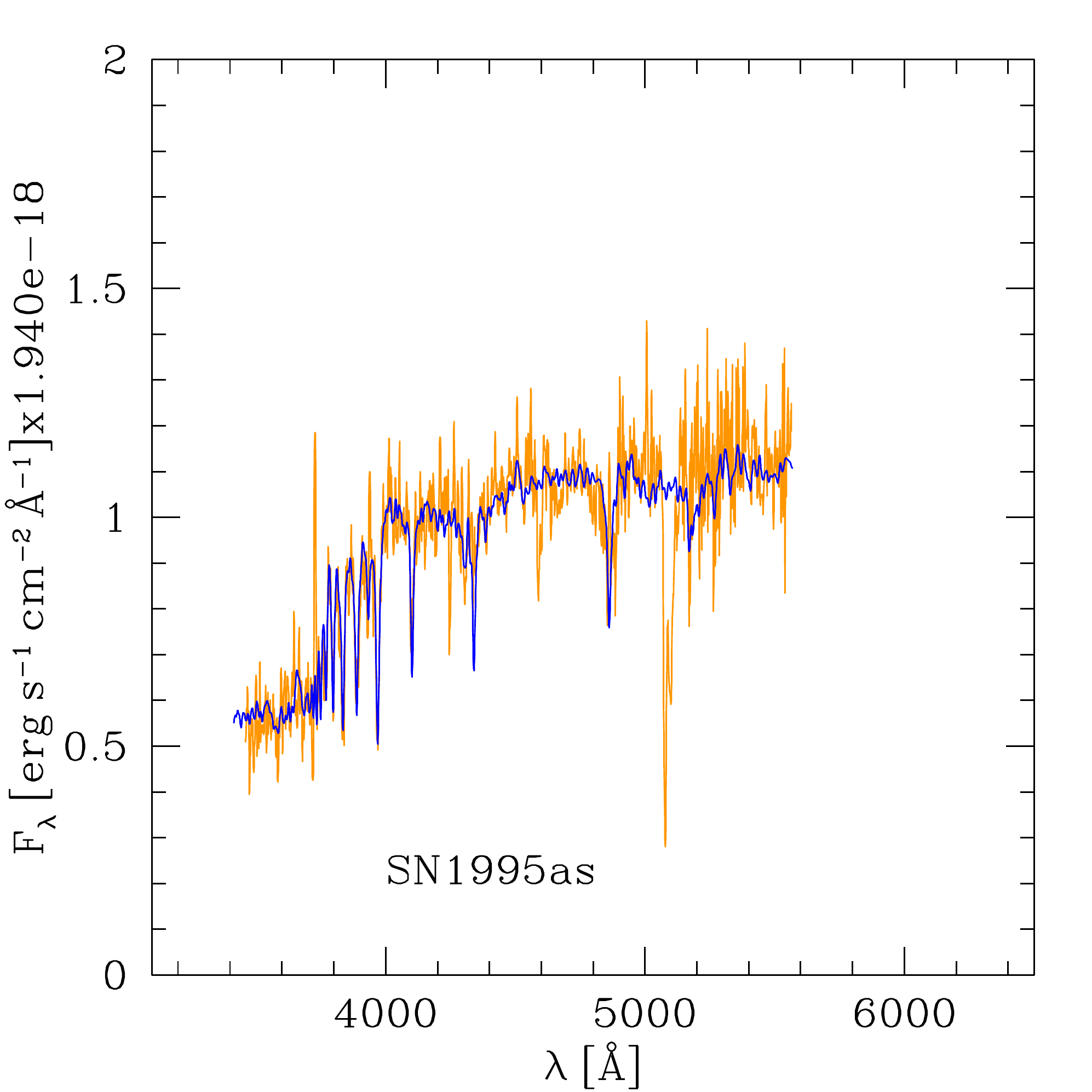}
\includegraphics[scale=0.2,angle=0]{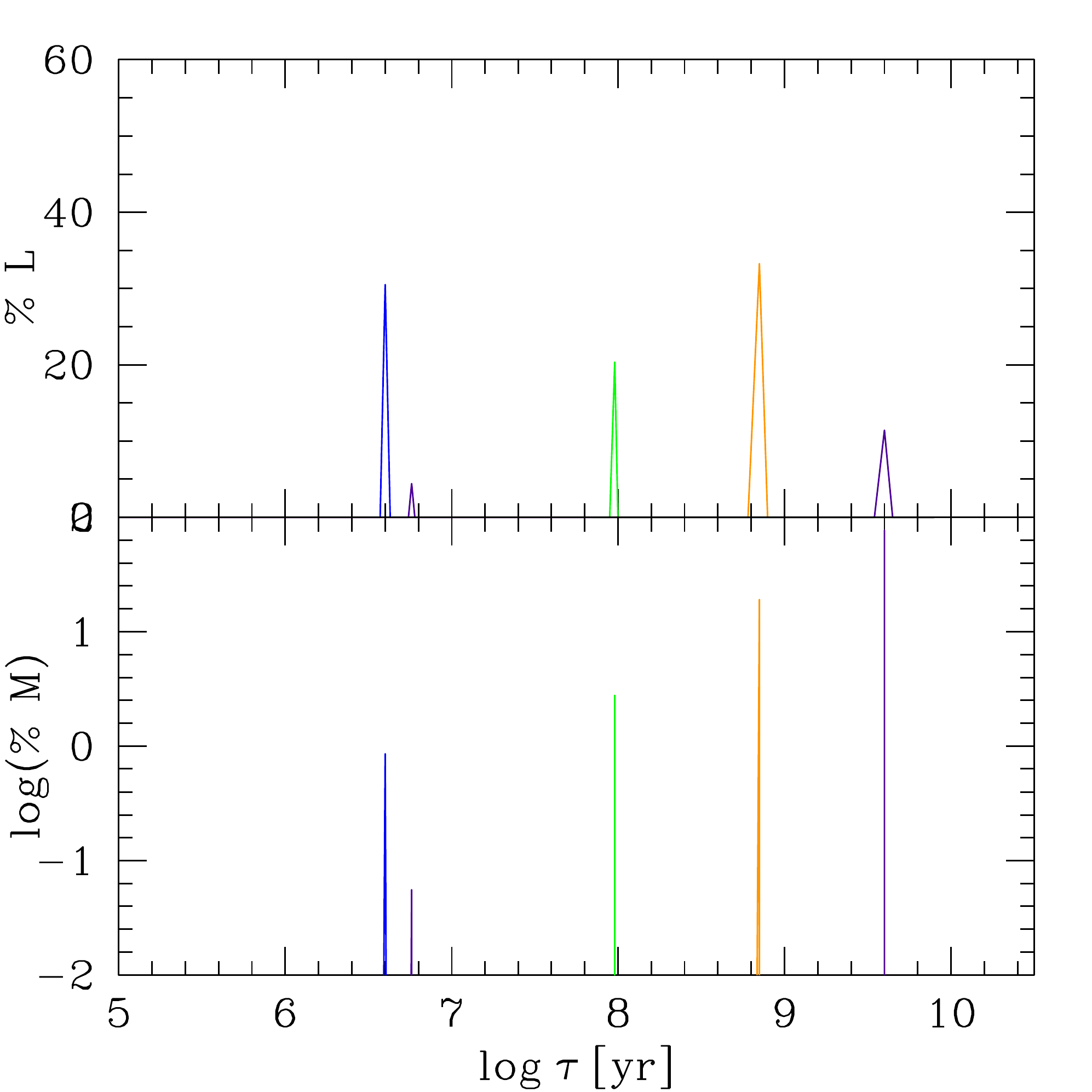}
\caption{Results produced by {\sc fado} for our six GTC-OSIRIS galaxies ordered by redshift, from top left to bottom right: SN2001iv, SN2005hw, SN 1995az, SN1999ff, SN1995ar, and SN1995as. The original spectra are drawn in orange while the fit from {\sc fado} is in blue line. In SN2001iv the green line is the fit obtained from {\sc fado} results by including the emission lines. The panels at the right of each spectrum show the contribution of each SSP, given as proportions of luminosity or stellar formed mass, plotted in different colors, as labelled in panel of SN2005hw, indicating  the metallicity of the corresponding SSP. 
}
\label{fig:resfado}
\end{figure*}

We use here as {\sl basis} the set of SSP calculated with the {\sc HR-PopStar} code from \citet{millan-irigoyen+2021}. For this work, we have developed a special set of models, with the IMF from \cite{Kroupa2002}, by changing the spectral library of normal stars from \citet{Coelho2014} used in the original work, by the one from \citet{Munari2005}. This last library has a shorter wavelength range than the original {\sc HR-pyPopStar}, with a range from 2500\,\AA\ to 10000\,\AA, (that is, without the ionizing part of spectra), but a wider range in metallicities,  $0.0001 < Z< 0.05$, which correspond, for a solar abundance $Z_{\sun}=0.02$ \citep{Asplund+09}, to a range 0.005 to 2.5 in units of solar metallicity (Z/Z$_{\sun}$). This point is important, since probably the high redshift galaxies in our sample are in the first phases of evolution, with low metallicities for a part of their stellar populations. More details about this new version of {\sc HR-PopStar} will be given in Mill{\'a}n-Irigoyen et al.(in preparation).

Since we are working with non-local galaxies, it is necessary to take into account the age of the Universe in each redshift. These galaxies could not have stellar populations older than that age of the Universe corresponding to their redshift. Thus, in the maximum redshift in our sample $z=0.9718$, the age of the Universe, using a flat cosmology with $\Omega_{M}=0.315$ and $H_{0}=70\,km\,s{-1}\,Mpc^{-1}$, is $t_{U}=5.744$\,Gyr ($\log{t_{U}}=9.76$), and, therefore, stellar populations older than this value can not exist in that COSMOS galaxy nor contribute to the spectra in the code {\sc fado}, while the SDSS galaxy with the minimum redshift $z=0.013$ in our sample, may have stellar populations in the whole range of ages of {\sc HR-PopStar} models, from 5\,Myr until $t_{U}=13.11$\,Gyr ($\log{t_{U}}=10.12$). Due to this, the set of SSP basis is variable for each galaxy, with the whole SSP set as {\sl basis} for nearby objects as this last galaxy, but reduced for the highest redshift ones. This age will be also used to represent our results about the star formation histories as a function of the evolutionary time in next section. 

\section{Results}
\label{sec:results}

\subsection{GTC-OSIRIS spectra results}\label{subsec:GTC results}

We start analyzing the results for our six galaxies observed with GTC-OSIRIS, since due to the small number we may show clearly the type of results that {\sc fado} gives.  As explained above, to run each {\sc fado} it was necessary to take into account the evolutionary time corresponding to each redshift. In this case, the closest galaxy (hosting SN2001iv at $z=0.397$) only has stellar populations as older as 9.05\,Gyr, $\log{\tau}=9.96$, while the farthest (SN1995as, $z=0.498$) only may have stellar populations younger than 8.3\,Gyr, $\log{\tau}=9.92$.

\begin{figure*}
\centering
\includegraphics[scale=0.35,angle=0]{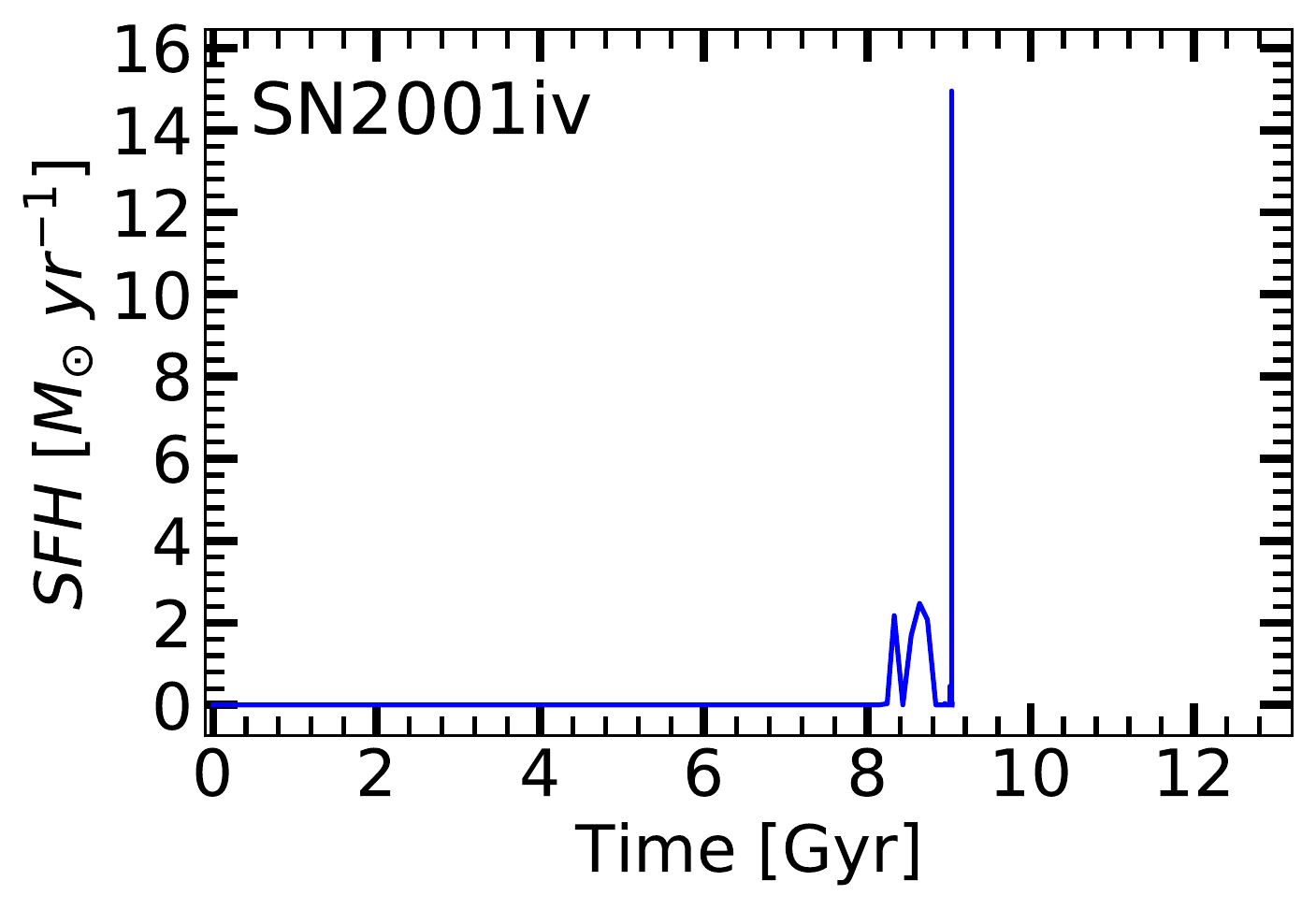}
\includegraphics[scale=0.35,angle=0]{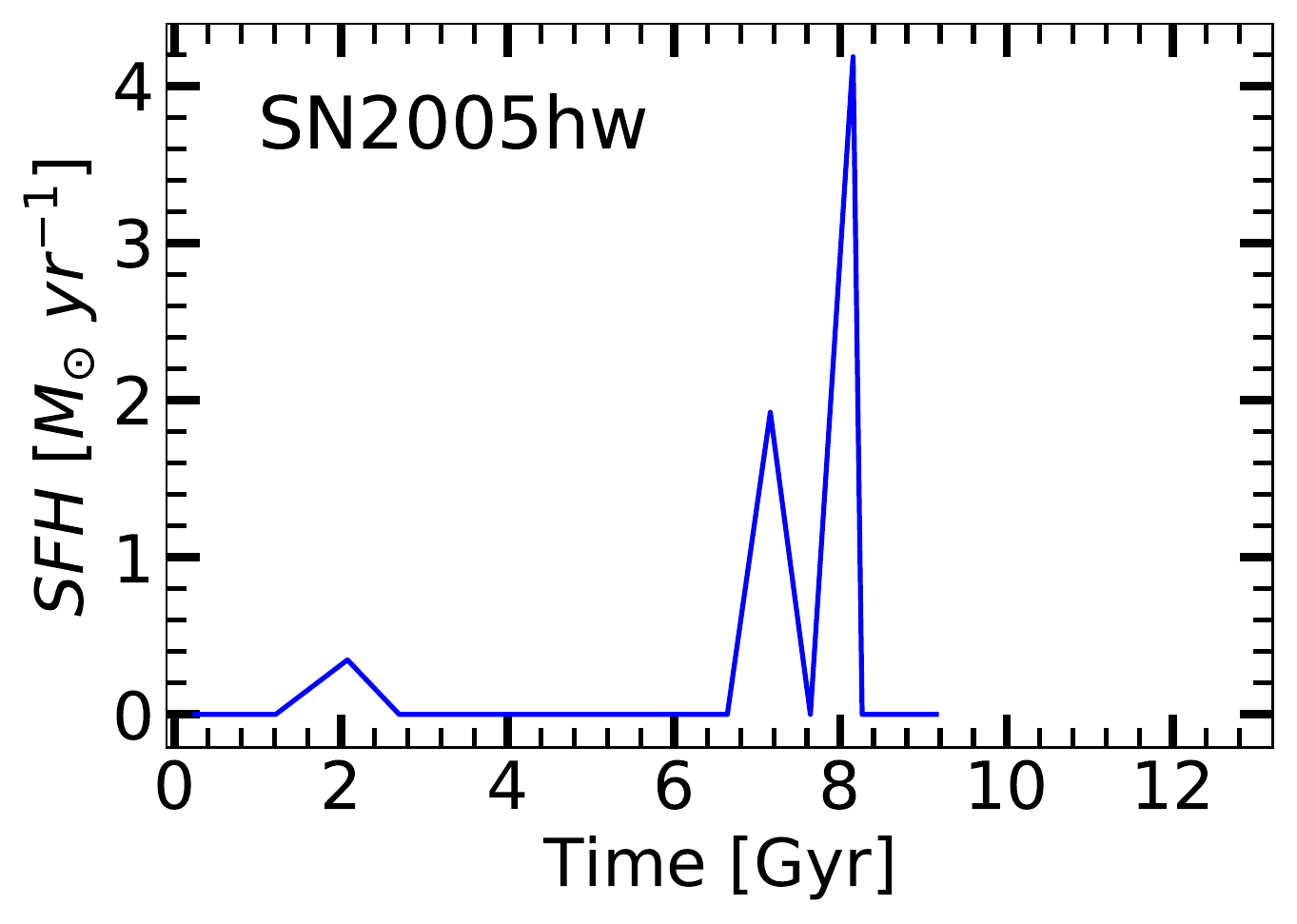}
\includegraphics[scale=0.35,angle=0]{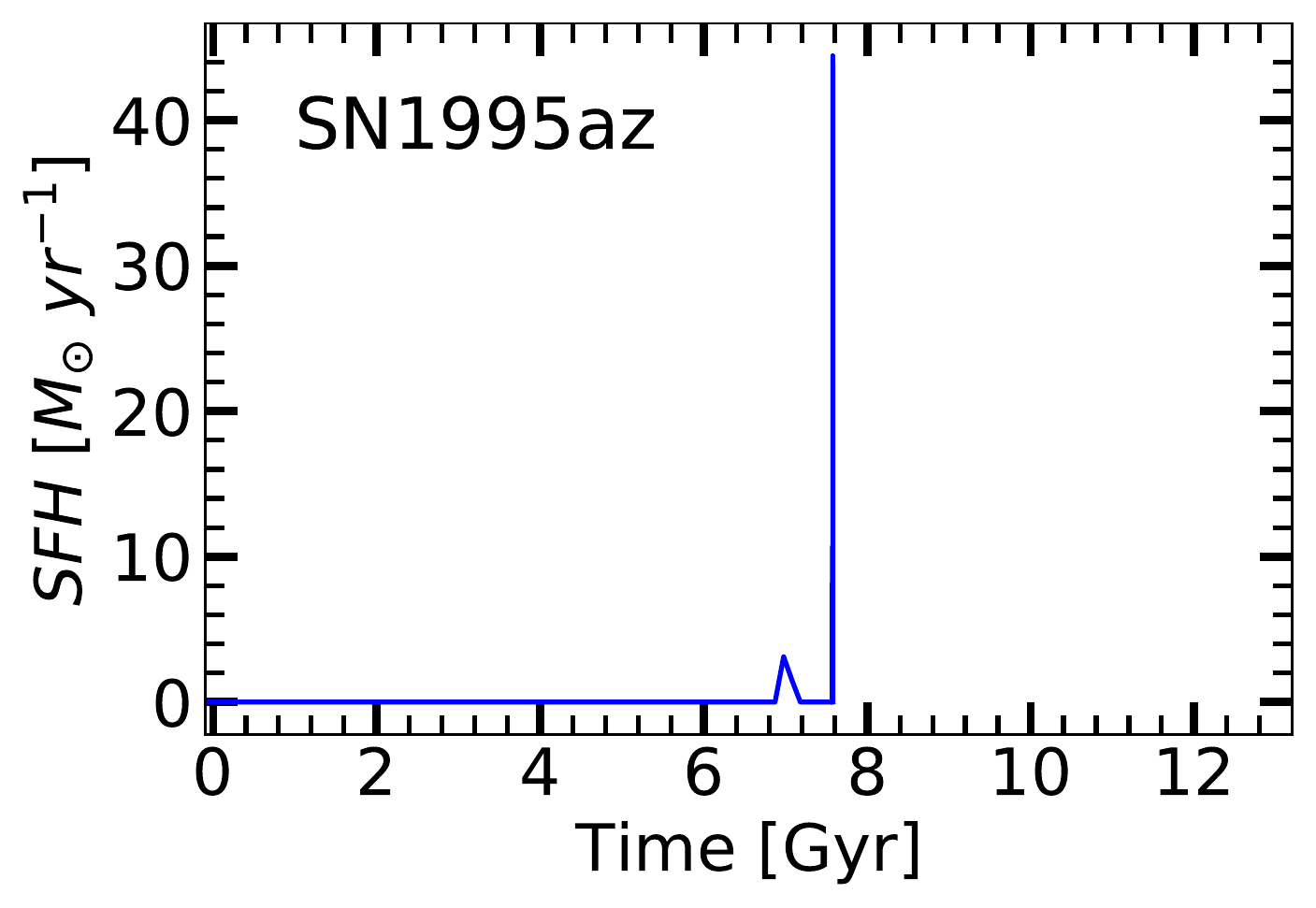}
\includegraphics[scale=0.35,angle=0]{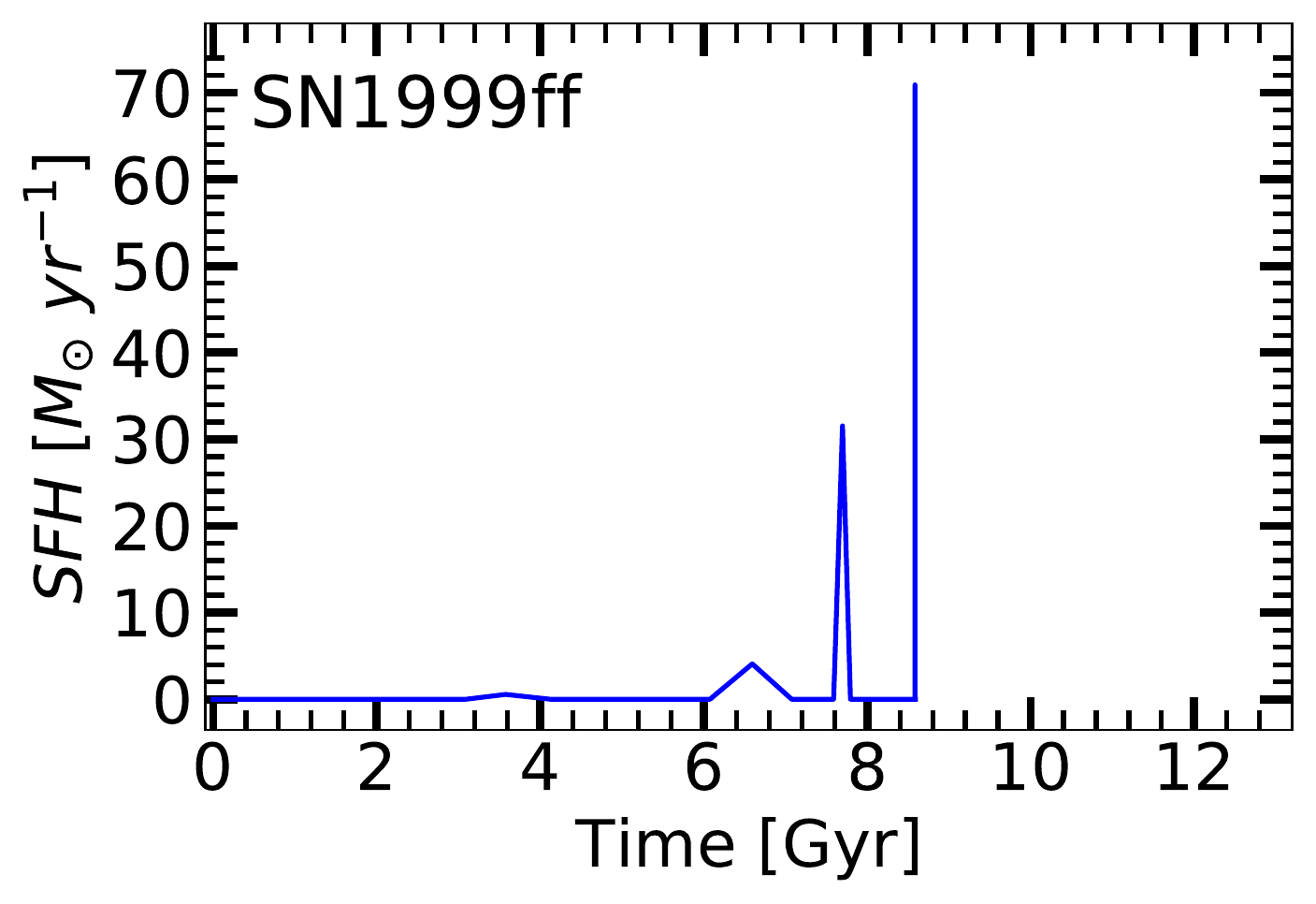}
\includegraphics[scale=0.35,angle=0]{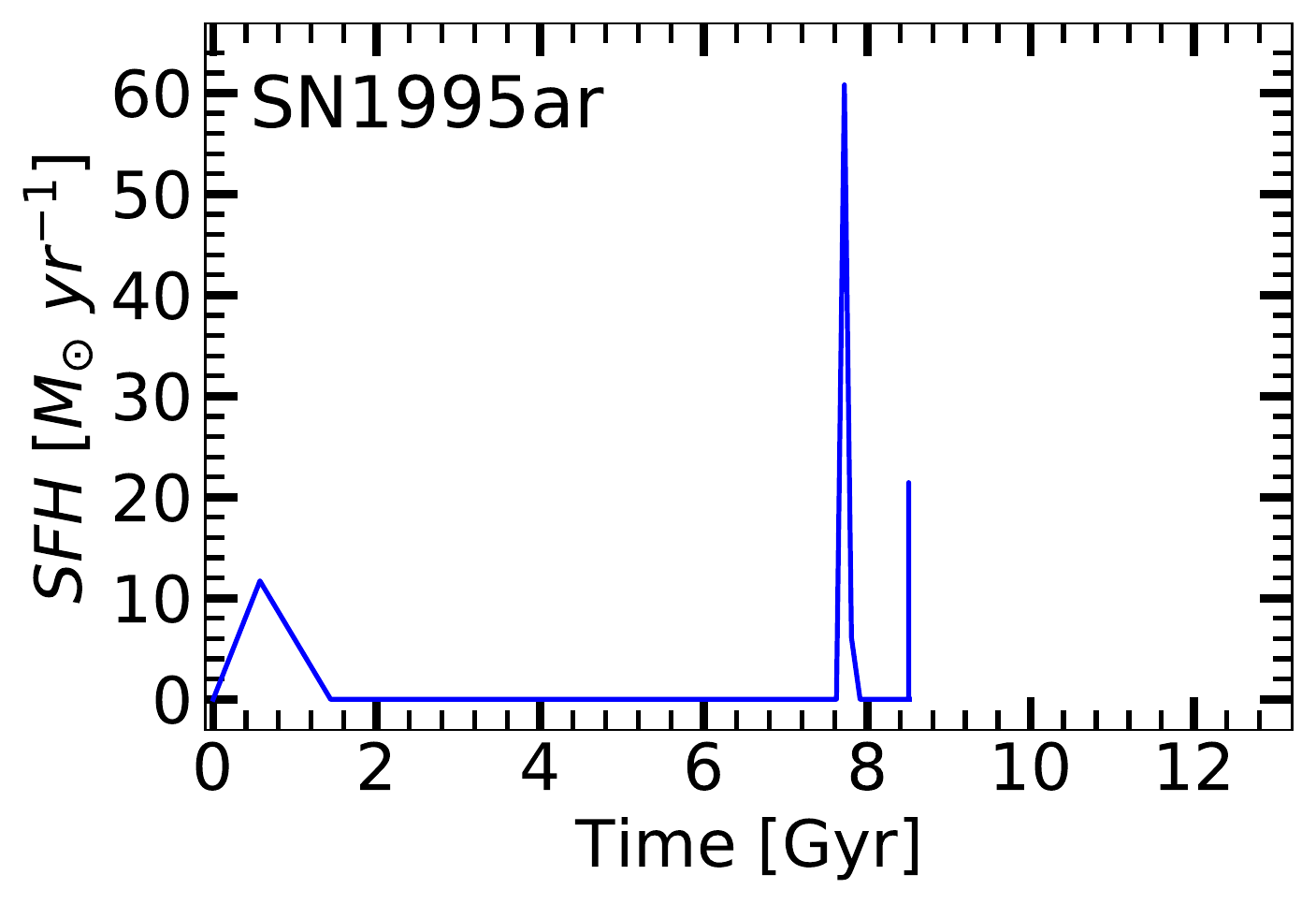}
\includegraphics[scale=0.35,angle=0]{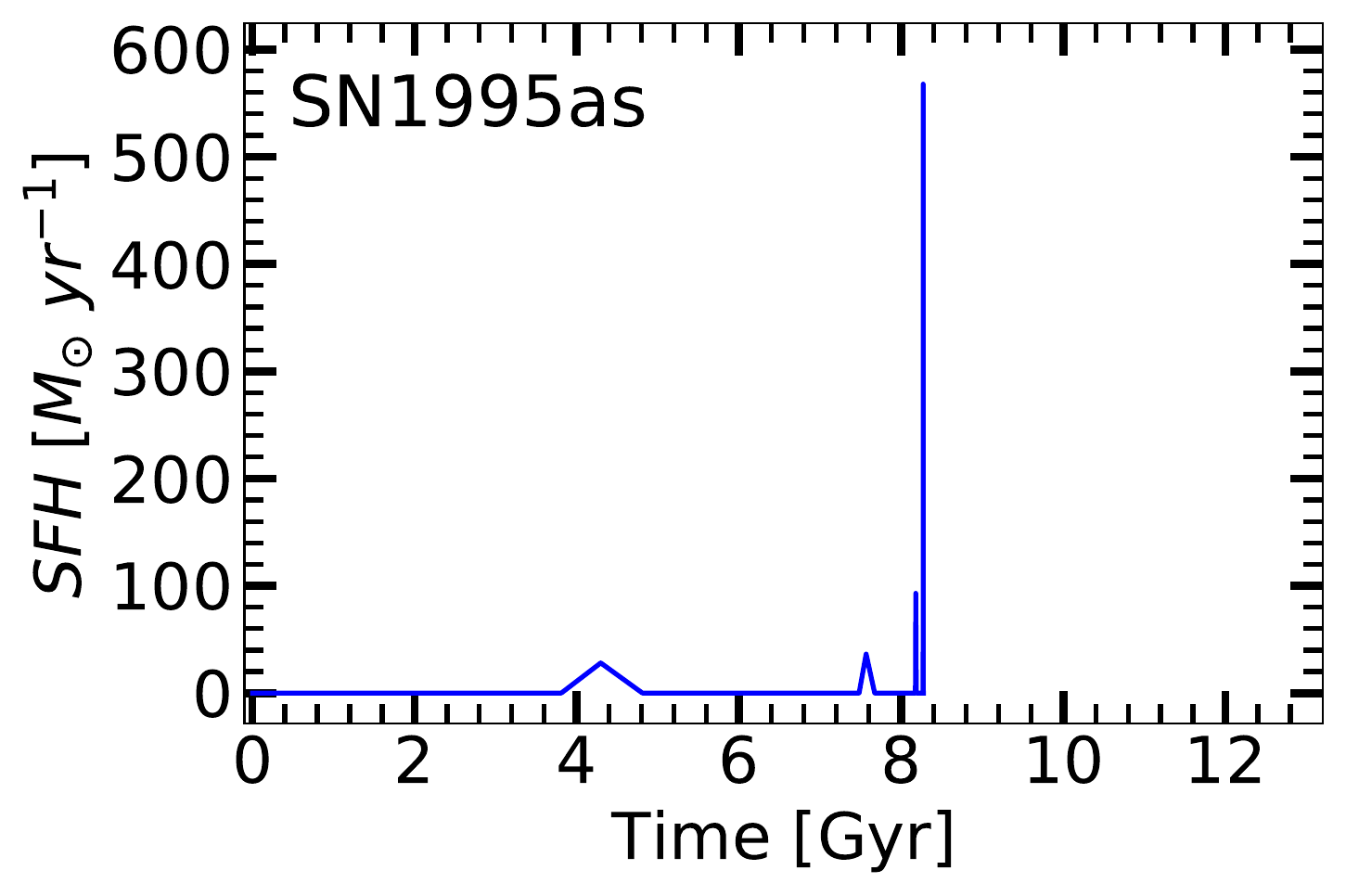}
\caption{Star formation histories for the six GTC-OSIRIS galaxies obtained from the stellar mass contributions of each SSP as given by {\sc fado} after the resulting spectra fitting.}
\label{fig:SFH}
\end{figure*}
We have analysed the 6 spectra observed in GTC-OSIRIS and reduced by us as described in section \ref{sec:data}. The code {\sc fado} produces for each spectrum a fit with a combination of SSPs, such as we show in Figure~\ref{fig:resfado}. 
For each galaxy, the plot shows the fit of the observed spectra, in orange line, obtained as a combination of several SSPs spectra, in blue line. The only galaxy that have emission lines is the host of the supernova SN2001iv. The code {\sc fado} do not show the total spectra, but it fits the fluxes of these emission lines and provide their intensities in a table. By using this information, we have computed the full spectral fitting for SN2001iv, shown as a green line in  the corresponding panel. Other galaxies seemed to have emission lines, for instance, SN1995as shows [\ion{O}{ii}], H$\beta$ and [\ion{O}{iii}], and SN1995ar and SN1999ff might have [\ion{O}{iii}], but all of them are in the limit of detection, and the code {\sc fado} is unable of measuring their intensities. Therefore, since the S/N is not high enough to take into account them, we have treated them as bad pixels, with mask for the analysis. Finally, the galaxy hosting SN2005hw shows lines that we consider as coming from a background object, and so we have also mask them for the {\sc fado} analysis.

The right plots in each panel shows the contribution of different metallicities and of different ages.
These contributions are given in percentages of light (top panels) or mass (bottom panels), shown by the lines in each ages with different color for different metallicities as labelled in panel corresponding to SN~2005hw. The stellar formed and actual mass, as other characteristics of the galaxy, are also given as {\sc fado} results. Therefore, it is possible to obtain  the star formation history (SFH) in each galaxy from this information. 

We show in Figure~\ref{fig:SFH} the resulting SFH for these six galaxies, which may also be considered as examples of the results obtained also for the other sets (we will give as an electronic Appendix the whole sample of spectra with their {\sc fado} fits and components). As we see in that Fig.~\ref{fig:SFH}, the six galaxies show a star formation burst near 8\,Gyr with other of some hundred of Myr in the time corresponding of each redshift. Only SN~2005hw has an old stellar populations, with a strong burst at a time 2\,Gyr. The other galaxies have other small peaks, as SN~1995ar at 1\,Gyr or, even smaller, SN1995~az at 3\,Gyr.

\begin{figure*}
\hspace{-1.5cm}
\includegraphics[scale=0.65]{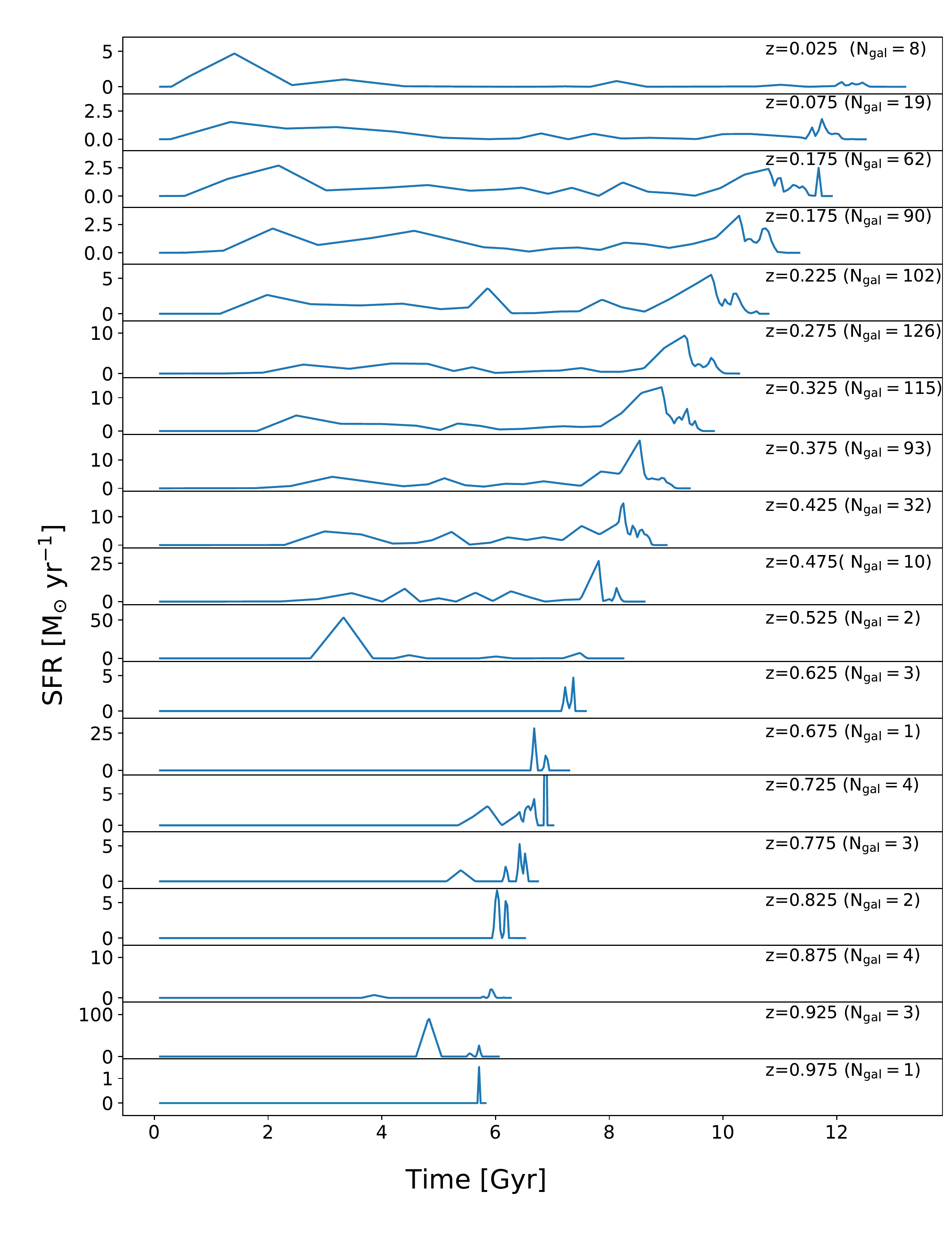}
\vspace{-0.7cm}
\caption{SFH for the whole sample of galaxies, binned by redshift steps of $\Delta z=0.05$ from $z=0.0$ to $z=1.0$. Each bin shows the number of galaxies as an indicator of the quality or dispersion of data within each one.}
\label{fig:SFH_whole_sample}
\end{figure*}

\begin{table*}
\centering
\caption{Results given by {\sc fado} for the sample of 680 galaxies with the corresponding information of their hosted SNe Ia. The complete table will be given in electronic format. Galaxy name in column (1), SN name in (2),  equatorial coordinates in (3) and (4), redshift in (5), the corresponding Universe age in (6), the weighted by mass and luminosity averaged age and metallicity, with their errors, in  columns (7) to (14), the stellar mass formed and the current one, in (15) and (16), the velocity and the dispersion velocity, in columns (17) to (20), the SNe Ia, stretch X1, color and distance modulus, with their corresponding errors, in columns (21) to (26). The sub-sample $S$ at which each object corresponds, 1 for SDSS, 2 for GTC-OSIRIS and 3 for zCOSMOS/MAGELLAN is in (27).}
\begin{tabular}{llccccccc}
\hline
Galaxy & ID & RA  &  DEC  & z & t$_{U}$ & $\langle\tau_{M}\rangle$ & $\langle\tau_{L}\rangle$ & $\langle Z_{M}\rangle$  \\
name & SN &  &  &  &  Gyr &  Gyr & Gyr  &  $\mathrm Z_{\sun}$  \\
(1) & (2) & (3) & (4) & (5) & (6) & (7) (8) & (9) (10) & (11) (12) \\
\hline
 1237663782589825148  &  8151      &    6.961914 &    -1.198235 & 0.0130 & 13.114  &  5.822 $\pm$   0.145 &  3.161 $\pm$ 0.086 &  0.658 $\pm$  0.023   \\
 1237663238739985091  & 17784      &   52.461662 &     0.056757 & 0.0371 & 12.790  &  5.733 $\pm$   0.118 &  2.282 $\pm$  0.050 &  0.865 $\pm$  0.023  \\
 1237678622234050898  & SN1995ar   &   15.335042 &     4.309389 & 0.4650 &  8.526  &  3.191 $\pm$   0.122 &  2.256 $\pm$  0.080 &  1.859 $\pm$  0.060   \\
 1237678622234051119  & SN1995as   &   15.397083 &     4.437444 & 0.4980 &  8.288  &  3.072 $\pm$   0.121 &  0.710 $\pm$  0.021 &  1.926 $\pm$  0.076  \\
 827945               & 05D2og     &  150.148636 &     2.127935 & 0.6163 &  7.511  &  0.309 $\pm$   0.012 &  0.168 $\pm$ 0.006 &  2.112 $\pm$  0.075  \\
 823980               & 06D2hh     &  150.287292 &     2.136736 & 0.6230 &  7.471  &  0.076 $\pm$   0.002 &  0.066 $\pm$  0.002 &  0.005 $\pm$  0.000  \\
\hline
\end{tabular}
\label{table:res-fado}
\end{table*}
\begin{table*}
\centering
\contcaption{Results given by {\sc fado} for the sample of 680 galaxies with the corresponding information of their hosted SNe Ia. The complete table will be given in electronic format.}
\begin{tabular}{ccccccccc}
\hline
$\langle Z_{L}\rangle$ & $\log{M_{formed}}$ & $\log{M_{current}}$ & $v$ & $\sigma_{*}$ & x1 & c & $\mu$ & S \\
$\mathrm Z_{\sun}$ &  M$_{\sun}$ & M$_{\sun}$ & km\,s$^{-1}$ &  km\,s$^{-1}$ &  mag &   &  mag & \\
(13) (14) & (15) & (16) & (17) (18) & (19) (20) & (21) (22) & (23) (24) & (25) (26) & (27) \\
\hline
0.693 $\pm$ 0.014 & 9.64 &  9.41  &  -1.11 $\pm$    73.43 &    81.05 $\pm$     0.00 & -1.25 $\pm$  0.14 &  0.03 $\pm$  0.03  & 34.55 $\pm$  0.27  &1\\
0.270 $\pm$ 0.005 &  9.50 &  9.29 &    23.07 $\pm$    74.63 &    97.60 $\pm$     0.10 & -0.80 $\pm$  0.11 &  0.05 $\pm$  0.03 & 36.11 $\pm$  0.20  &1\\
1.940 $\pm$  0.070 & 10.06 &  9.88 &  -11.96 $\pm$  2686.00 &   394.00 $\pm$    10.17 &  0.42 $\pm$  0.72 &  0.38 $\pm$  0.19 & 41.83 $\pm$  0.60  &2\\
0.571 $\pm$  0.014 & 10.26 & 10.09 &   -56.17 $\pm$   300.80 &   198.30 $\pm$    71.82 &  0.26 $\pm$  1.39 &  0.00 $\pm$  0.20 & 42.98 $\pm$  0.64  &2\\
1.282 $\pm$  0.039 &  9.60 &  9.52 & -341.50 $\pm$   186.00 &   146.20 $\pm$     0.00 & -1.04 $\pm$  0.11 & -0.28 $\pm$  0.01 & 43.47 $\pm$  0.16  &3\\
0.005 $\pm$  0.000 &  9.52 &  9.47 &  -296.80 $\pm$   152.70 &   335.90 $\pm$     0.01 &  0.02 $\pm$  0.37 & -0.06 $\pm$  0.02 & 42.97 $\pm$  0.18  &3\\
\hline
\end{tabular}
\end{table*}
\normalsize

\subsection{Star formation histories}
Following the same procedure, we have run {\sc fado} for the other two subsamples of SDSS and COSMOS, obtaining the star formation histories for the rest of galaxies of our sample. In order to see evolutionary effects, we plot in Figure~\ref{fig:SFH_whole_sample} these SFHs. Since the number of galaxies is high, we have computed the SFH binned in each redshift with bin sizes of 0.050. Thus, each line in a redshift $z$ is representing the averaged SFH for galaxies within that bin, and the color scale at the right indicates the intensity of this SFR. It is clear in that figure how the SFR stops at the age of the Universe corresponding to each redshift, only reaching the present time those galaxies located at $z\sim 0$. We observe that only the bins with low redshift ($z<0.425$) have significant star formation at early times. We also observe that the nearby galaxies have a more continuous SFH, with stellar populations created since the first Gyrs of the Universe evolution. As the redshift increases, the SFH is more as in bursting mode, more concentrated in the time in which we observe them and without star formation at the early times. This shape as an {\sl V} shape in the Figure may be indicating the process of galaxy formation as mergers: The present time galaxies may be formed by mergers of two or more galaxies, thus presenting SF at all times, or at the moments when the mergers took place, while the high redshift galaxies are showing only the SF burst occurring in each individual galaxy. However, it is also possible that we are observing an statistical effect, since the bins with SDSS galaxies have been calculated with a larger number of objects than the ones at larger redshift, sometimes with only one galaxy in the bin. This, added to the best conditions of observation for the nearby galaxies, may be producing SFH with larger uncertainties in the far galaxies compared with the closer ones.

\subsection{The fundamental plane and other correlations}

Besides to recovered the SFH shown in the previous subsection, {\sc fado} gives some galaxy characteristics, as the stellar age weighted by mass or luminosity, $\langle\tau_{M}\rangle$, and $\langle\tau_{L}\rangle$, the stellar metallicity, also weighted by mass or luminosity, $\langle Z_{M}/Z_{\sun}\rangle$ and $\langle Z_{L}/Z_{\sun}\rangle$, both in units of $Z_{\sun}$, the dispersion velocity, $\sigma_{\star}$, and the stellar mass, $\mathrm M_{formed}$, formed in each age or metallicity bin as well as the current stellar mass, $\mathrm M_{current}$, that still remains. 

We give a summary of these results in Table~\ref{table:res-fado}, where for each galaxy hosting a SN Ia identified in column 1, giving the SN name in column 2, we give the equatorial coordinates in columns 3 and 4, the redshift in column 5, with the corresponding Universe age at that redshift in column 6, the resulting weighted by mass and luminosity averaged age, with their errors, in columns 7 to 10, and metallicity for their stellar populations in columns 3 and 4, the stellar mass formed and the current one (existing in the observation time) in solar masses, in logarithmic scale, in columns 5 and 6, the velocity and the dispersion velocity, in columns 7 and 8, and the information for their SNe Ia, stretch X1, color and distance modulus, with their corresponding errors, in columns 9, 10 and 11. In column 12 we give the Universe age for the used cosmology. The last column 13 indicates the sub-sample $S$ at which each object corresponds, 1 for SDSS, 2 for GTC-OSIRIS and 3 for zCOSMOS/MAGELLAN.

\begin{figure*}
\centering
\includegraphics[scale=0.38]{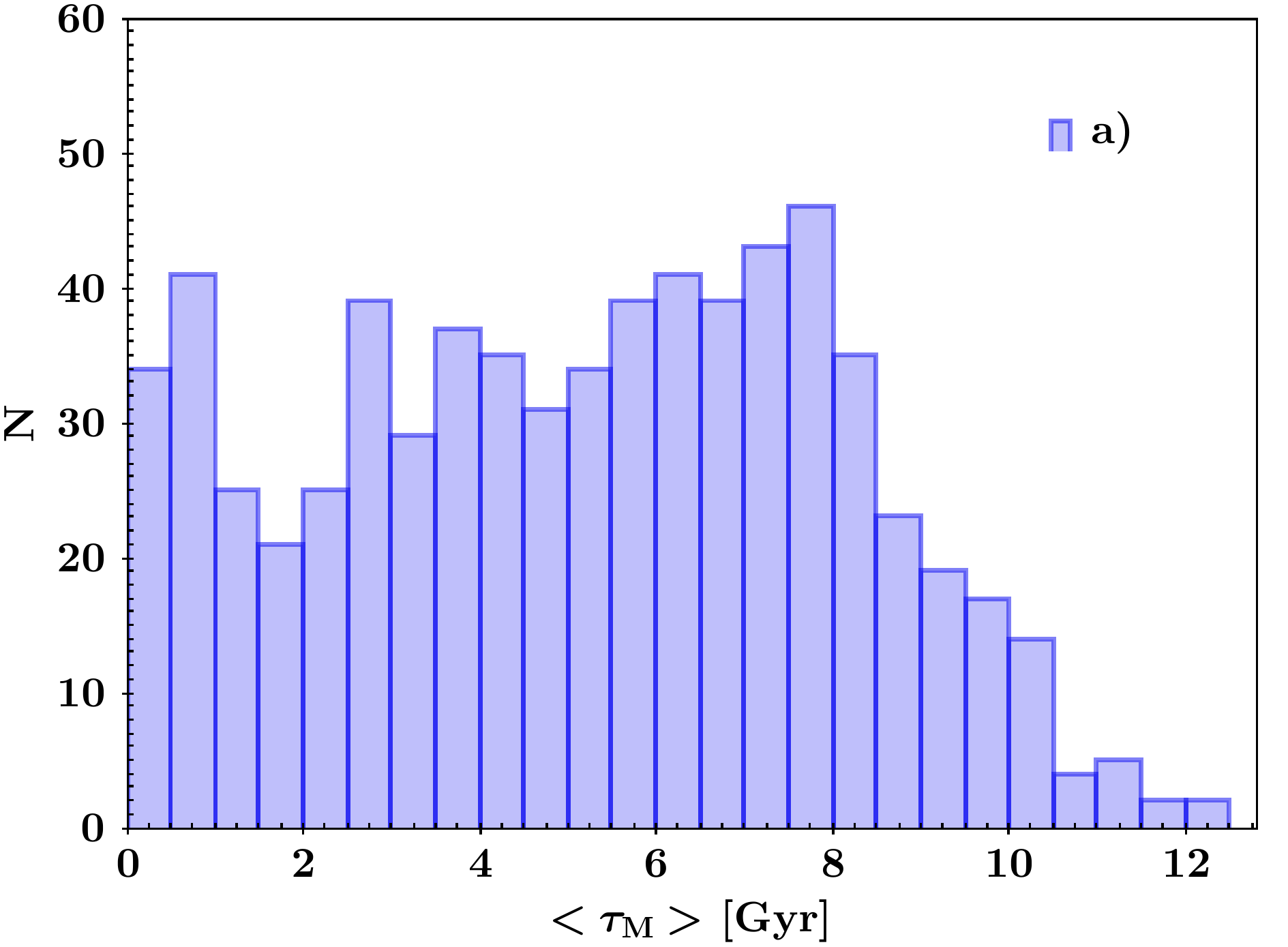}
\includegraphics[scale=0.38]{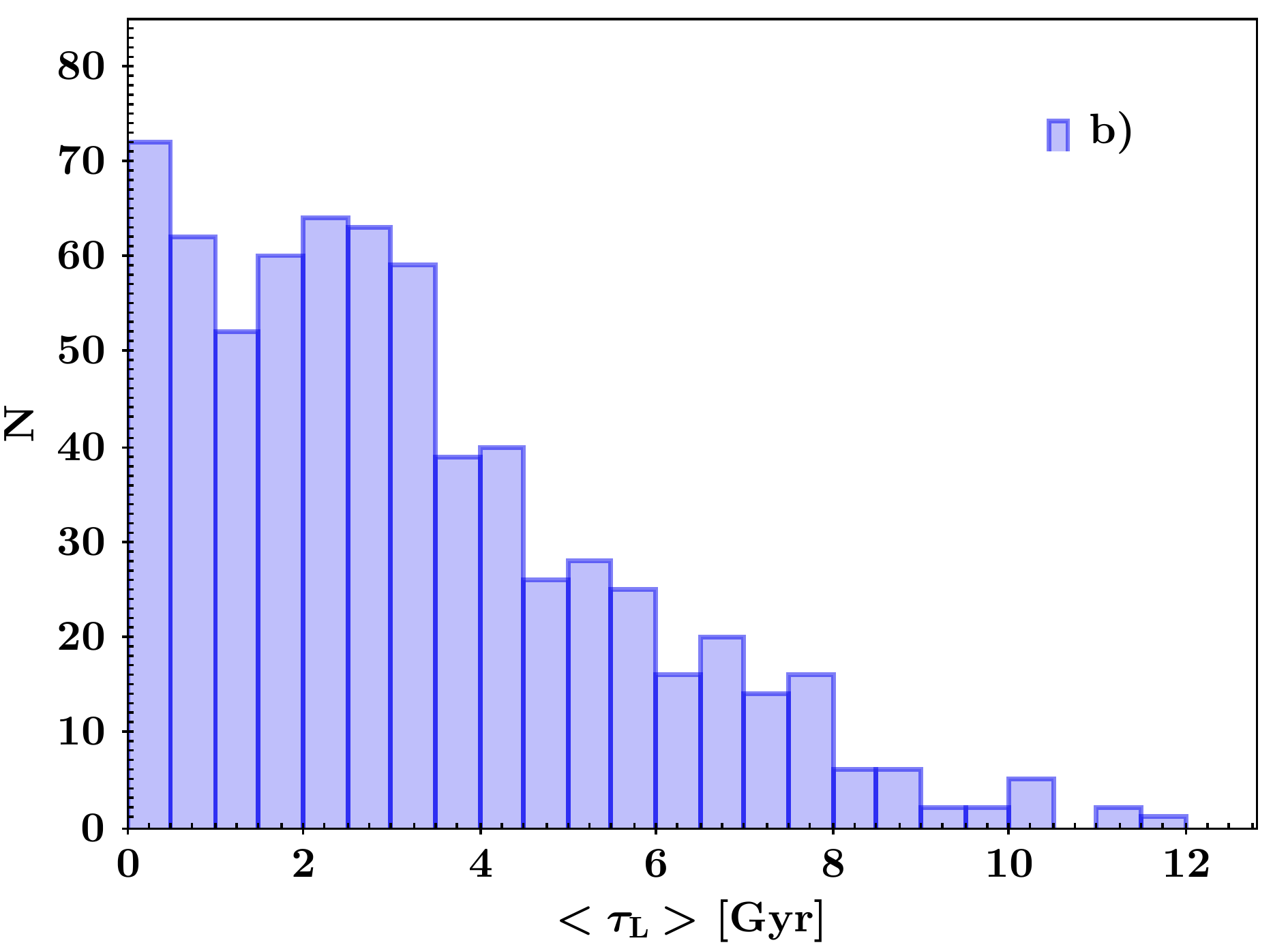}
\includegraphics[scale=0.38]{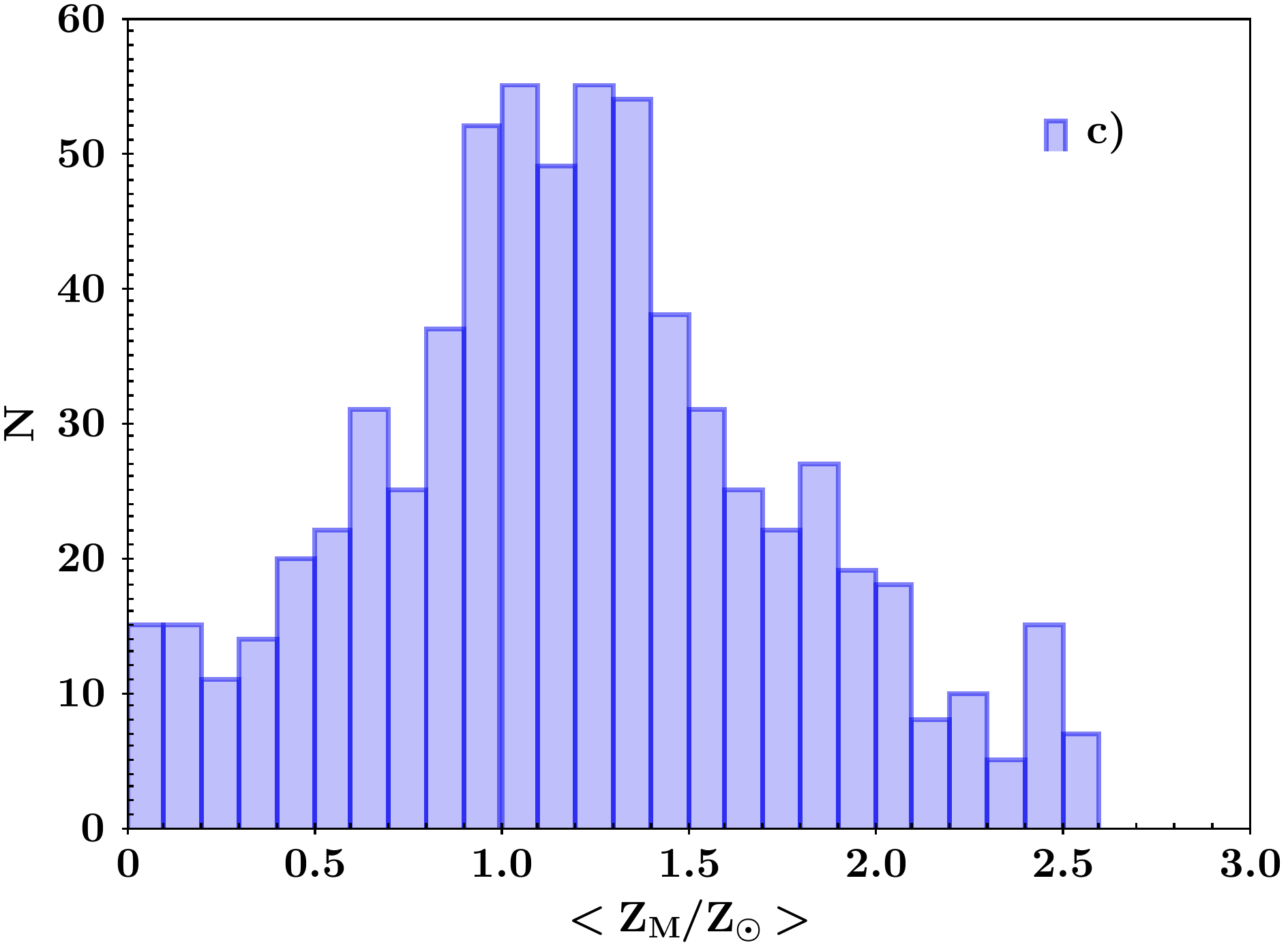}
\includegraphics[scale=0.38]{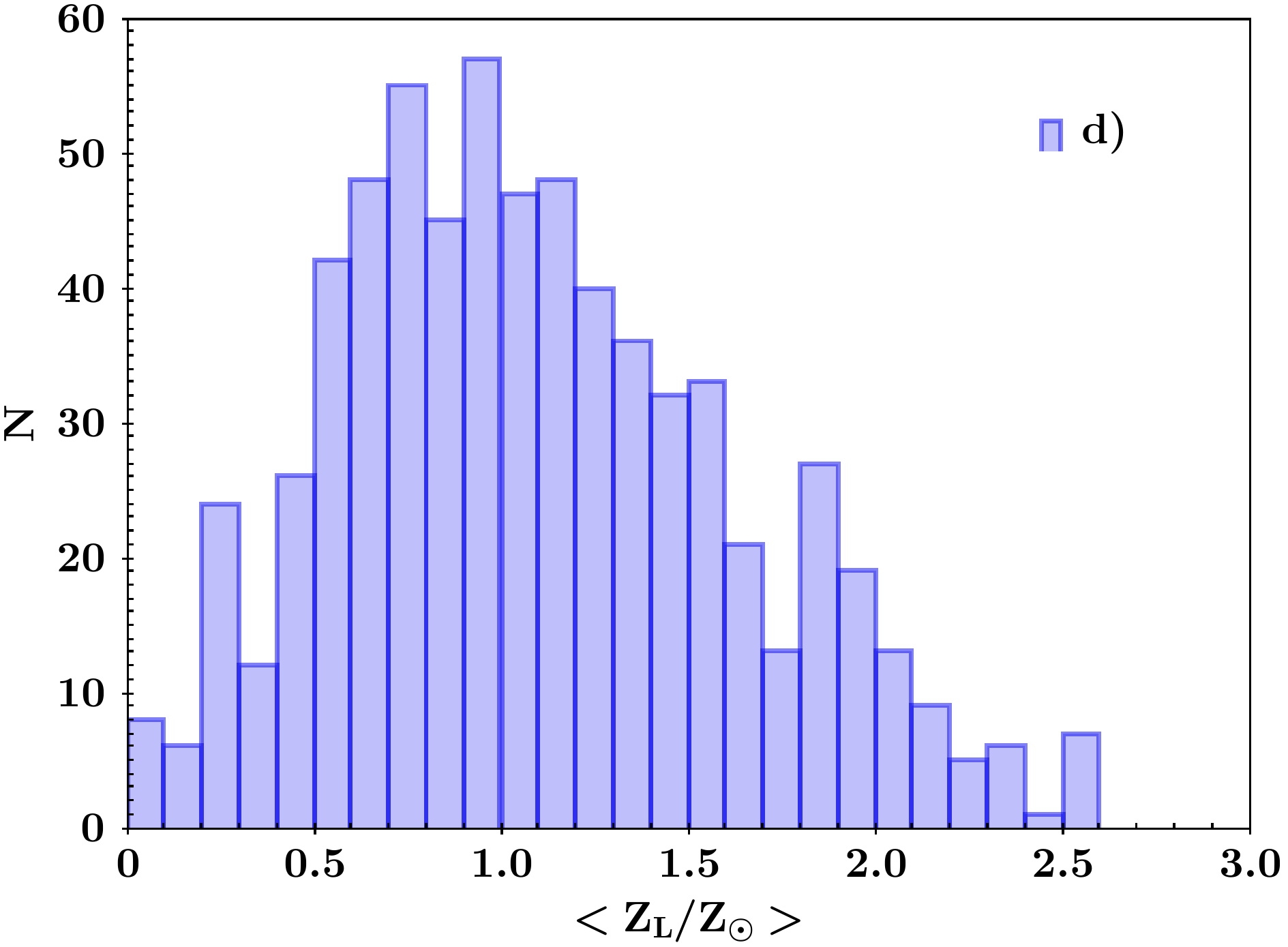}
\includegraphics[scale=0.38]{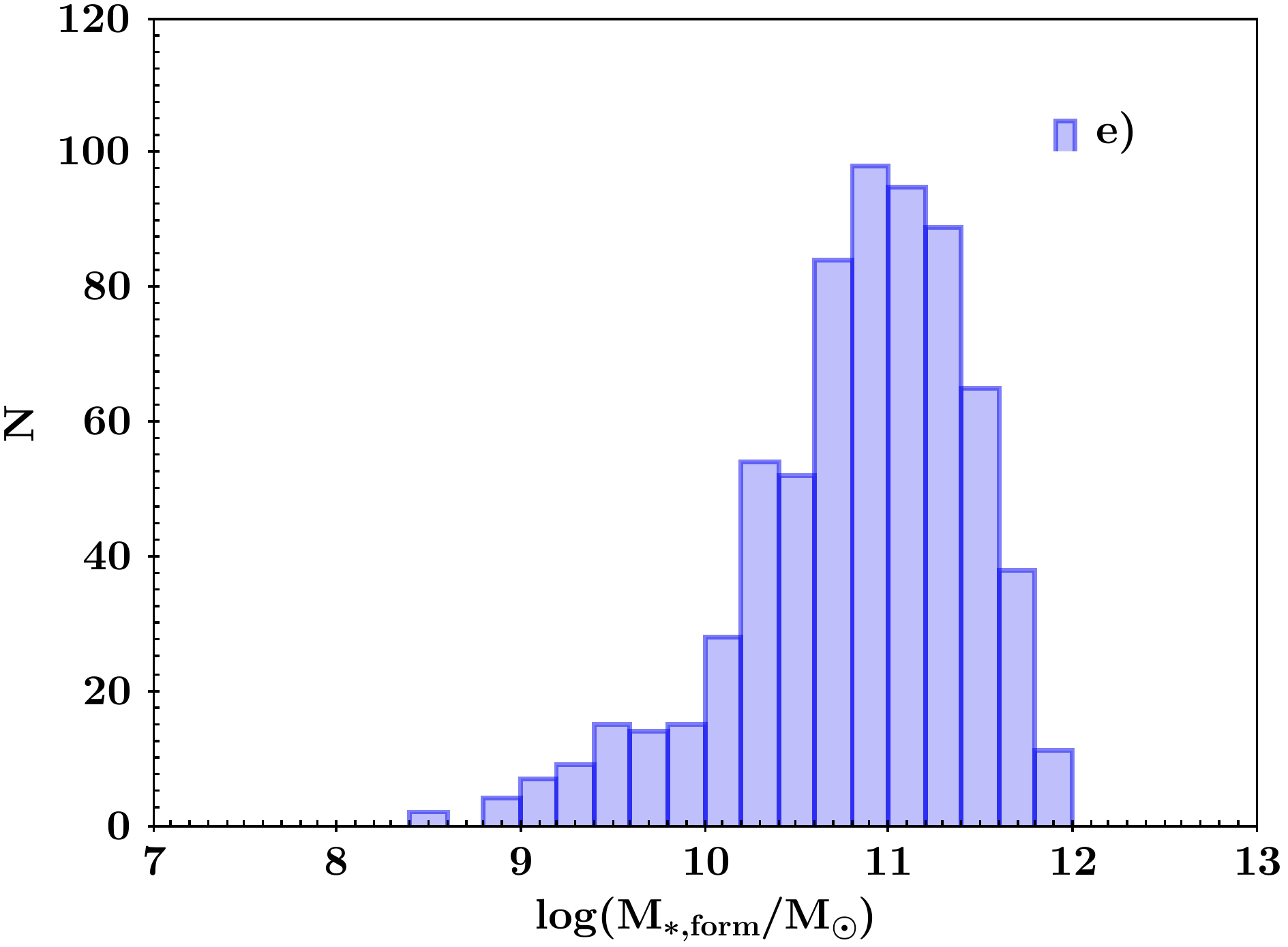}
\includegraphics[scale=0.38]{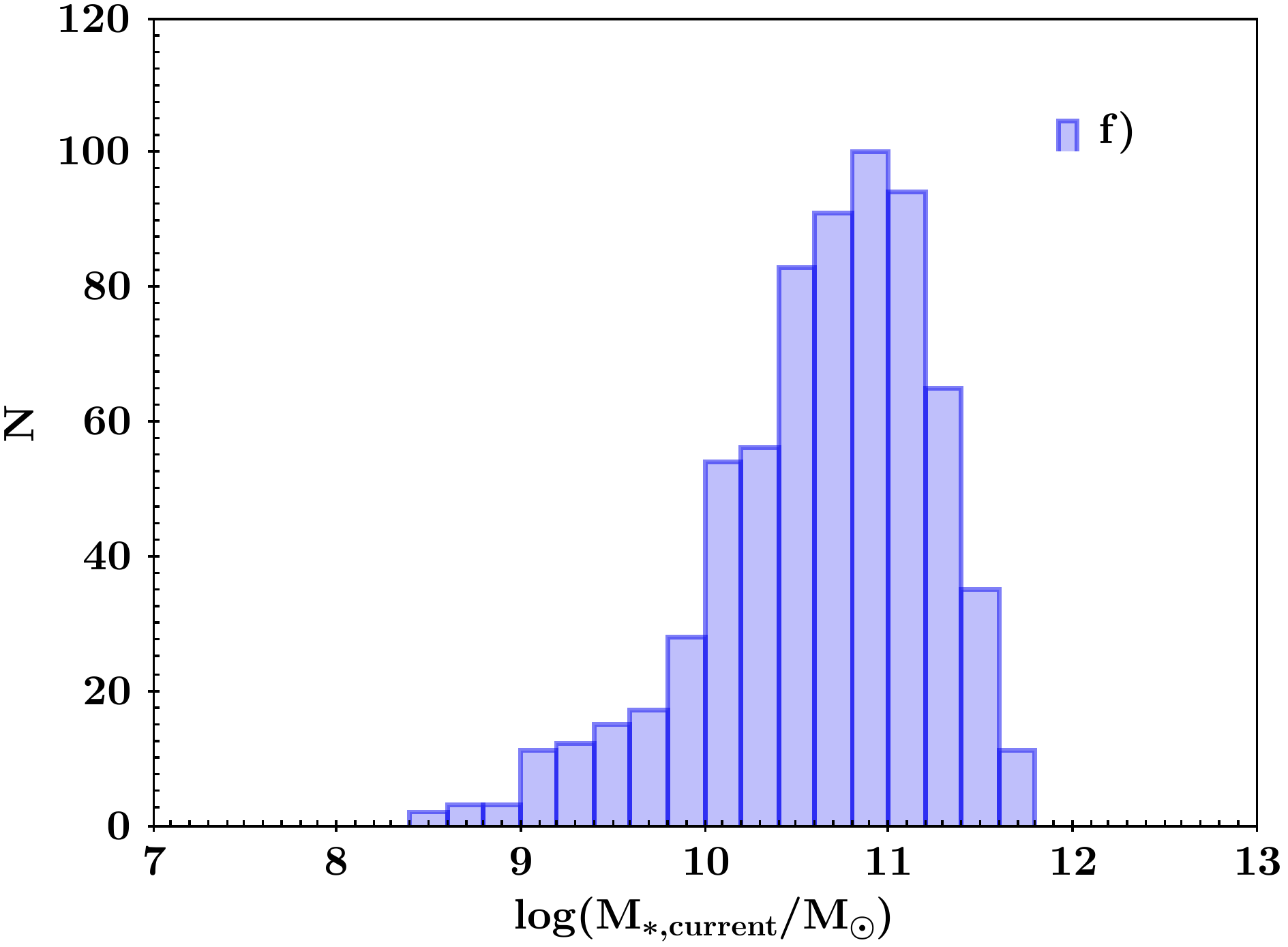}
\includegraphics[scale=0.38]{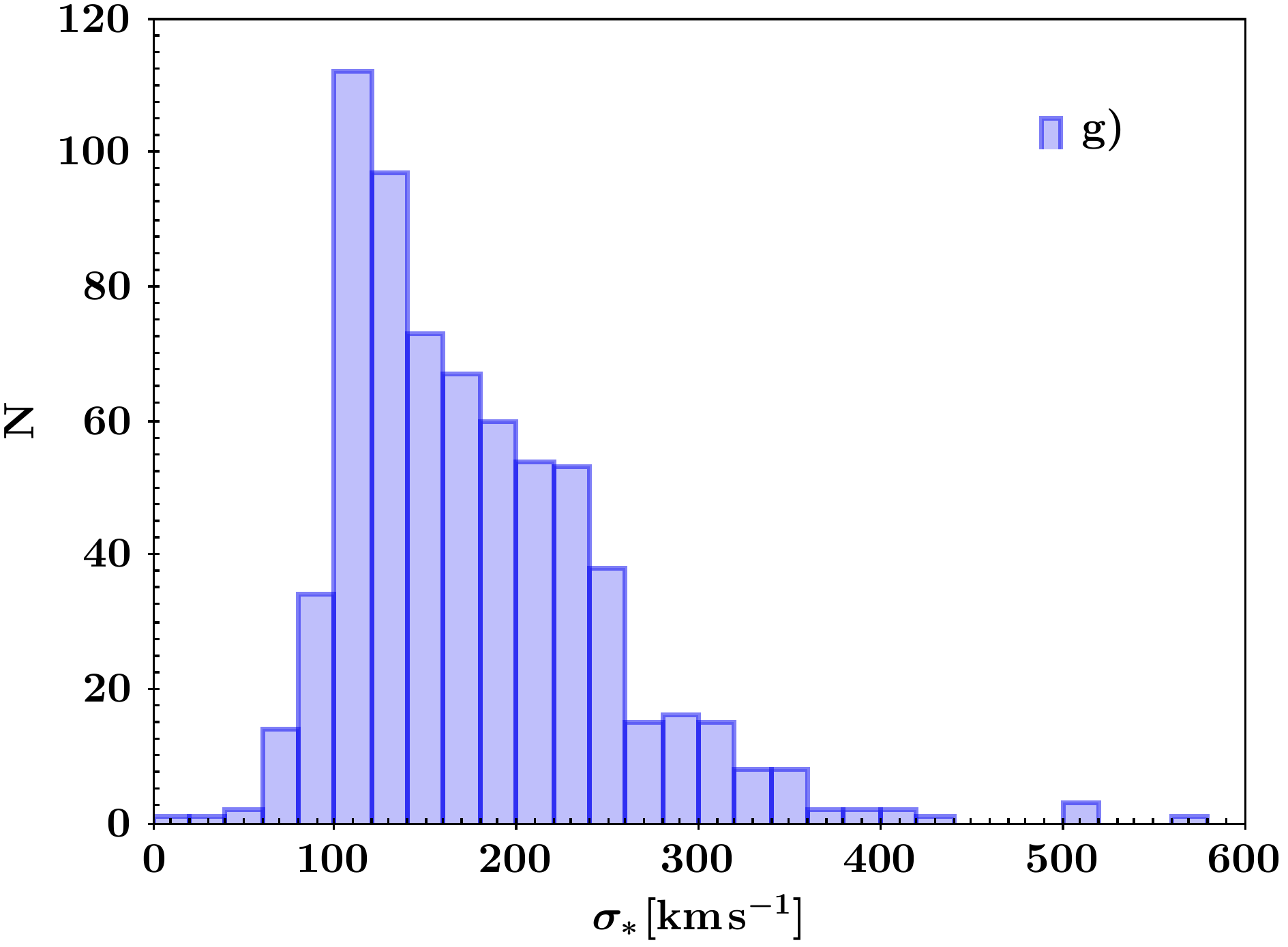}
\includegraphics[scale=0.38]{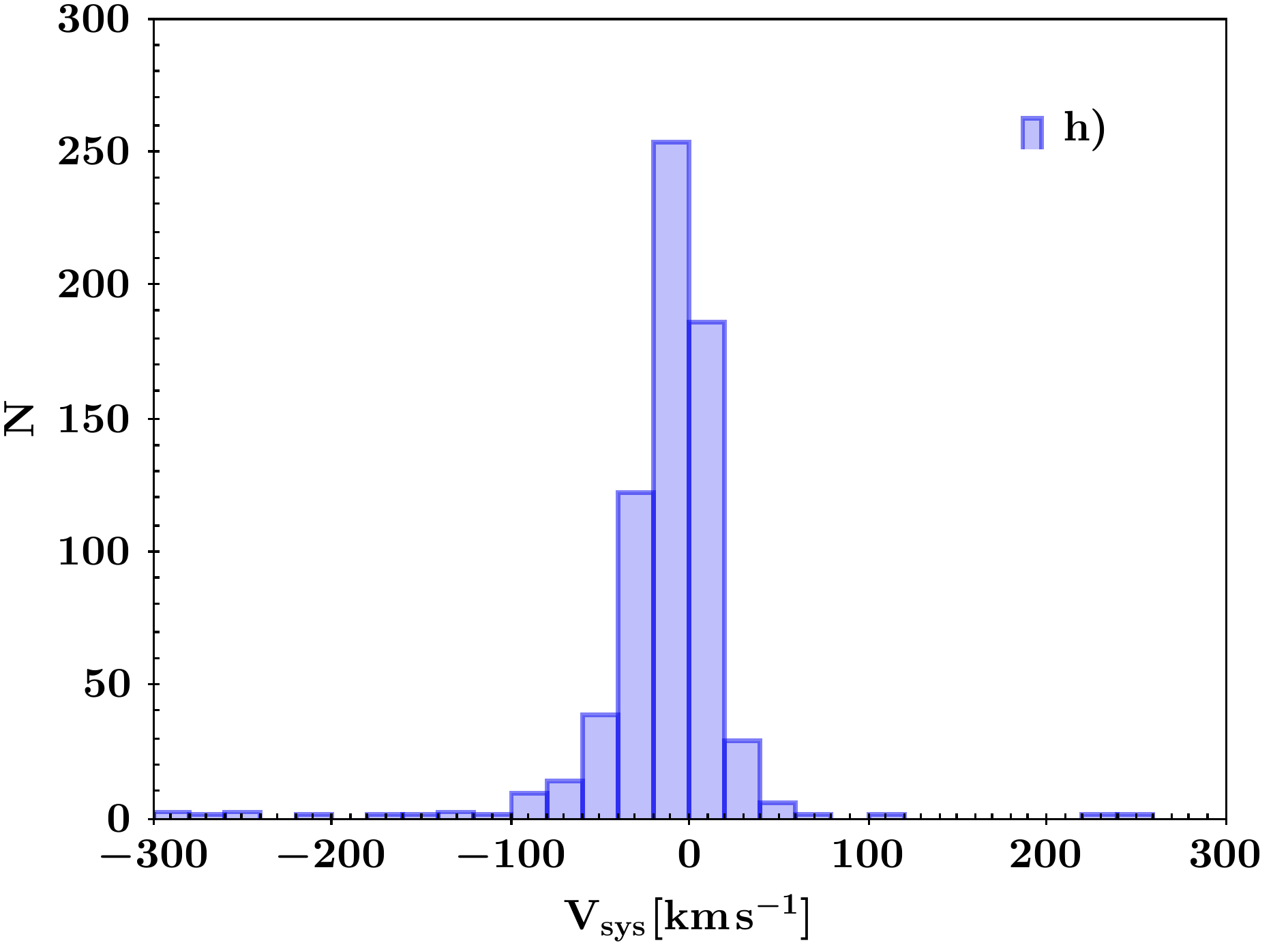}
\caption{Distributions of {\sc fado} results for our sample of 680 galaxies. First row panels: averaged age $\langle\tau_{M}\rangle$ and $\langle\tau_{L}\rangle$, weighted by mass and by luminosity, respectively, at left and right, in Gyr;  Second row panels, averaged metallicity in units of solar abundance, $\langle Z_{M}/Z_{\sun}\rangle$ and $\langle Z_{L}/Z_{\sun}\rangle$, again weighted by mass and by luminosity, respectively,  at left and right; third row panels, the formed and current stellar mass in solar mass units and logarithmic scale, $\log{M_{*,\mathrm formed}}$ and $\log{M_*,{\mathrm current}}$; four row bottom panels: dispersion velocity and systemic velocity, $\sigma$ and $v_{sys}$, both in km\,s$^{-1}$ units.}
\label{fig:hist}
\end{figure*}

Now, we analyze the results obtained from {\sc fado} for these quantities: formed stellar mass, averaged metallicity and age, as $\rm M_{formed}$, $\langle Z_{M}/Z_{\sun}\rangle$, or $\langle Z_{L}/Z_{\sun}\rangle$, $\langle \tau_{M}\rangle$, or $\langle\tau_{L}\rangle$, and dispersion velocity $<\sigma>$. 
First, we plot the results in term of histograms of the found characteristics in Figure~\ref{fig:hist}. In the first row, we show the distribution of ages
weighted by mass and by luminosity, respectively, at left and right, in Gyr. We see that the second method provides younger ages than the method weighting by mass, the maximum being in $\sim$ 3\,Gyr instead of 8\,Gyr as in the first panel. Second row panels show the averaged metallicity in units of solar abundance, $\langle Z_{M}/Z_{\sun}\rangle$ and $\langle Z_{L}/Z_{\sun}\rangle$, also weighted by mass and by luminosity, respectively,  at left and right. Similarly, we see that the weighted by mass metallicities behaves as a Gaussian with a maximum around $\sim 1.00-1.25$ solar metallicities, while the weighted by luminosity metallicities are biased towards metal-poorer objects with the maximum being lower than solar. 
\begin{figure*}
\includegraphics[scale=0.42]{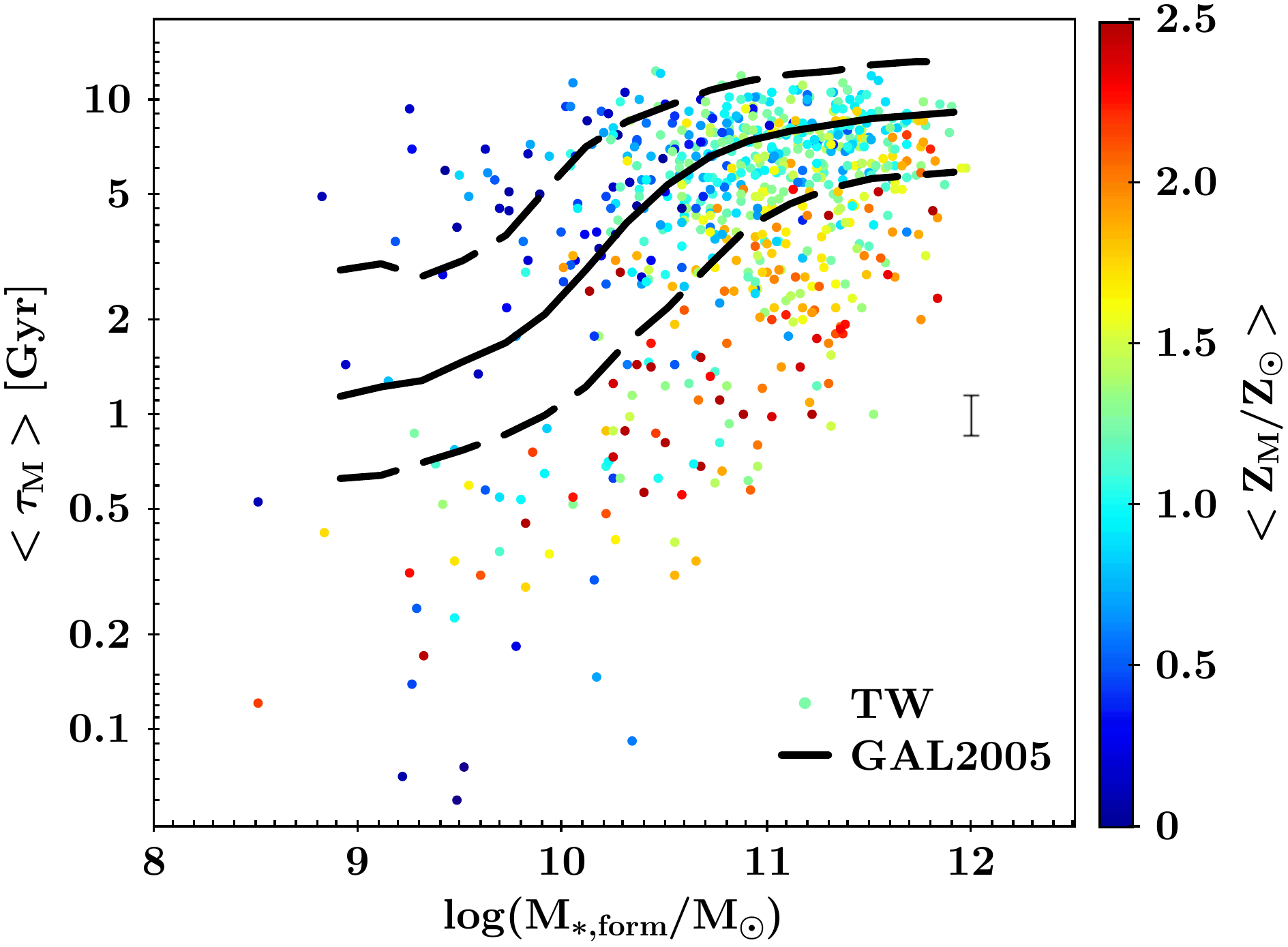}
\includegraphics[scale=0.42]{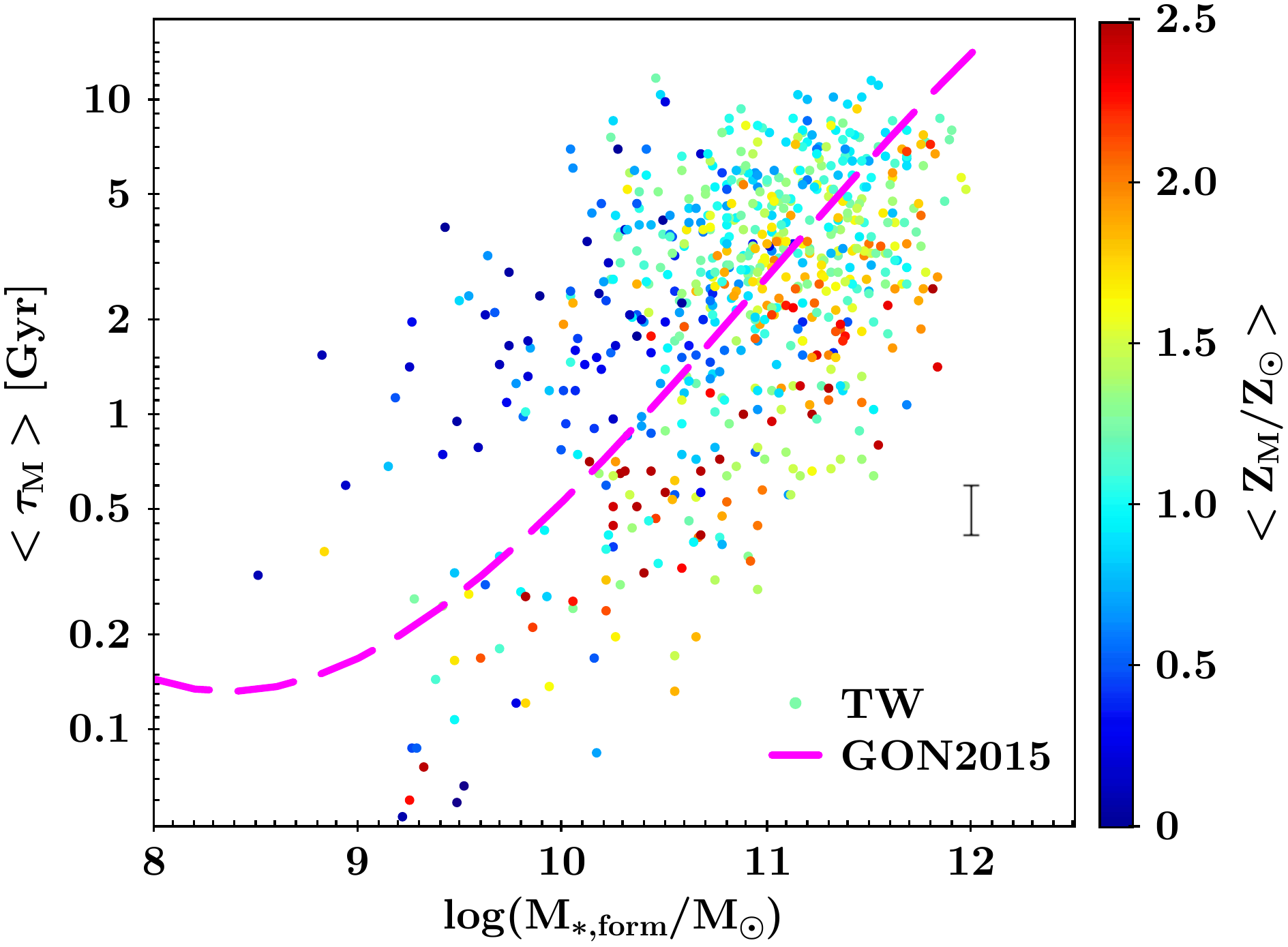}
\includegraphics[scale=0.42]{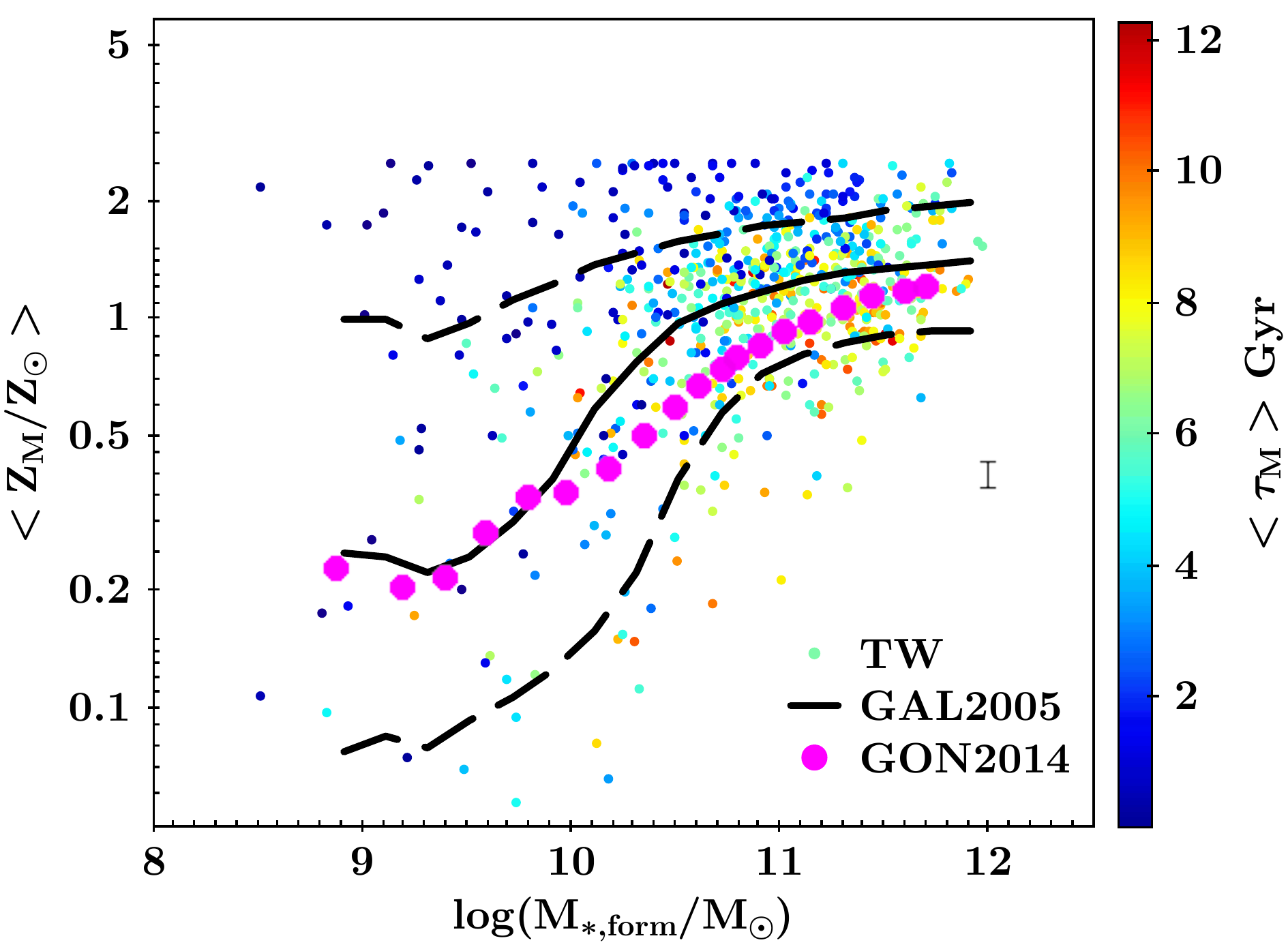}
\includegraphics[scale=0.42]{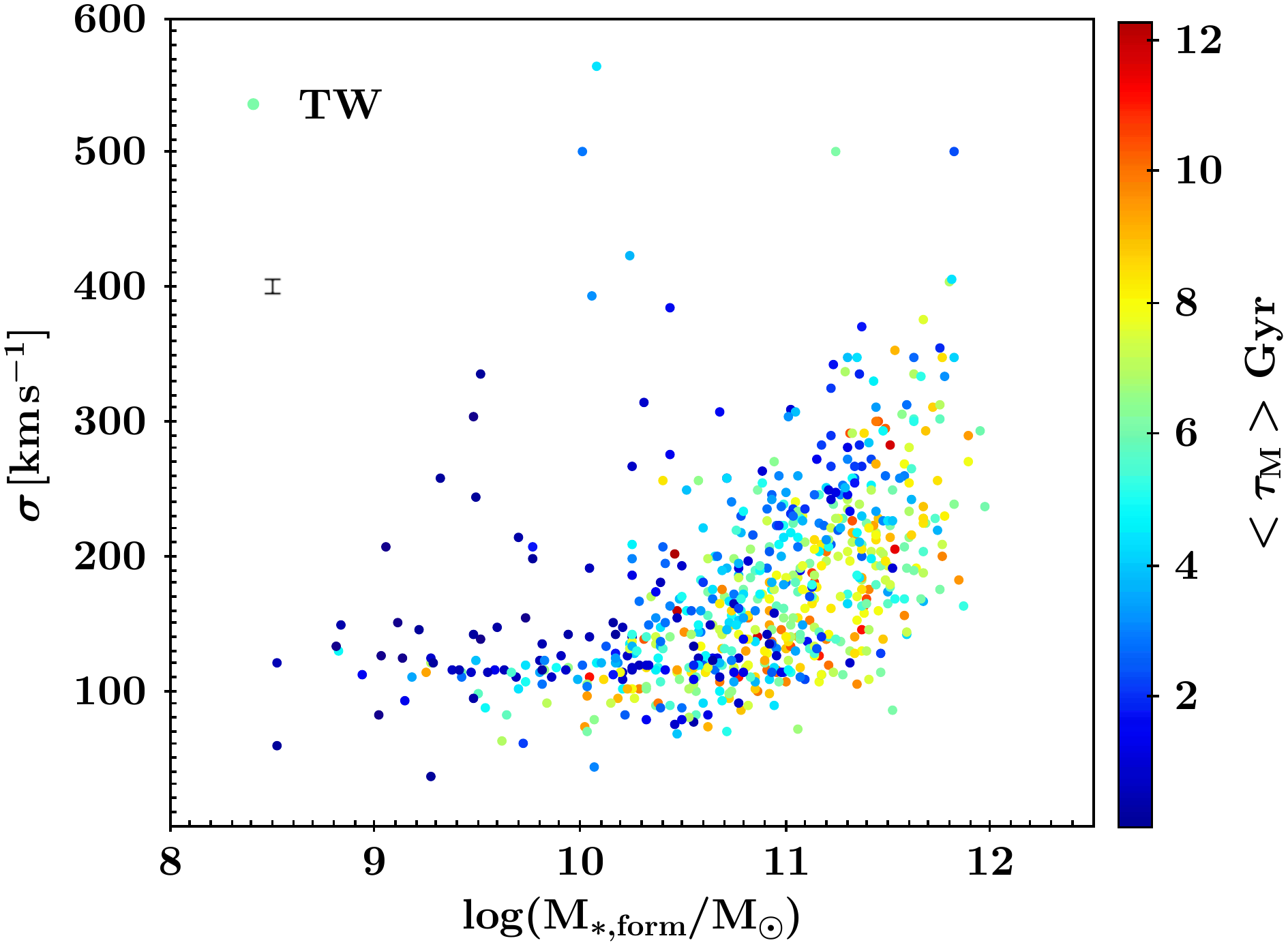}
\caption{Different quantities estimated from {\sc fado} as a function of the formed stellar mass $\log{(M_{\mathrm formed}/M_{\sun})}$: a) the averaged weighted by stellar mass age, as $\langle Z_{M}/Z_{\sun}\rangle$,  b) the averaged weighted by luminosity age, as $\langle Z_{L}/Z_{\sun}\rangle$, c) the weighted by mass averaged stellar metallicity, $\langle Z_{M}/Z_{\sun}\rangle$, and d) the dispersion velocity $\langle\sigma\rangle$,  Solid and dashed black lines are the fits found by GAL2005 while the magenta dashed line and dots are the fits to the CALIFA galaxies results by GON2015 and GON2014, as labelled (see text for meaning of references).}
\label{fig:correlations}
\end{figure*}
In third row panels, we have the formed and current stellar mass in solar mass units and logarithmic scale, $\log{M_{\mathrm formed}}$ and $\log{M_{\mathrm current}}$ with very similar curves; Finally, four row bottom panels show the dispersion velocity and systemic velocity, $\sigma$ and $v_{sys}$, both in km\,s$^{-1}$ units.
The first of these two indicates the stellar mass. The second one is an indicator of the errors in the redshift estimates, with large values for the highest objects of the zCOSMOS/MAGELLAN sub-sample.

In Figure~\ref{fig:correlations}, we
analyze the classical mass metallicity, and mass age relations. In each panel we indicate the mean error of the age with an error bar. In top panels, we have the relations of the weighted by mass age and the weighted by luminosity age with the stellar mass, $\langle \tau_{M}\rangle$ and $\langle\tau_{L}\rangle$ {\sl vs}  $\log{\mathrm M_{*,form}}$. The color is given in the right side scale and refers to the weighted by mass metallicity $\langle Z_{M}/Z_{\sun}\rangle$. The solid lines at the left panels are the resulting curves by \citet[][hereafter GAL2005]{gallazzi2005} with a black line limited by dashed lines representing the limits of 16-84\% around this averaged trend. We see that our results show a larger dispersion with lower ages than those authors for the low mass galaxies. In the right panel, we plot the age weighted by luminosity, compared with the fit found with the results from \citet[][hereafter GON2015]{gonzalezdelgado2015} --as a magenta dashed line-- computed in the same way for the CALIFA survey galaxies.
When we use $\langle\tau_{L}\rangle$, we obtain lower values than using $\langle\tau_{M}\rangle$, reproducing well the the CALIFA galaxies findings\footnote{In CALIFA works, the usual age is measured as an averaged of the logarithmic weighted by luminosity values, that is $\langle\log{\tau_{L}}\rangle$, while we are representing the logarithmic of the weighted by luminosity averaged age, $\log{\langle\tau_{L}\rangle}$. That may produce some differences in the results}. 

In next left panel at the second row, we show the mass-metallicity relation $\langle Z_{M}/Z_{\sun}\rangle$ {\sl vs}  $\mathrm M_{*,formed}$; we see the well known correlation with a trend to have more metal-rich galaxies in the high stellar mass region;  In this case, the color scale depends on the weighted by mass stellar age, given at the right of the panel. We have again drawn the findings from \citet{gallazzi2005} with a black line limited by dashed lines representing the limits of 16-84\% around this averaged.  We have also plot the trends obtained by \citet[][hereafter GON2014]{gonzalezdelgado2014} for CALIFA galaxies using the code {\sc starlight}, drawn as magenta dots. 
In the last panel, at the right of the second row, we show the classical dependence of the dispersion velocity with the formed stellar mass, the color scale indicating the averaged weighted by mass stellar age. 
\begin{figure}
\centering
\includegraphics[scale=0.4,angle=0]{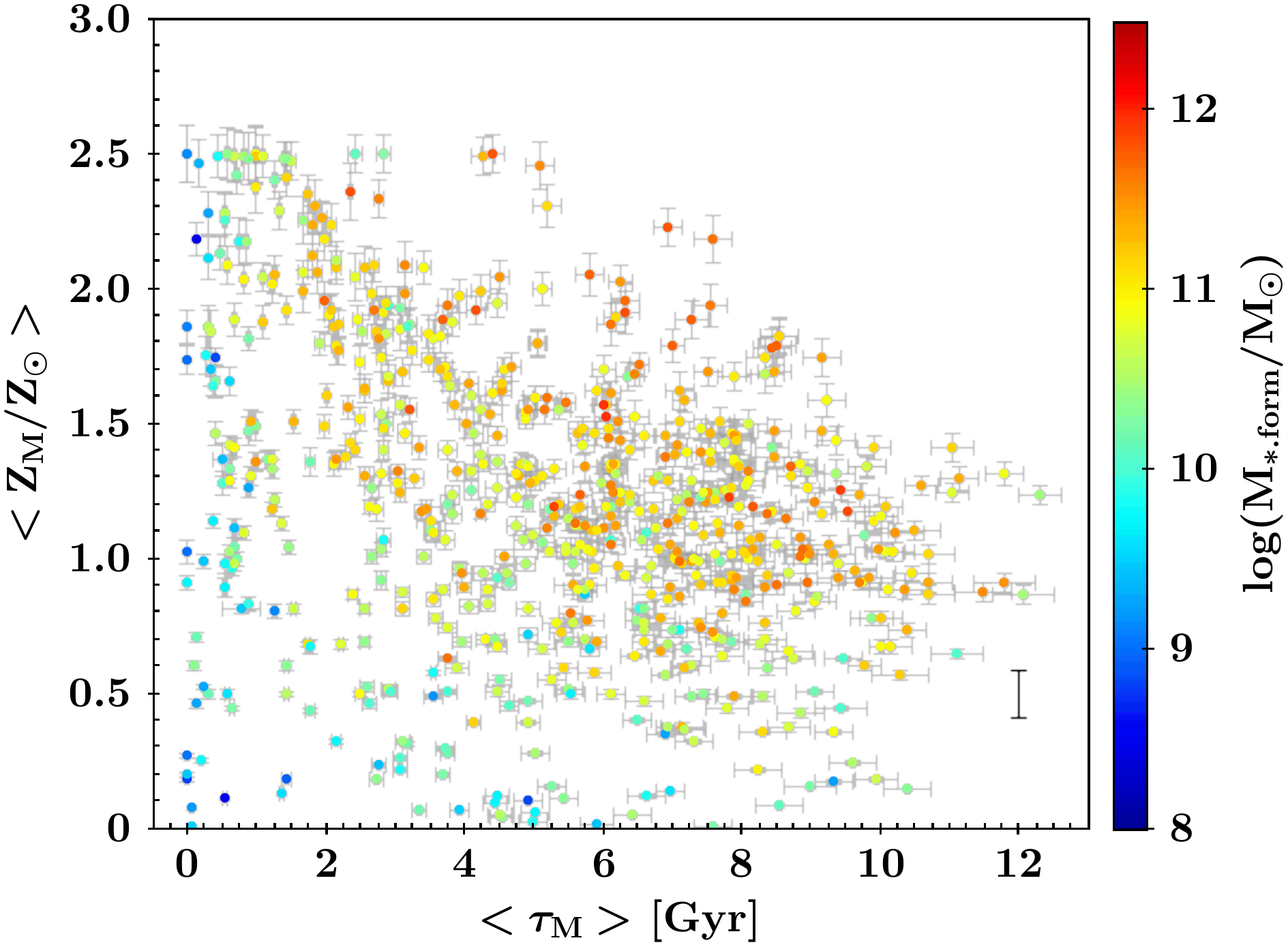}
\caption{Age-metallicity relation, given by $\tau_{M}$ as a function of the metallicity  $\langle Z_{M}/Z_{\sun}\rangle$, obtained for the stellar populations of our galaxies. The errors are included for all points. The color scale indicates the formed stellar mass.}
\label{fig:age-z}
\end{figure}

In all cases the dispersion of results is quite large. Furthermore, a underlying correlation age-metallicity might be biasing the results. Therefore, we plot the age {\sl vs} the metallicity in Fig.~\ref{fig:age-z}. We see that there is a trend, with old stellar populations in the metal-poor regions, but not a correlation since a large cloud of points appears with young ages for metal-rich and metal-poor objects. Therefore, the classical age-metallicity degeneracy does not appear so clear in this work.

Our method uses {\sc fado} to obtain these parameters for the stellar populations by using as {\sl basis} the {\sc HR-pyPopStar} models that are specially devoted to the computation of the youngest stellar populations, thus by including the most recent stellar models library for the hot stars as WR, O and B stars.  In turn,  \citet{gallazzi2005} used the \citet{bc03} models to fit some specific spectral absorption lines in order to obtain averaged stellar age and metallicity, $\langle\tau_{L}\rangle$  and $\langle Z_{M}/Z_{\sun}\rangle$, by estimating the formed stellar mass from the continuum luminosity in $z$-band, $L_{z}$. \citet{gonzalezdelgado2014,gonzalezdelgado2015} used the code {\sc starlight}, similar to {\sc fado},  with the SEDs computed by their own group based in isochrones by Padova and Geneva (depending on the age) and the stellar libraries from Granada and MILES \citep{gonzalezdelgado2005,martins2005,vazdekis2010,falcon-barroso2011}. 
Our results follow the same trends as these works found for the CALIFA survey galaxies, once the differences in IMF and in the computation of parameters weighted by mass or luminosity are taken into account. Our averaged stellar ages show to be younger for a given stellar mass, specially for the lower stellar mass galaxies, compared with GON2015. This result is actually interesting since these points correspond to the most metal-poor galaxies, such as we see in the third panel of $\langle Z_{M}/Z_{\sun}\rangle$, {\sl vs} the formed stellar mass as $\log{(M_{*.formed}/M_{\sun})}$, where there exist points below the lowest line from GAL2005. We interpret these differences as the result of this improvement in the SEDs of the youngest SSPs in {\sc HR-pyPopStar}. Probably,  this use of HR spectra for the fit is also the cause of the almost nonexistent age-metallicity degeneracy. On one hand, these new  models are better to estimate the stellar ages of the low metal low mass galaxies. On the other hand, the HR spectra have a large number of stellar absorption lines when the metallicity is solar or super-solar, which does not appear when the metallicity is low, even at old ages  \citep[see details in spectra from][]{millan-irigoyen+2021}. This allows to better discriminate the age and the metallicity of the stellar populations in our galaxies.

\begin{figure}
\centering
\includegraphics[scale=0.3,angle=-90]{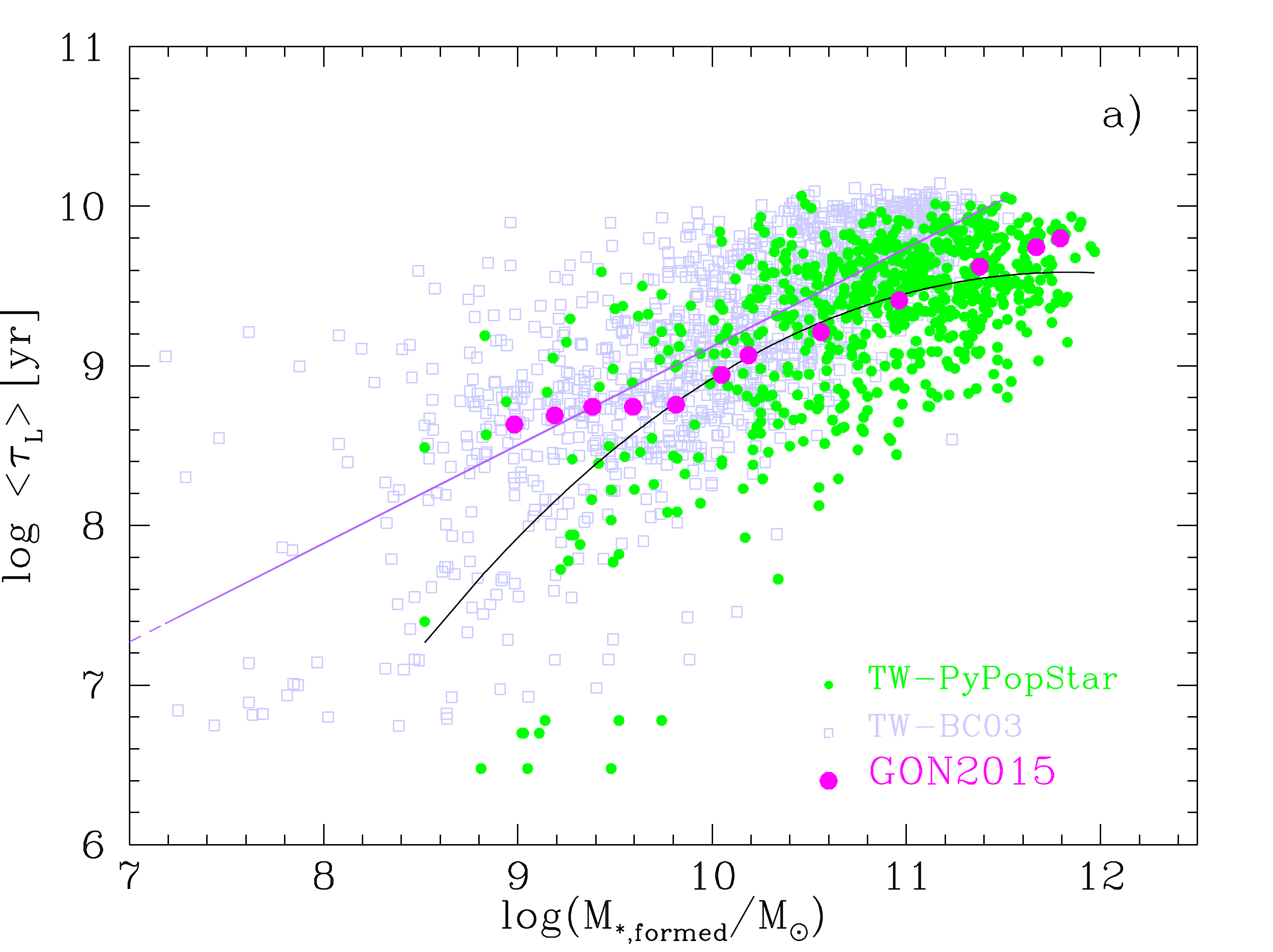}
\includegraphics[scale=0.3,angle=-90]{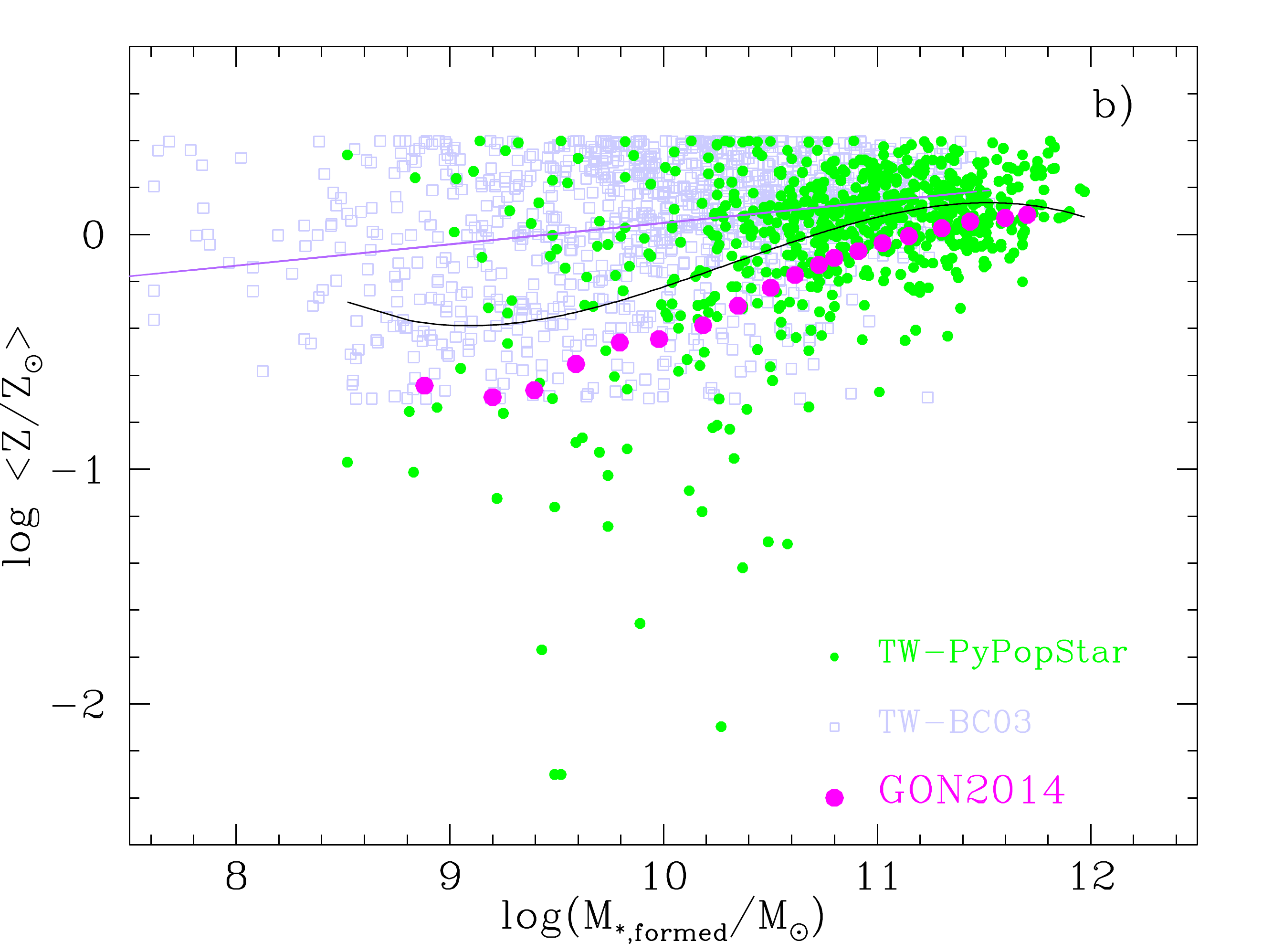}
\caption{Comparison of results by using {\sc fado} with different SSP bases, {\sc HR-PyPopStar} in green dots and BC07 in grey squares: Top) Weighted by luminosity averaged age $\langle\tau_{L}\rangle$; Bottom) weighted by mass averaged metallicity, $\langle Z_{M}/Z_{\sun}\rangle$, {\sl vs} the formed stellar mass as $\log{(M_{*.formed}/M_{\sun})}$. In both plots we have also included the results from GON2014 and GON2015 as in previous figure~\ref{fig:correlations} as labelled.}
\label{BC03-PyPoStar}
\end{figure}

\begin{figure}
\centering
\includegraphics[scale=0.42]{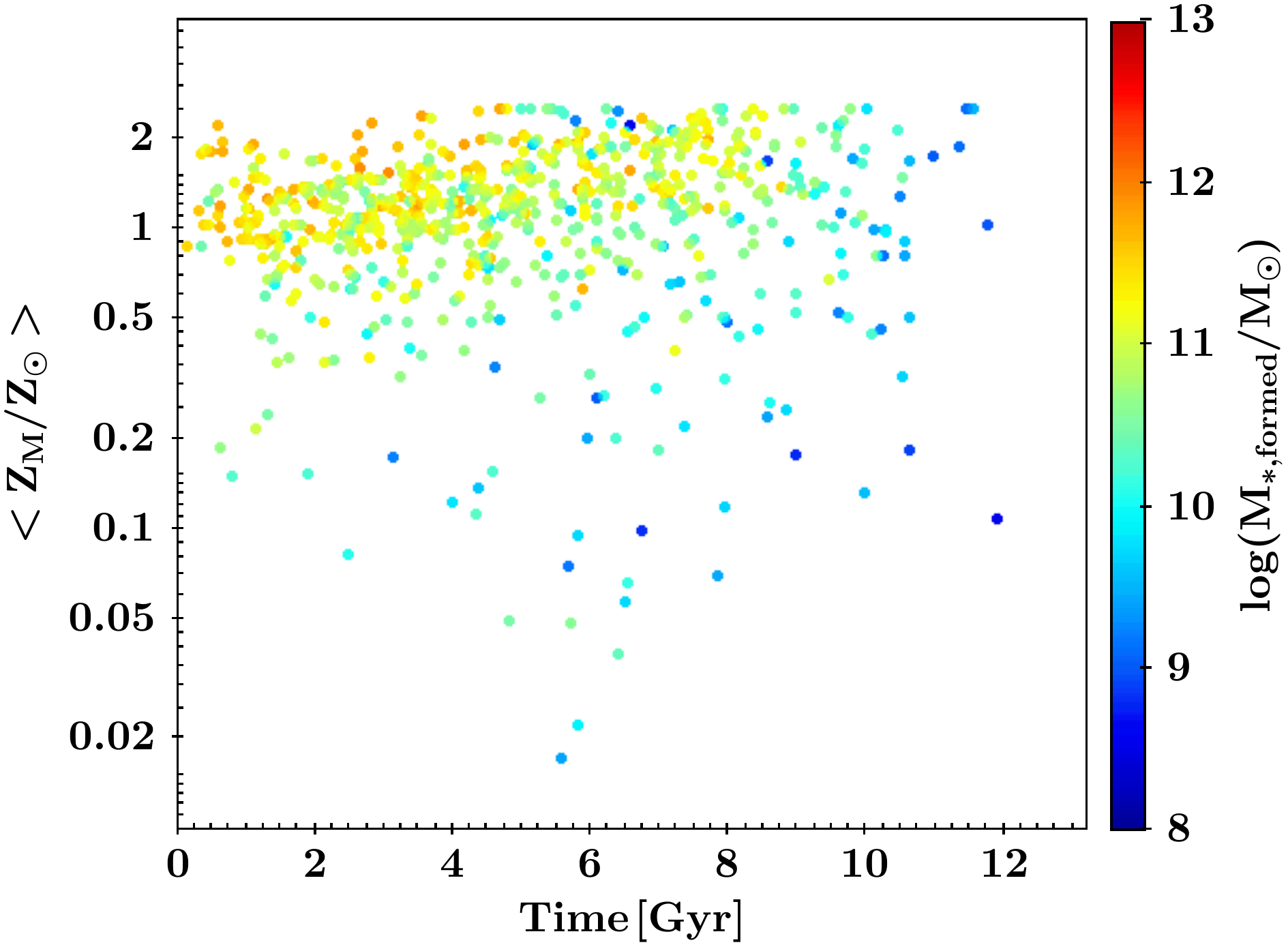}
\includegraphics[scale=0.42]{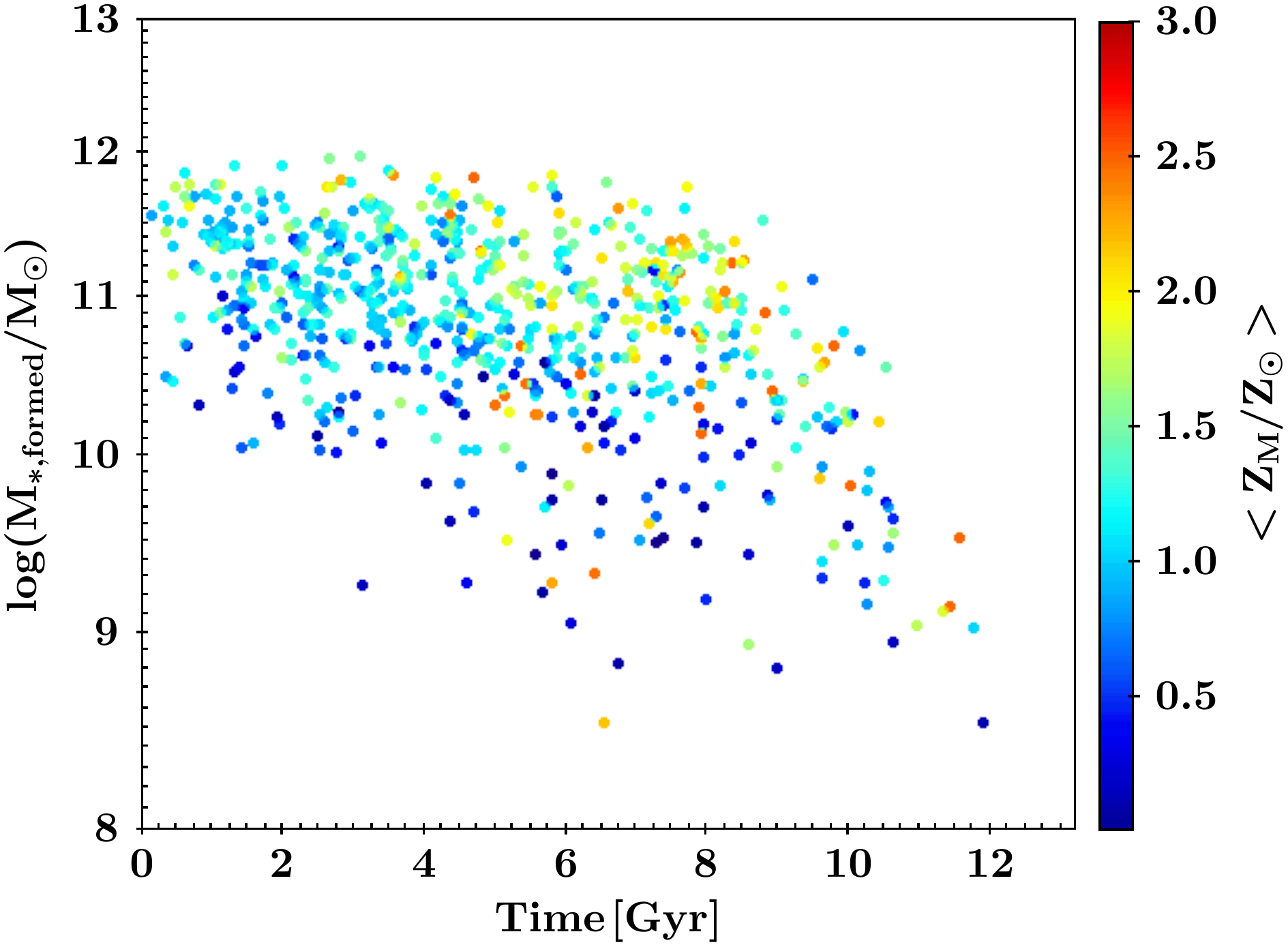}
\includegraphics[scale=0.42]{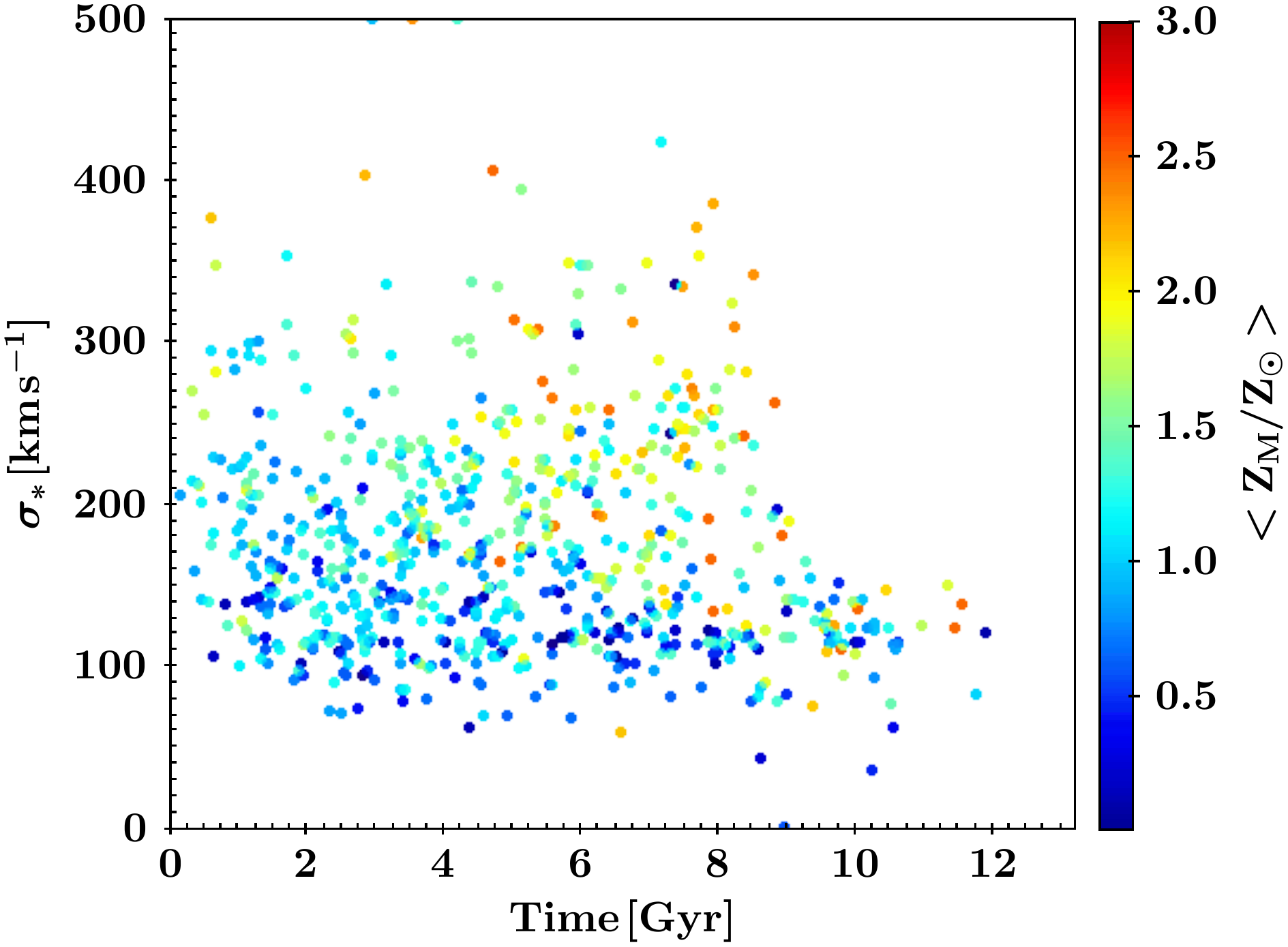}
\caption{Time evolution of the formed stellar mass, metallicity and dispersion velocity.}
\label{fig:M_and_Z_vs_time}
\end{figure}

In order to check if this statement may be due to our new set of SSPs from {\sc HR-pyPopStar} used as basis or if this is an effect of selection of the own code {\sc fado}, we have done  the same calculations for SDSS galaxies, by using now the set of spectra from BC03 included as default with the code release. We plot our results in Fig.~\ref{BC03-PyPoStar} with the BC03 results as grey dots and the {\sc HR-PyPoStar} results as green squares. We have included a least-squares straight line fitted for our results obtained with BC03 as {\sl basis} (Trying with a two slope fit or a two grade polynomial function does not produce large differences over this fit). We have also calculated a polynomial function for our results computed with {\sc HR-pyPopStar}. We over-plot the binned results from GON2015 for CALIFA galaxies as magenta dots. In the bottom panel, results from GON2014 are again represented as the large magenta dots as binned values. We see that, effectively, by using {\sc HR-pyPopStar} spectra, both, age and metallicity result to be lower than those obtained by using BC03, being similar as the GON2014 and GON2015. This demonstrate the importance of using the adequate SSP spectra for young stellar populations within a code as {\sc fado} or {\sc starlight}.
In both figures we have also added our best fits to our results with a polynomial function:
\begin{equation}
\label{pol}
y=ax^{3}+bx^{2}+cx+d,
\end{equation}
where $x=\log{(M_{*,formed}/M_{\sun})}$, and $y=\log{\tau_{L}}$ and $\log{(Z_{M}/Z_{\sun})}$, respectively. The values of $a,b$ and $c$ are given in Table~\ref{table: misfits}.

\begin{table}
\caption{Coeficients of the polynomical functions fitted to our results as Eq.~\ref{pol} drawn as black solid lines in Figure~\ref{BC03-PyPoStar}.}
\label{table: misfits}
\begin{tabular}{lcccc}
\hline
 & a & b & c & d \\
\hline
$\log{\tau_{L}}$  & 0.009529 & -0.516657& 8.229098 & -31.234329 \\
$\log{(Z_{M}/Z_{\sun})}$ & -0.073128 & 2.26080 & -22.974613 & 76.569707\\
\hline
\end{tabular}
\end{table}

Since we have the age of the Universe for each redshift, we may also represent our results in terms of the evolutionary time. Thus, we have the evolution along the time in Figure~\ref{fig:M_and_Z_vs_time}. We see in the top panel the evolution of the metallicity $\langle Z_{M}/Z_{\sun}\rangle$ with time with a color scale showing the formed stellar mass as $\log{(M_{*.formed}/M_{\sun})}$. We  show clearly that the enrichment history of the universe is not homogeneous, with the massive galaxies showing a very smooth evolution, almost a flat line, while the low mass galaxies show a larger dispersion and an apparent stronger increase along the time. In the middle panel, we show the evolution of the formed stellar mass, $\log{(M_{*.formed}/M_{\sun})}$, indication of the star formation rate in galaxies. We see that  the stellar mass formed in each galaxy decreases with time, as expected from the evolution of the SFR of the Universe in this range of redshift/time. There exists a dependence with the metallicity too, being the low metallicity galaxies below the most metal ones, that is showing a correlation mass-Z for all times, with low Z objects creating lower stellar mass than the high-rich galaxies. The trend given by these low metallicity galaxies is continuously decreasing, as a straight line, while the most metal ones show something similar to a plateau until 8\,Gyr with a stronger decreasing afterwards.
The dispersion velocity is presented in the bottom panel also showing lower values for the less metal rich galaxies.

\section{Hubble diagram and metallicity dependence}
\label{sec: HR}

Finally, to see if there is any correlation between SNe~Ia parameters and their host galaxy properties, the main aim of this work, we measure the Hubble residual (HR), the difference between the SN~Ia distance modulus ($\mu_{obs}$) and the corresponding $\mu$ of a flat $\Lambda$CDM cosmology with $\Omega_{M}=0.315$ and $H_{0}$=70 km s$^{-1}$ Mpc$^{-1}$.
In order to compute the distance modulus for the supernova, we use the SALT2 model \cite{salt2} and the Tripp formula,
\begin{equation}\label{Eq:mu_GTC}
\mu_{\rm obs}= -2.5 \log x_0 - M + \alpha x_1 - \beta c,
\end{equation}
where $x_0$ is the normalization factor (from which one can obtain the
apparent magnitude at maximum brightness in the B band; $m_B$), $x_1$ and $c$ are the stretch and the color of the SN~Ia light-curve, respectively, $M$ is the average magnitude of the SALT2  for a $x_1 =c=0$ SNe~Ia at 10\,kpc, and $\alpha$ and $\beta$ are nuisance parameters that represent the luminosity-width and luminosity-color relations, which result from minimizing the difference between $\mu_{obs}$ and the distance modulus $\mu$ of the assumed fiducial cosmology.

The light-curve parameters of the low and intermediate redshift SDSS SNe~Ia are directly taken from \cite{Sako+2018}.
For the 6 intermediate redshift SNe Ia from Union 2.1 (whose galaxies were observed with GTC) and the 20 high redshift SNe Ia from SNLS, we obtained the light-curve parameters by fitting their light-curves using {\sl sncosmo} \citep{barbary_kyle_2022_6363879} and the SALT2 model. 
The nuisance parameters
($M$, $\alpha$, $\beta$) used here are those from \cite{Sako+2018}, ($-$29.967, 0.187 $\pm$ 0.009, 2.89 $\pm$ 0.09), since our focus is finding possible correlations between SNe~Ia parameters and the properties of their hosts, and it is not our objective to determine the best cosmological parameters driven by our dataset.

\begin{figure}
\centering
\vspace{-1cm}
\includegraphics[scale=0.55]{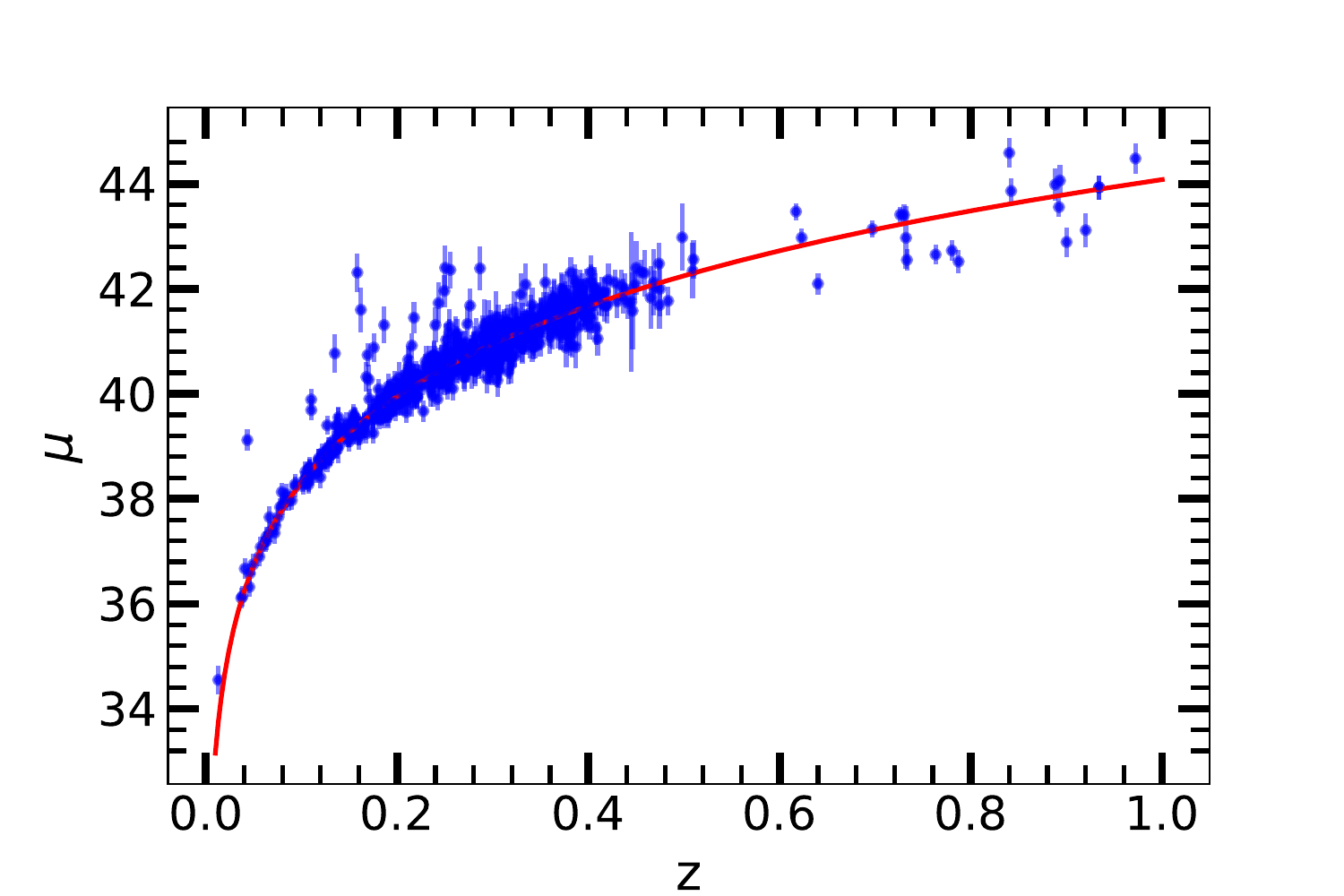}
\vspace{-0.5cm}
\caption{Hubble diagram using the supernovas included in this work as blue dots, and a flat lambda-CDM cosmological model with $H_0=70\,km\,s^{-1}\,Mpc^{-1}$ and $\Omega_{M} = 0.315$ drawn as a red line. The dots for which $|\mu_{obs}-\mu|> 0.1$ will be discarded in our next analysis (see text).
}
\label{fig:Hubble diagram}
\end{figure}

The Hubble Diagram is presented in Figure~\ref{fig:Hubble diagram}, where blue dots are the results obtained from the above Eq.\ref{Eq:mu_GTC} by using the values from Table~\ref{table:res-fado}, and the red line corresponds to the assumed cosmological model. Once we have made the Hubble Diagram and computed the Hubble Residuals, HR, we discard the outliers of our sample, which we consider as those ones with $HR<-1$ or $1<HR$. We have discarded 14 outliers, so the total number of objects that we have used for the analysis is 664.

\begin{figure}
\centering
\includegraphics[scale=0.28,angle=-90]{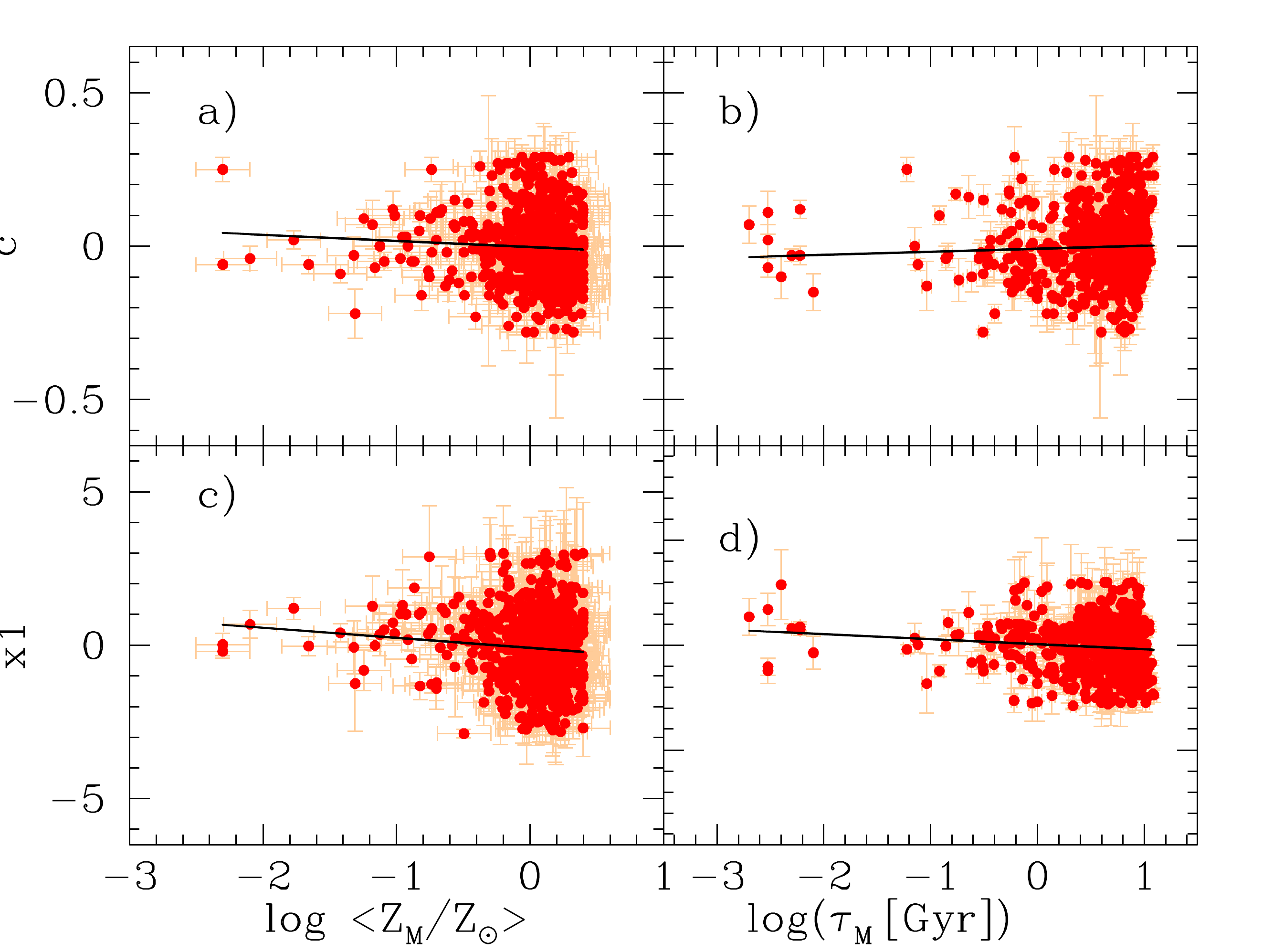}
\caption{The characteristics of SN Ia, $c$, and $x1$ as a function of the stellar metallicity $\langle Z_{M}/Z_{\sun}\rangle$ and of the stellar age $\tau_{M}$ as red dots. The error bars are plotted in orange for sake of clarity. }
\label{fig: fig14}
\end{figure}

We have checked the possible dependence of the SN~Ia parameters, $c$, and $x1$ on the stellar metallicity and age, as presented in Figure~\ref{fig: fig14} looking for the existence of any correlation. The least-squares straight lines are over-plotted as a black line in each panel. The slopes of the relationships between $c$  {\sl vs} $\log{(\langle Z_{M}/Z_{\sun}\rangle)}$, and {\sl vs} $\log{(\langle\tau_{M}\rangle)}$ are: $-0.023\pm 0.013$, and  $0.012\pm 0.0080$\,dex$^{-1}$, respectively. Their significance are $1.79\sigma$ and $1.47\sigma$ respectively, implying that the effect of the galactic parameters on the color is small. For $x1$, as a function of the same parameters, $\log{(\langle Z_{M}/Z_{\sun}\rangle)}$ and $\log{(\langle\tau_{M}\rangle)}$, the least-squares straight lines give slopes  $-0.36 \pm 0.14$ and $-0.240\pm 0.086$\,dex$^{-1}$, and significance of $2.73 \sigma$ and $2.81\sigma$, respectively. This last correlation was also found by other authors of the literature, such as \citet{Johansson+2013}. Our slope, however, is much lower than their value for the dependence on the age: $-1.86$ \,dex$^{-1}$ for these authors, against $-0.24$\,dex$^{-1}$ in our result. Probably, the low number of objects of their sample compared with the ours is the reason of these discrepancies. Moreover, their method, by using the stellar spectral indices, is different than the technique of {\sc fado}, what also may explain the variation in this slope. In fact, we find a steeper slope (although with smaller significance) in the relation with the metallicity compared with the one of the age. Given the values of these slopes, we may say that we do not find any strong correlation between color $c$ and the stellar age or metallicity, but there exists a slight one between $x1$ and the stellar parameters. The dependence of the stretch parameter $x1$ on stellar metallicity and age maybe is suggesting a relation between the scenario of the binary system (SD/DD) to produce a SN Ia, and the corresponding DTD, and the stellar populations where these systems formed.

\begin{table}
\caption{Summary of the fits of the $x1$ and $c$ with age and metallicity and their significance.}
\label{Table:x1_c_results}
\begin{tabular}{lcc}
 \hline
 Relation &  \multicolumn{2}{c}{Parameter} \\
          &  slope & significance\\ 
 \hline
c {\sl vs} $\log{<Z_{M}>}$  &  -0.0023 $\pm$ 0.013 & 1.79\\ 
c {\sl vs } $\log{<\tau_{M}>}$  &  0.012 $\pm$ 0.008 & 1.47\\ 
x1 {\sl vs} $\log{<Z_{M}>}$ &  -0.36 $\pm$  0.14 & 2.74\\ 
x1 {\sl vs } $\log{<\tau_{M}>}$ & -0.240 $\pm$ 0.086 &  2.81\\ 
\hline
\end{tabular}
\end{table}

\begin{figure}
\centering
\hspace{-0.3cm}
\includegraphics[scale=0.4]{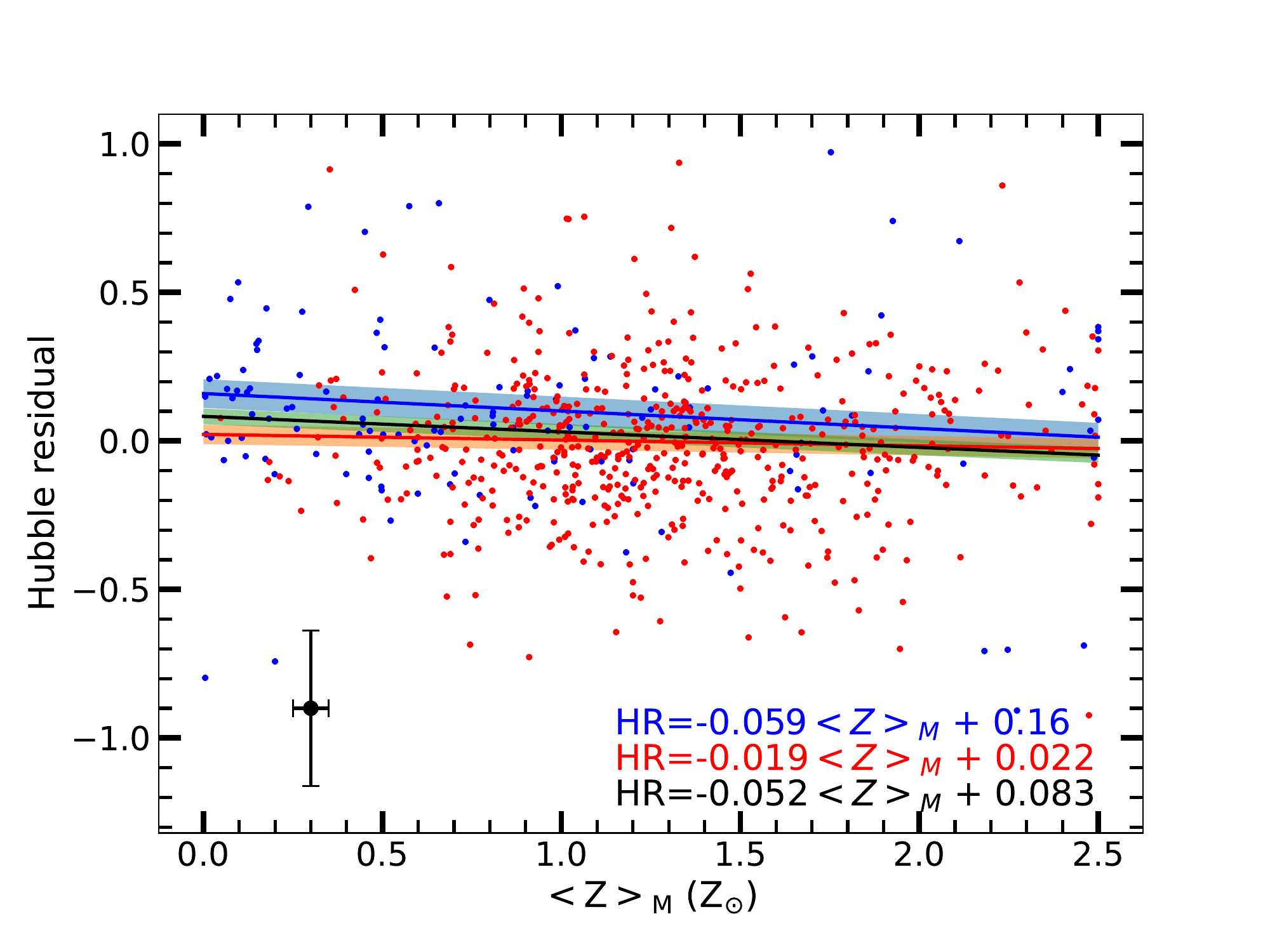}
\hspace{-0.5cm}
\includegraphics[scale=0.4]{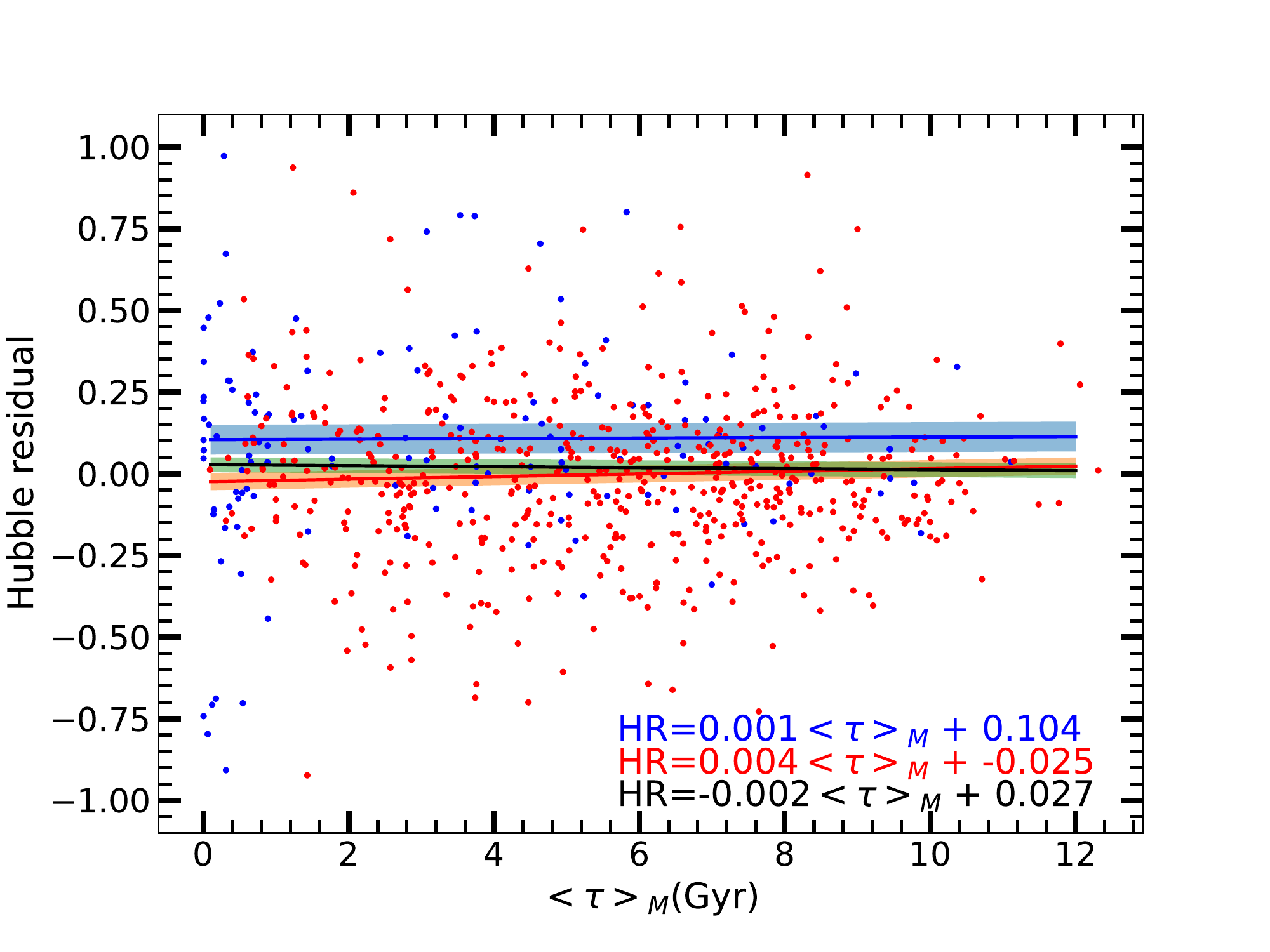}
\vspace{-0.1cm}

\includegraphics[scale=0.4]{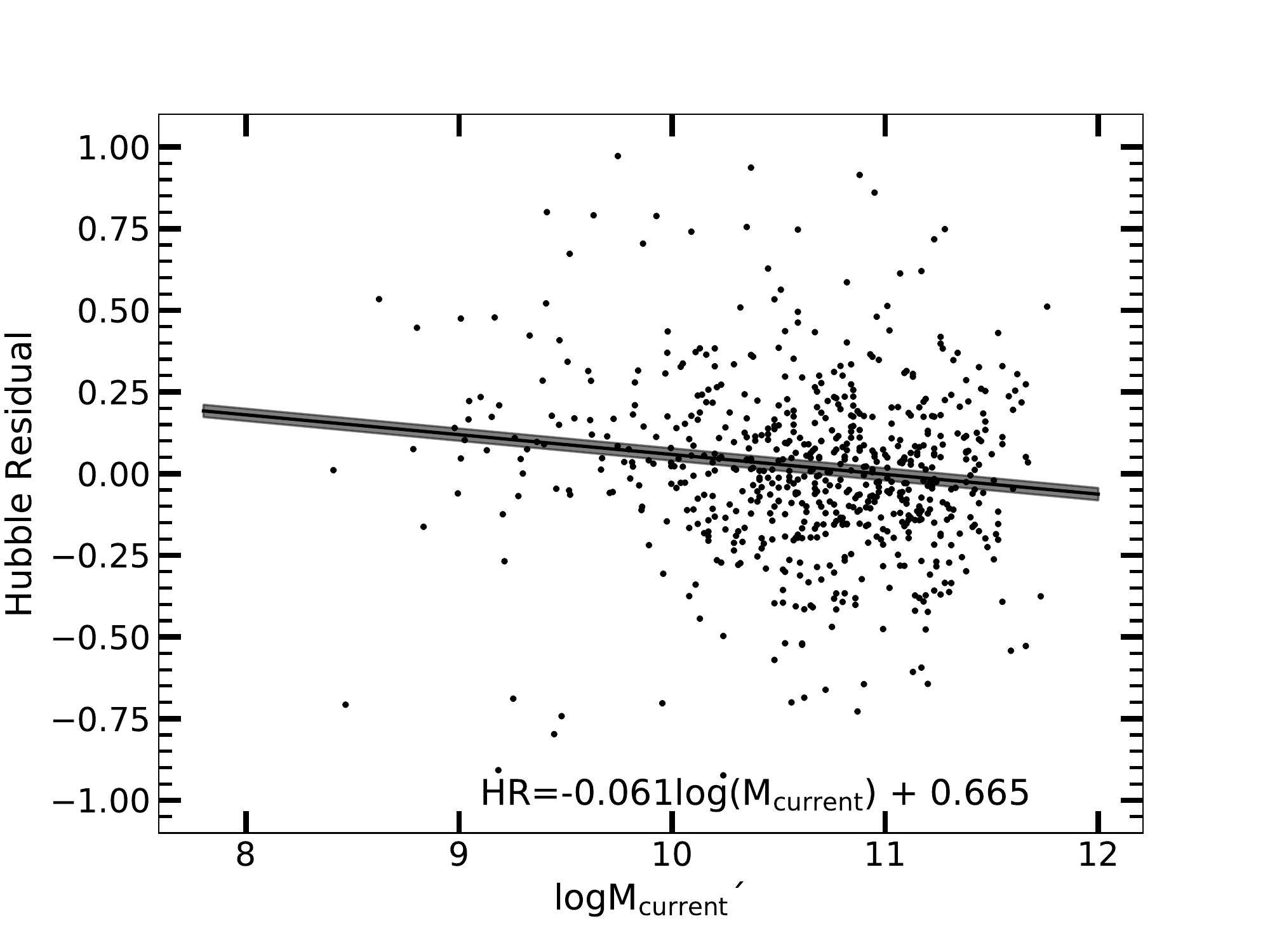}
\caption{Hubble residuals HR of the SN Ia supernovae as a function of: top) the stellar metallicity $\langle(Z_{M}/Z_{\sun})\rangle$ for the whole sample; middle) the stellar age, $\langle\tau_{M}\rangle$  and the current mass, $\langle\log(M_{current})\rangle$. The solid black line represents in both panels the corresponding least-squares fit to all shown points. Red and blue lines are the obtained fits for the massive and low mass galaxies, respectively (see text for details).}
\label{fig:Hubble_residual_fits}
\end{figure}

The resulting HR is then represented as a function of the same stellar population parameters, metallicity $\langle(Z_{M}/Z_{\sun})\rangle$, age 
$\langle\tau_{L}\rangle$, and created stellar mass in Figure \ref{fig:Hubble_residual_fits}.
We have performed a fit with a least squares straight line in each one. The linear fits including the 664 selected above galaxies are drawn as black lines in both panels. We have found a clear correlation for HR as a function of the mean stellar metallicity $\langle Z_{M}/Z_{\sun} \rangle$ with a slope of -0.0528\,mag and a significance of 2.69$\sigma$. We found no correlation between the Hubble residual and the averaged age when measured as $\langle\tau_{M}\rangle$. However, when we represent HR as a function of  $\langle\tau_{L}\rangle$, we do find a correlation with an slope of -0.0104\,mag. This is a result difficult to interpret, since the weighted by light age is giving information mainly about the youngest hot stars, that should have nothing to do with the older progenitor stars of SNe~Ia. Finally, we have also found a strong correlation between the HR and the current stellar mass $\log{M_{current}}$ with a slope of -0.061\,mag and a significance of 3.359, similar as other works in the literature. This result, taking into account the mass-metallicity relation, may be due to the metallicity itself, probably the main driver of the  relationship of bottom panel of Fig.\ref{fig:Hubble_residual_fits}.

\begin{figure}
\centering
\includegraphics[scale=0.4]{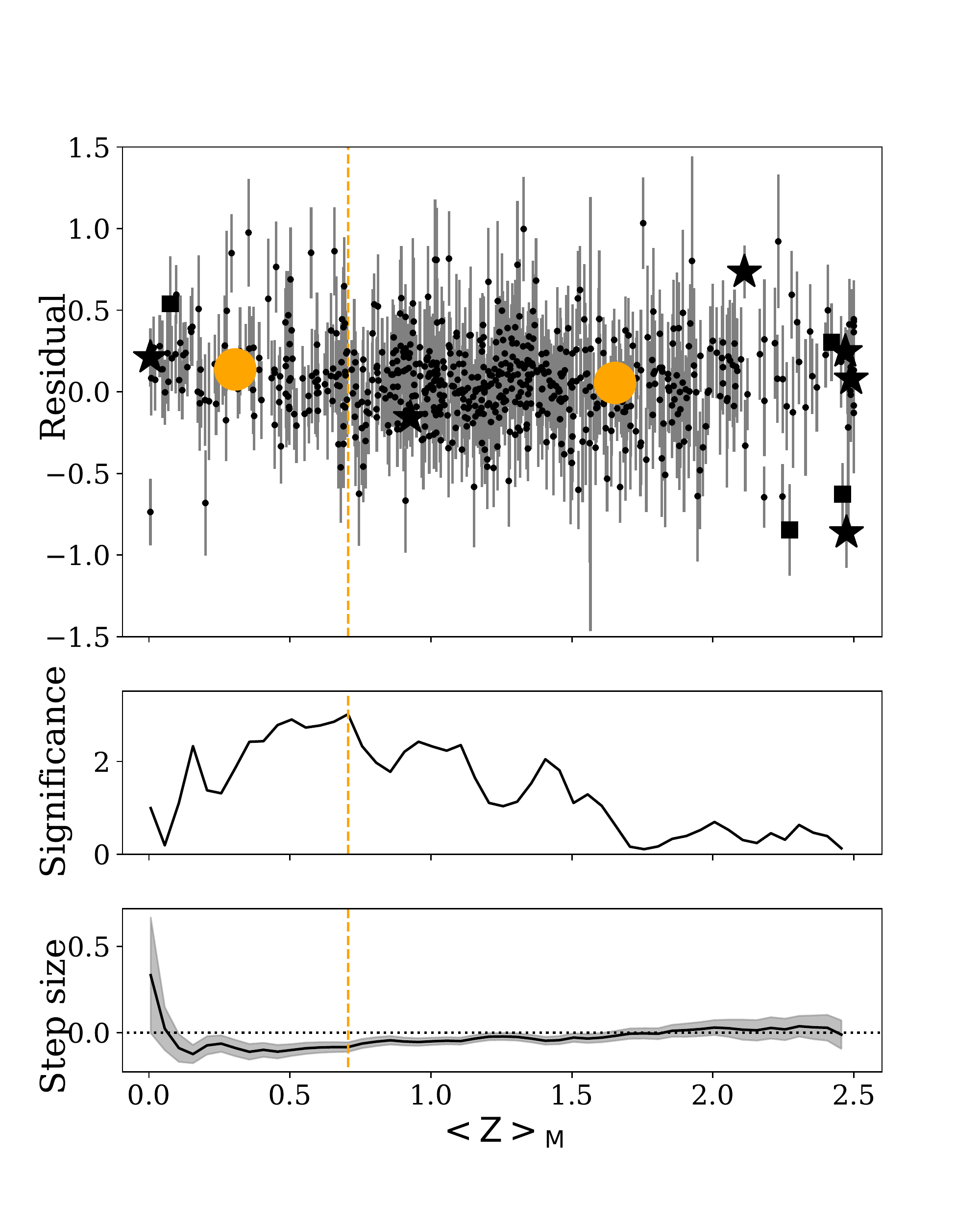}
\caption{The significance and step magnitude for different values of $\log{M_{split}}$. In the top panel, we show the Hubble residual of SDSS (circle), GTC-OSIRIS (stars) and zCOSMOS-MAGELLAN (square); in the middle panel the significance of the step in $\sigma$  is shown; and the bottom panel displays the magnitude of the step at each location with the grey shaded region showing the uncertainty. The orange dashed lines shows the step that gives the maximum significance.}
\label{fig:Significance}
\end{figure}

We have also looked for the presence of a difference in the these correlations when low-mass and high-mass galaxies are considered separately. In order to do this, we first need to classify the galaxies in {\it low} and {\it high} mass galaxies, searching first for the characteristic stellar mass, $M_{split}$, that best divides the sample in two groups. 
For that, we do different sets of 2 bins, we compute the mean and standard deviation of each bin to compute the significance of that point using,
\begin{equation}
    {\rm significance} = \frac{|\mu_1-\mu_2|}{\sqrt{\sigma^2_1+\sigma_2^2}},
\end{equation}
where $\mu_1$ is the mean of the left side, $\mu_2$ is the mean of the right side, $\sigma_1$ is the standard deviation of the left side and $\sigma_2$ is the standard deviation of the right side.
We move $\log{M_{split}}$ in steps of 0.05 dex in the range [9.0,11.4] in order to find the $M_{split}$ with the best significance. The results can be seen in Figure~\ref{fig:Significance}. We found that at $\log{M_{split}} = 10.2$ dex the significance is the highest. 

\small
\begin{table}
\caption{Summary of the fits for all the sample, low mass and high mass galaxies and their significance.}
\label{Table:Hubble_Residual_results}
\begin{tabular}{lcc}
 \hline
 HR vs &  \multicolumn{2}{c}{Parameter} \\
 Data   &  slope & significance\\ 
 \hline
$<\tau_{M}>$ low mass               &  0.00 $\pm$ 0.01 & 0.085\\ 
$<\tau_{M}>$ high mass              &  0.004 $\pm$ 0.004 & 0.961\\ 
$<\tau_{M}>$ all sample             & -0.001 $\pm$ 0.004 & 0.373\\ 
$<\tau_{L}>$ low mass               &  0.00 $\pm$ 0.01 & 0.770\\
$<\tau_{L}>$ high mass              &  -0.006 $\pm$ 0.005 & 1.315\\ 
$<\tau_{L}>$ all sample             & -0.010 $\pm$ 0.004 & 2.430\\ 
 $<Z_{M}>$ low mass                 & -0.06 $\pm$ 0.04 & 1.414\\ 
$<Z_{M}>$ high mass                 & -0.02 $\pm$ 0.02 & 0.883\\ 
$<Z_{M}>$ all sample                & -0.05 $\pm$ 0.02 & 2.686\\ 
$log(<Z_{M}>)$ all sample           & -0.06 $\pm$ 0.03 & 2.080\\
$M_{*,\mathrm current}$ all sample  & -0.06 $\pm$ 0.02 & 3.359\\ 
\hline
\end{tabular}
\end{table}
\normalsize

Then, we repeat the calculations of the least squares straight lines for the two groups low-mass and high-mass galaxies, separately, for both stellar parameters,
$\langle(Z_{M}/Z_{\sun})\rangle$  and $\langle \tau_{L}\rangle$. We found that low-mass galaxies have a similar trend to the fit that includes all galaxies in the HR {\sl vs} $\langle(Z_{M}/Z_{\sun})\rangle$, although with less significance (1.41$\sigma$). However, the high-mass galaxies fit has a smaller slope and not much significance. 
We do not find any correlations for the ages weighted by mass in low-mass galaxies nor in high-mass galaxies. 
Nevertheless, we find that the correlation of the HR with the age weighted by light  $\langle \tau_{L}\rangle$  has a slope of $0.0104 \pm 0.0044$. We find no significant difference between low mass, that has a slope of $-0.007 \pm 0.014$, and high mass, whose slope is $-0.006 \pm 0.005$. 
These results are summarised in Table~\ref{Table:Hubble_Residual_results}.

\begin{figure}
\centering
\includegraphics[scale=0.45]{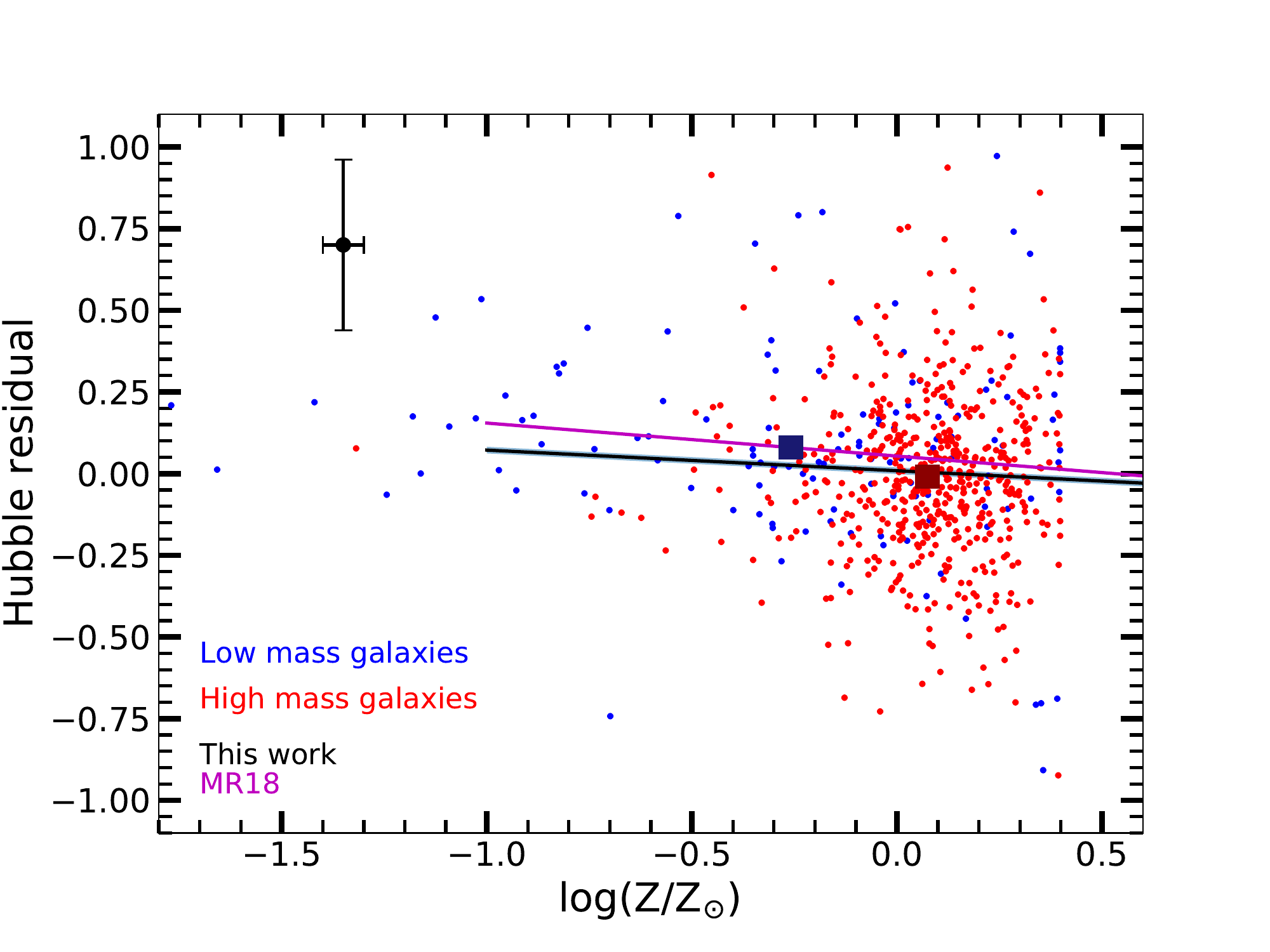}
\caption{HR results as a function of the stellar metallicity in logarithmic scale $\log{\langle(Z_{M}/Z_{\sun})\rangle}$ compared with the fit found by
\citep{2018MNRAS.476..307M} by assuming $\log{(Z/Z_{\sun})}= 12+\log{(O/H)}-8.67$. Blue and red dots are our results with color indicating if they are of low or high mass bins respectively, (see text for details). We also show with two large squares the averaged values for these two bins in mass separated by $\log{M_{split}}$.
}
\label{fig:Manu_Comparison}
\end{figure}
Finally, we compare our results with the previous work of \citet[][hereafter MR18]{2018MNRAS.476..307M}, who found a correlation between the HR and the oxygen abundance of the gas, measured as $12+\log(O/H)$. We have converted their equation given in terms of $12+\log(O/H)$ to another using the total metallicity $Z$, by using the solar abundances from \citep{Asplund+09} and the classical expression $log(Z/Z_{\sun})=12+log(O/H)-8.67$. We show these results in Figure~\ref{fig:Manu_Comparison}, where our resulting HR are drawn as a function of the $\log{\langle (Z_{M}/Z_{\sun})\rangle}$ with blue and red color small dots for the low and high mass  galaxies using the limit between them as the one we have previously computed. 
The large squares are the averaged values for these both groups. The black line is our least squares straight line, which has a slope of -0.061\,mag\,dex$^{-1}$.  MR18 found a steeper slope than ours: -0.185 \,mag\,dex$^{-1}$, but with less significance (1.52\,$\sigma$ {\sl vs} 2.08\,$\sigma$) than in the present work. This difference may be caused by the different metallicity parameter used in both works,  stellar metallicity or gas abundances from emission lines. This is neccesary to take into account that these lines
can only be obtained when there exists ionized gas in the region of the galaxy, thus limiting the study to galaxies with star-forming regions populated by young stellar populations.  

\begin{figure} 
\centering
\includegraphics[scale=0.55]{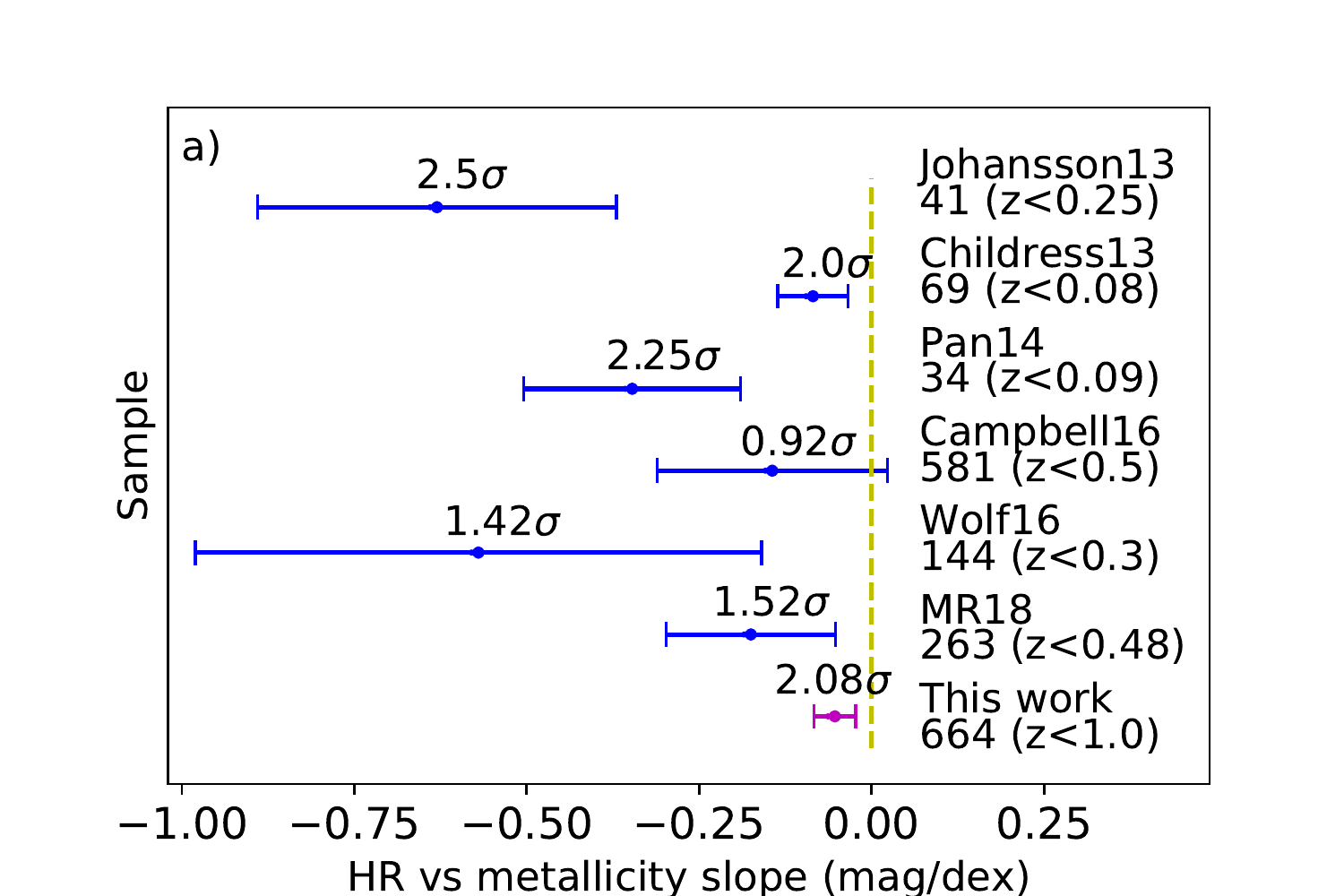}
\includegraphics[scale=0.55]{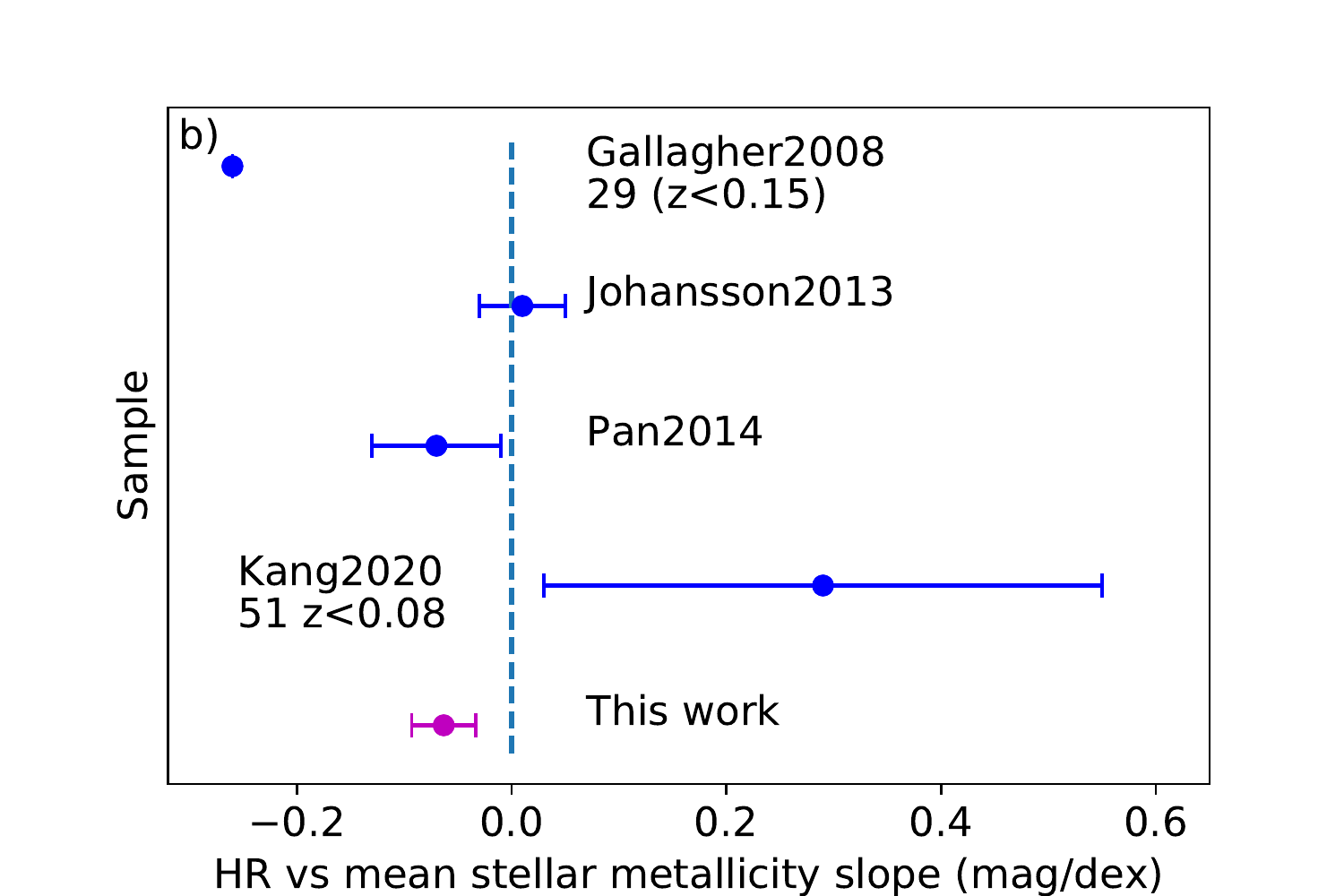}
\caption{Comparison of this work result with other previous studies: a) Using the slope of the HR {\sl vs} the gas Oxygen abundance $12+log(O/H)$; b) Using the slope of the HR {\sl vs} the weighted by mass stellar metallicity $\log{(\langle Z_{M}/Z_{\sun})}\rangle$.
}
\label{fig:Comparacion slopes}
\end{figure}

Other possible differences reside in the number of studied objects and in the redshift range. We show in Figure~\ref{fig:Comparacion slopes} a comparison with other works that have studied this dependence of the HR on metallicity of SNe~Ia host galaxies. In panel a), results obtained from emission line analyses and their corresponding gas oxygen abundances by \citet{Johansson+2013,Childress+2013,Pan+2014,Campbell+2016} and \citet{Wolf+2016} are shown. In panel b), we show results obtained with stellar metallicities of galaxies, as ours, by \citet{Johansson+2013} and \citet{Pan+2014}. In both panels, our result from this work is included. We have also indicated in the top panel, the number of objects and the maximum redshift of each set (the information of two works included in the bottom panel is already shown in the top panel). 
  
Comparing with other works in panel a), \citet{Childress+2013} obtain a similar result to us with a negative but small slope, while \citet{Campbell+2016} does not find a clear correlation in their data. However, \citet{Johansson+2013} and \citet{Wolf+2016} find a steeper negative slope than us. Finally, \citet{Pan+2014} find an intermediate value between the most negative one and our value. We see that our sample has a larger number objects, even when compared with  \citet{Campbell+2016} with a number similar as our (664 {\sl vs} 581). Besides that, our redshift range reaching $z=1$ is wider than any other previous work. 

Concerning panel b), \citet{Johansson+2013} did not find a correlation between the mean stellar metallicity and the HR, but \citet{Pan+2014} do find a negative slope although with little significance  ($\sim 1.1\sigma$). In turn, \citet{Kang+2020} find a positive slope, with a little significance too $\sim 1.1\sigma$. Our result is more precise due to the amount of galaxies we have, 664 in our work compared to 51 in \citet{Kang+2020}, that is the one of the literature with the biggest sample among studies using similar technique.

The most important consequence of this result is that SNe~Ia located in metal-rich galaxies would have lower HR than those that appear in metal-poor galaxies. This implies that it there would be an overestimation of the distance modulus, $\mu$, at high metallicity and an underestimation at low metallicity. The cosmology would need a new parameter to fit the data adequately.

\section{Conclusions}\label{sec:conclusions}

We have expanded our project to study galaxies that hosted SNe~Ia reaching further redshift (z$\sim$ 0.4-1.0) than in previous works where the maximum redshift reaches $z\le 0.5$ \citep{2018MNRAS.476..307M,Campbell+2016}. In order to do so, we have analysed 680 SN~Ia host galaxies: 654 from the SDSS survey, 6 in the redshift range (z $\sim$ 0.4-0.5) that we have observed with GTC optical spectroscopy, and a sample of 20 high-$z$ galaxies obtained from COSMOS.  This way, the number of objects is high compared with other similar studies which analyze the metallicity dependence in the SN Ia luminosity.

After selecting the galaxies by S/N and the SN~Ia parameters, we have analysed them using the code {\sc fado} to obtain the star formation histories (SFH), the averaged stellar metallicities and ages, weighted by mass or luminosity, $(\langle Z_{M}/Z_{\sun})\rangle$, $(\langle Z_{L}/Z_{\sun})\rangle$, $(\langle \tau_{M}\rangle$,  and $(\langle \tau_{L}\rangle$, as the systematic velocity, $v_{sys}$ and the dispersion velocity $\sigma_{star}$, and the totally formed and present stellar masses.

Then, we have compared the mass-metallicity,  age-mass, and the mass-velocity dispersion relations. In the case of mass-metallicity relation, our results show more dispersion than \citep{gallazzi2005} and \citep{gonzalezdelgado2014} because we obtain a larger number of galaxies in the low metal content region. 

In the case of the mass-age relation, we find lower mean ages than \citep{gallazzi2005}. This might be due to the difference in the SSP used as the basis for the analysis. We interpret this difference in the improvement of the SSPs models from {\sc HR-pyPopStar} we have used as bases, which have a better treatment of young stars. Thus, this may be an indication that {\sc HR-pyPopStar} models could be better to estimate stellar masses, ages and metallicities. 

We have finally wanted for any correlation between the SNe~Ia parameters and the characteristics of their host galaxies. We measured Hubble residuals HR using a standard flat $\Lambda$CDM cosmology. From the whole sample of 680 galaxies, we have selected 664 for which $|HR| <1.0$\,mag in order to analysis the HR results and their dependence on the stellar metallicity or age. We have found a linear correlation between the HR and the stellar metallicity, $\langle(Z_{M}/Z_{\sun})\rangle$: $HR=-0.059 \langle(Z_{M}/Z_{\sun})\rangle+0.16$. We have also looked for a correlation with the averaged stellar age weighted by mass $\langle\tau_{M}\rangle$ or weighted by light $\langle\tau_{L}\rangle$, finding no correlation for the first one, but a linear correlation for the second one with a slope of $-0.0104(\pm 0.0044)$. We have also divided our sample in low- and high-mass galaxies by getting the mass that has the largest significance in HR dividing the group in two. We have made a linear fit for low-mass and high-mass galaxies separately. For the metallicity, we find that low mass stars follow a similar trend to the fit of all galaxies, while high mass galaxies do not have a very correlation.

Finally, we have compared our results with the ones found by other authors in the literature. For that purpose, we have performed a correlation of HR with the metallicity measured in logarithmic scale $\log{\langle (Z/Z_{\sun})\rangle}$, obtaining a slope of -0.063$\pm$0.031\,mag\,dex$^{-1}$. We have compared with works that have measured the metallicity of the gas phase \cite{Johansson+2013, Childress+2013, Pan+2014, Campbell+2016, Wolf+2016} and MR18 finding a similar slope to \cite{Childress+2013} and smaller than the rest of authors. Moreover, other works have measured the correlation between the stellar metallicity and HR, \cite{Gallagher+2008, Johansson+2013, Pan+2014, Kang+2020}. We find a correlation with a slope similar to \citet{Pan+2014} and with the highest significance compared with all the others.

As a generic result, we see that the slope show in Figure~\ref{fig:Comparacion slopes} is steeper when it is measured as a function of the Oxygen abundances than the result when stellar metallicities are used. Taking into account that the SN Ia need a time from the creation of their progenitors until the explosion, it is more realistic to use the stellar metallicities, (even though these are not the metallicity with which they were formed, too), than the gas abundances, which represent the final state of the galaxy evolution. In this sense, the fact that the HR do not depend on the age when the averaged weighted by mass stellar values are taken, but depend on the weighted by light ages, would have to be interpret in the same way. The weighted by light age is giving information on the youngest stellar populations, which have not had time to create a SN Ia. Therefore the no-dependence on the weighted by mass age is a more important fact when we study the variation of the SN Ia luminosity with their own initial conditions.

\section{Acknowledgements}

This work is part of the grants I+D+i AYA2016-79724-C4-3-P, MDM-2017-0737 and PID2019-107408GB-C41, which have been funded by Ministerio de Ciencia e Innovación and Agencia Estatal de Investigaci{\'o}n  (MCIN/AEI/10.13039/501100011033). I.M. acnowledges support from the
Ministerio de Econom{\'\i}a y Competitividad under the program CFP-Unidad de Excelencia  Mar\'ia de Maeztu MDM-2015-0509.
L.G. acknowledges financial support from the Spanish Ministerio de Ciencia e Innovaci\'on (MCIN), the Agencia Estatal de Investigaci\'on (AEI) 10.13039/501100011033, and the European Social Fund (ESF) "Investing in your future" under the 2019 Ram\'on y Cajal program RYC2019-027683-I and the PID2020-115253GA-I00 HOSTFLOWS project, from Centro Superior de Investigaciones Cient\'ificas (CSIC) under the PIE project 20215AT016, and the program Unidad de Excelencia Mar\'ia de Maeztu CEX2020-001058-M.
J.M.G. thank FCT that supported this work via Fundo Europeu de Desenvolvimento Regional (FEDER) through COMPETE2020 - Programa Operacional Competitividade e Internacionalização (POCI) through the research grants UID/FIS/04434/2019,
UIDB/04434/2020 and UIDP/04434/2020. J.M.G. is supported by the DL 57/2016/CP1364/CT0003 contract and acknowledges the previous support by the fellowships CIAAUP-04/2016-BPD in the context of the FCT project UID/FIS/04434/2013 and POCI-01-0145-FEDER-007672, and SFRH/BPD/66958/2009 funded by FCT and POPH/FSE (EC).
Based on observations made with the Gran Telescopio Canarias (GTC), installed at the Spanish Observatorio del Roque de los Muchachos of the Instituto de Astrofísica de Canarias, in the island of La Palma.
This research has made use of the HST-COSMOS database, operated at CeSAM/LAM, Marseille, France.

\section{Data Availability}
\begin{enumerate}
\item Table 2 will be given in electronic format as an ASCII file.  
There we show results given by {\sc fado} for the sample of 680 galaxies with the corresponding information of their hosted SNe Ia with 29 columns:
\begin{itemize}
\item (1) Galaxy name 
\item (2) SN name 
\item (3) Right ascension RA 
\item (4) Declination DEC 
\item (5) Redshift $z$, 
\item (6) Universe age $t_{U}$, 
\item (7) The weighted by mass age $\langle\tau_{M}\rangle$ in Gyr
\item (8) The corresponding error of (7)
\item (9) The weighted by light age $\langle\tau_{L}\rangle$ in Gyr
\item (10) The corresponding error of (9)
\item (11) The weighted by mass metallicity $\langle Z_{M}\rangle$ in $Z_{\sun}$ units
\item (12) The corresponding error of (11)
\item (13) The weighted by light metallicity $\langle Z_{L}\rangle$ in $Z_{\sun}$ units
\item (14) The corresponding error of (13)
\item (15) The formed stellar mass, $\log{M_{*,formed}}$, in solar masses $M_{\odot}$ units and logarithmic scale
\item (16) The current stellar mass, $\log{M_{*,current}}$, in solar masses $M_{\odot}$ units and logarithmic scale
\item (17) The stellar velocity $v$ in km\,s$^{-1}$ units
\item (18) The corresponding error of (19)
\item (19) The dispersion velocity $\sigma$ in km\,s$^{-1}$ units
\item (20) The corresponding error of (21)
\item (21) The stretch parameter $x1$ of the supernova light curve
\item (22) The corresponding error of (23)
\item (23) The color of the supernova $c$
\item (24) The corresponding error of (24)
\item (25) the distance modulus of the SN $\mu$ in mag
\item (26) The corresponding error of (27)
\item (27) The indicator $S$ of the subsample: 1 for SDSS, 2 for GTC-OSIRIS and 3 for zCOSMOS/MAGELLAN 
\end{itemize} 

\item An electronic catalogue with all SFH for the 680 galaxies resulting from the code {\sc fado} will be available. This is a table with three columns: 
\begin{itemize}
\item (1) The name of the galaxy
\item (2) The evolutionary time $t$ in Gyr units
\item (3) The star formation rate SFR in M$_{\sun}$\,yr$^{-1}$ units
\end{itemize}
\item The reduced spectra for the six galaxies observed in GTC will be also given as an electronic table. The table consists in three columns:
\begin{itemize}
\item (1) The name of the galaxy
\item (2) The Wavelength $\lambda$ in \AA  units.
\item (3) The rest-frame flux in erg\,s$^{-1}$\,cm$^{-2}$\,\AA 
\end{itemize}
\end{enumerate}

\bibliographystyle{mnras}
\bibliography{Bibliography.bib} 

\begin{thebibliography}{}
\makeatletter
\relax
\def\mn@urlcharsother{\let\do\@makeother \do\$\do\&\do\#\do\^\do\_\do\%\do\~}
\def\mn@doi{\begingroup\mn@urlcharsother \@ifnextchar [ {\mn@doi@}
  {\mn@doi@[]}}
\def\mn@doi@[#1]#2{\def\@tempa{#1}\ifx\@tempa\@empty \href
  {http://dx.doi.org/#2} {doi:#2}\else \href {http://dx.doi.org/#2} {#1}\fi
  \endgroup}
\def\mn@eprint#1#2{\mn@eprint@#1:#2::\@nil}
\def\mn@eprint@arXiv#1{\href {http://arxiv.org/abs/#1} {{\tt arXiv:#1}}}
\def\mn@eprint@dblp#1{\href {http://dblp.uni-trier.de/rec/bibtex/#1.xml}
  {dblp:#1}}
\def\mn@eprint@#1:#2:#3:#4\@nil{\def\@tempa {#1}\def\@tempb {#2}\def\@tempc
  {#3}\ifx \@tempc \@empty \let \@tempc \@tempb \let \@tempb \@tempa \fi \ifx
  \@tempb \@empty \def\@tempb {arXiv}\fi \@ifundefined
  {mn@eprint@\@tempb}{\@tempb:\@tempc}{\expandafter \expandafter \csname
  mn@eprint@\@tempb\endcsname \expandafter{\@tempc}}}

\bibitem[\protect\citeauthoryear{{Ahumada} et~al.,}{{Ahumada}
  et~al.}{2020}]{2020ApJS..249....3A}
{Ahumada} R.,  et~al., 2020, \mn@doi [\apjs] {10.3847/1538-4365/ab929e}, \href
  {https://ui.adsabs.harvard.edu/abs/2020ApJS..249....3A} {249, 3}

\bibitem[\protect\citeauthoryear{{Asplund}, {Grevesse}, {Sauval}  \&
  {Scott}}{{Asplund} et~al.}{2009}]{Asplund+09}
{Asplund} M.,  {Grevesse} N.,  {Sauval} A.~J.,   {Scott} P.,  2009, \mn@doi
  [\araa] {10.1146/annurev.astro.46.060407.145222}, \href
  {https://ui.adsabs.harvard.edu/abs/2009ARA&A..47..481A} {47, 481}

\bibitem[\protect\citeauthoryear{Barbary et~al.,}{Barbary
  et~al.}{2022}]{barbary_kyle_2022_6363879}
Barbary K.,  et~al., 2022, SNCosmo, \mn@doi{10.5281/zenodo.6363879}, \url
  {https://doi.org/10.5281/zenodo.6363879}

\bibitem[\protect\citeauthoryear{{Betoule} et~al.,}{{Betoule}
  et~al.}{2014a}]{betoule2014}
{Betoule} M.,  et~al., 2014a, \mn@doi [\aap] {10.1051/0004-6361/201423413},
  \href {https://ui.adsabs.harvard.edu/abs/2014A&A...568A..22B} {568, A22}

\bibitem[\protect\citeauthoryear{{Betoule} et~al.,}{{Betoule}
  et~al.}{2014b}]{2014A&A...568A..22B}
{Betoule} M.,  et~al., 2014b, \mn@doi [\aap] {10.1051/0004-6361/201423413},
  \href {https://ui.adsabs.harvard.edu/abs/2014A&A...568A..22B} {568, A22}

\bibitem[\protect\citeauthoryear{{Bravo}, {Dom{\'\i}nguez}, {Badenes},
  {Piersanti}  \& {Straniero}}{{Bravo} et~al.}{2010}]{bravo2010}
{Bravo} E.,  {Dom{\'\i}nguez} I.,  {Badenes} C.,  {Piersanti} L.,   {Straniero}
  O.,  2010, \mn@doi [\apjl] {10.1088/2041-8205/711/2/L66}, \href
  {https://ui.adsabs.harvard.edu/abs/2010ApJ...711L..66B} {711, L66}

\bibitem[\protect\citeauthoryear{{Brout} et~al.,}{{Brout}
  et~al.}{2019}]{2019ApJ...874..150B}
{Brout} D.,  et~al., 2019, \mn@doi [\apj] {10.3847/1538-4357/ab08a0}, \href
  {https://ui.adsabs.harvard.edu/abs/2019ApJ...874..150B} {874, 150}

\bibitem[\protect\citeauthoryear{{Bruzual} \& {Charlot}}{{Bruzual} \&
  {Charlot}}{2003}]{bc03}
{Bruzual} G.,  {Charlot} S.,  2003, \mnras, 344, 1000

\bibitem[\protect\citeauthoryear{{Campbell}, {Fraser}  \& {Gilmore}}{{Campbell}
  et~al.}{2016}]{Campbell+2016}
{Campbell} H.,  {Fraser} M.,   {Gilmore} G.,  2016, \mn@doi [\mnras]
  {10.1093/mnras/stw115}, \href
  {https://ui.adsabs.harvard.edu/abs/2016MNRAS.457.3470C} {457, 3470}

\bibitem[\protect\citeauthoryear{{Cepa}}{{Cepa}}{2010}]{Cepa2010}
{Cepa} J.,  2010, in Highlights of Spanish Astrophysics V. p.~15,
  \mn@doi{10.1007/978-3-642-11250-8\_2}

\bibitem[\protect\citeauthoryear{{Cepa} et~al.,}{{Cepa}
  et~al.}{2005}]{Cepa2005}
{Cepa} J.,  et~al., 2005, in {Hidalgo-G{\'a}mez} A.~M.,  {Gonz{\'a}lez} J.~J.,
  {Rodr{\'\i}guez Espinosa} J.~M.,   {Torres-Peimbert} S.,  eds,  Revista
  Mexicana de Astronomia y Astrofisica Conference Series Vol. 24, Revista
  Mexicana de Astronomia y Astrofisica Conference Series. pp~1--6

\bibitem[\protect\citeauthoryear{{Childress} et~al.,}{{Childress}
  et~al.}{2013}]{Childress+2013}
{Childress} M.,  et~al., 2013, \mn@doi [\apj] {10.1088/0004-637X/770/2/108},
  \href {https://ui.adsabs.harvard.edu/abs/2013ApJ...770..108C} {770, 108}

\bibitem[\protect\citeauthoryear{{Cid Fernandes} \& {Gonz{\'a}lez
  Delgado}}{{Cid Fernandes} \& {Gonz{\'a}lez Delgado}}{2010}]{starlight2}
{Cid Fernandes} R.,  {Gonz{\'a}lez Delgado} R.~M.,  2010, \mn@doi [\mnras]
  {10.1111/j.1365-2966.2009.16153.x}, \href
  {https://ui.adsabs.harvard.edu/abs/2010MNRAS.403..780C} {403, 780}

\bibitem[\protect\citeauthoryear{{Cid Fernandes} et~al.,}{{Cid Fernandes}
  et~al.}{2009}]{starlight}
{Cid Fernandes} R.,  et~al., 2009, in Revista Mexicana de Astronomia y
  Astrofisica Conference Series. pp 127--132 (\mn@eprint {arXiv} {0802.0849})

\bibitem[\protect\citeauthoryear{{Coelho}}{{Coelho}}{2014}]{Coelho2014}
{Coelho} P.~R.~T.,  2014, \mn@doi [\mnras] {10.1093/mnras/stu365}, \href
  {https://ui.adsabs.harvard.edu/abs/2014MNRAS.440.1027C} {440, 1027}

\bibitem[\protect\citeauthoryear{{D'Andrea} et~al.,}{{D'Andrea}
  et~al.}{2011a}]{2011ApJ...743..172D}
{D'Andrea} C.~B.,  et~al., 2011a, \mn@doi [\apj] {10.1088/0004-637X/743/2/172},
  \href {http://adsabs.harvard.edu/abs/2011ApJ...743..172D} {743, 172}

\bibitem[\protect\citeauthoryear{{D'Andrea} et~al.,}{{D'Andrea}
  et~al.}{2011b}]{dAndrea2011}
{D'Andrea} C.~B.,  et~al., 2011b, \mn@doi [\apj] {10.1088/0004-637X/743/2/172},
  \href {https://ui.adsabs.harvard.edu/abs/2011ApJ...743..172D} {743, 172}

\bibitem[\protect\citeauthoryear{{Falc{\'o}n-Barroso},
  {S{\'a}nchez-Bl{\'a}zquez}, {Vazdekis}, {Ricciardelli}, {Cardiel}, {Cenarro},
  {Gorgas}  \& {Peletier}}{{Falc{\'o}n-Barroso}
  et~al.}{2011}]{falcon-barroso2011}
{Falc{\'o}n-Barroso} J.,  {S{\'a}nchez-Bl{\'a}zquez} P.,  {Vazdekis} A.,
  {Ricciardelli} E.,  {Cardiel} N.,  {Cenarro} A.~J.,  {Gorgas} J.,
  {Peletier} R.~F.,  2011, \mn@doi [\aap] {10.1051/0004-6361/201116842}, \href
  {https://ui.adsabs.harvard.edu/abs/2011A&A...532A..95F} {532, A95}

\bibitem[\protect\citeauthoryear{{Galbany}, {Smith}, {Duarte-Puertas}  \&
  {Gonz\'alez-Gait\'an}}{{Galbany} et~al.}{2022}]{galbany22}
{Galbany} L.,  {Smith} M.,  {Duarte-Puertas} S.,   {Gonz\'alez-Gait\'an} S.,
  2022, A\&A

\bibitem[\protect\citeauthoryear{{Gallagher}, {Garnavich}, {Berlind},
  {Challis}, {Jha}  \& {Kirshner}}{{Gallagher} et~al.}{2005}]{Gallagher+2005}
{Gallagher} J.~S.,  {Garnavich} P.~M.,  {Berlind} P.,  {Challis} P.,  {Jha} S.,
    {Kirshner} R.~P.,  2005, \mn@doi [\apj] {10.1086/491664}, \href
  {https://ui.adsabs.harvard.edu/abs/2005ApJ...634..210G} {634, 210}

\bibitem[\protect\citeauthoryear{{Gallagher}, {Garnavich}, {Caldwell},
  {Kirshner}, {Jha}, {Li}, {Ganeshalingam}  \& {Filippenko}}{{Gallagher}
  et~al.}{2008}]{Gallagher+2008}
{Gallagher} J.~S.,  {Garnavich} P.~M.,  {Caldwell} N.,  {Kirshner} R.~P.,
  {Jha} S.~W.,  {Li} W.,  {Ganeshalingam} M.,   {Filippenko} A.~V.,  2008,
  \mn@doi [\apj] {10.1086/590659}, \href
  {https://ui.adsabs.harvard.edu/abs/2008ApJ...685..752G} {685, 752}

\bibitem[\protect\citeauthoryear{{Gallazzi}, {Charlot}, {Brinchmann}, {White}
  \& {Tremonti}}{{Gallazzi} et~al.}{2005}]{gallazzi2005}
{Gallazzi} A.,  {Charlot} S.,  {Brinchmann} J.,  {White} S. D.~M.,   {Tremonti}
  C.~A.,  2005, \mn@doi [\mnras] {10.1111/j.1365-2966.2005.09321.x}, \href
  {https://ui.adsabs.harvard.edu/abs/2005MNRAS.362...41G} {362, 41}

\bibitem[\protect\citeauthoryear{{Gomes} \& {Papaderos}}{{Gomes} \&
  {Papaderos}}{2017}]{fado}
{Gomes} J.~M.,  {Papaderos} P.,  2017, \aap, 603, A63

\bibitem[\protect\citeauthoryear{{Gonz{\'a}lez Delgado}, {Cervi{\~n}o},
  {Martins}, {Leitherer}  \& {Hauschildt}}{{Gonz{\'a}lez Delgado}
  et~al.}{2005}]{gonzalezdelgado2005}
{Gonz{\'a}lez Delgado} R.~M.,  {Cervi{\~n}o} M.,  {Martins} L.~P.,  {Leitherer}
  C.,   {Hauschildt} P.~H.,  2005, \mn@doi [\mnras]
  {10.1111/j.1365-2966.2005.08692.x}, \href
  {https://ui.adsabs.harvard.edu/abs/2005MNRAS.357..945G} {357, 945}

\bibitem[\protect\citeauthoryear{{Gonz{\'a}lez Delgado} et~al.,}{{Gonz{\'a}lez
  Delgado} et~al.}{2014}]{gonzalezdelgado2014}
{Gonz{\'a}lez Delgado} R.~M.,  et~al., 2014, \mn@doi [\apjl]
  {10.1088/2041-8205/791/1/L16}, \href
  {https://ui.adsabs.harvard.edu/abs/2014ApJ...791L..16G} {791, L16}

\bibitem[\protect\citeauthoryear{{Gonz{\'a}lez Delgado} et~al.,}{{Gonz{\'a}lez
  Delgado} et~al.}{2015}]{gonzalezdelgado2015}
{Gonz{\'a}lez Delgado} R.~M.,  et~al., 2015, \mn@doi [\aap]
  {10.1051/0004-6361/201525938}, \href
  {https://ui.adsabs.harvard.edu/abs/2015A&A...581A.103G} {581, A103}

\bibitem[\protect\citeauthoryear{{Gupta} et~al.,}{{Gupta}
  et~al.}{2011}]{2011ApJ...740...92G}
{Gupta} R.~R.,  et~al., 2011, \mn@doi [\apj] {10.1088/0004-637X/740/2/92},
  \href {http://adsabs.harvard.edu/abs/2011ApJ...740...92G} {740, 92}

\bibitem[\protect\citeauthoryear{{Guy}}{{Guy}}{2007}]{salt2}
{Guy} J. e.~a.,  2007, \aap, 466, 11

\bibitem[\protect\citeauthoryear{{Guy} et~al.,}{{Guy} et~al.}{2010}]{Guy+2010}
{Guy} J.,  et~al., 2010, \mn@doi [\aap] {10.1051/0004-6361/201014468}, \href
  {https://ui.adsabs.harvard.edu/abs/2010A&A...523A...7G} {523, A7}

\bibitem[\protect\citeauthoryear{{Hicken} et~al.,}{{Hicken}
  et~al.}{2009}]{2009ApJ...700..331H}
{Hicken} M.,  et~al., 2009, \mn@doi [\apj] {10.1088/0004-637X/700/1/331}, \href
  {http://adsabs.harvard.edu/abs/2009ApJ...700..331H} {700, 331}

\bibitem[\protect\citeauthoryear{{Howell} et~al.,}{{Howell}
  et~al.}{2009}]{2009ApJ...691..661H}
{Howell} D.~A.,  et~al., 2009, \mn@doi [\apj] {10.1088/0004-637X/691/1/661},
  \href {http://adsabs.harvard.edu/abs/2009ApJ...691..661H} {691, 661}

\bibitem[\protect\citeauthoryear{{Johansson} et~al.,}{{Johansson}
  et~al.}{2013}]{Johansson+2013}
{Johansson} J.,  et~al., 2013, \mn@doi [\mnras] {10.1093/mnras/stt1408}, \href
  {https://ui.adsabs.harvard.edu/abs/2013MNRAS.435.1680J} {435, 1680}

\bibitem[\protect\citeauthoryear{{Jones} et~al.,}{{Jones}
  et~al.}{2018}]{Jones+2018}
{Jones} D.~O.,  et~al., 2018, \mn@doi [\apj] {10.3847/1538-4357/aae2b9}, \href
  {https://ui.adsabs.harvard.edu/abs/2018ApJ...867..108J} {867, 108}

\bibitem[\protect\citeauthoryear{{Kang}, {Lee}, {Kim}, {Chung}  \&
  {Ree}}{{Kang} et~al.}{2020}]{Kang+2020}
{Kang} Y.,  {Lee} Y.-W.,  {Kim} Y.-L.,  {Chung} C.,   {Ree} C.~H.,  2020,
  \mn@doi [\apj] {10.3847/1538-4357/ab5afc}, \href
  {https://ui.adsabs.harvard.edu/abs/2020ApJ...889....8K} {889, 8}

\bibitem[\protect\citeauthoryear{{Kelly}, {Hicken}, {Burke}, {Mandel}  \&
  {Kirshner}}{{Kelly} et~al.}{2010}]{2010ApJ...715..743K}
{Kelly} P.~L.,  {Hicken} M.,  {Burke} D.~L.,  {Mandel} K.~S.,   {Kirshner}
  R.~P.,  2010, \mn@doi [\apj] {10.1088/0004-637X/715/2/743}, \href
  {http://adsabs.harvard.edu/abs/2010ApJ...715..743K} {715, 743}

\bibitem[\protect\citeauthoryear{{Kim}, {Kang}  \& {Lee}}{{Kim}
  et~al.}{2019}]{Kim+2019}
{Kim} Y.-L.,  {Kang} Y.,   {Lee} Y.-W.,  2019, \mn@doi [Journal of Korean
  Astronomical Society] {10.5303/JKAS.2019.52.5.181}, \href
  {https://ui.adsabs.harvard.edu/abs/2019JKAS...52..181K} {52, 181}

\bibitem[\protect\citeauthoryear{{Konishi} et~al.,}{{Konishi}
  et~al.}{2011}]{konishi2011}
{Konishi} K.,  et~al., 2011, arXiv e-prints, \href
  {https://ui.adsabs.harvard.edu/abs/2011arXiv1101.4269K} {p. arXiv:1101.4269}

\bibitem[\protect\citeauthoryear{{Kroupa}}{{Kroupa}}{2002}]{Kroupa2002}
{Kroupa} P.,  2002, \mn@doi [Science] {10.1126/science.1067524}, \href
  {https://ui.adsabs.harvard.edu/abs/2002Sci...295...82K} {295, 82}

\bibitem[\protect\citeauthoryear{{Lampeitl} et~al.,}{{Lampeitl}
  et~al.}{2010}]{2010ApJ...722..566L}
{Lampeitl} H.,  et~al., 2010, \mn@doi [\apj] {10.1088/0004-637X/722/1/566},
  \href {http://adsabs.harvard.edu/abs/2010ApJ...722..566L} {722, 566}

\bibitem[\protect\citeauthoryear{{Lara-L{\'o}pez}, {L{\'o}pez-S{\'a}nchez}  \&
  {Hopkins}}{{Lara-L{\'o}pez} et~al.}{2013}]{laralopez2010}
{Lara-L{\'o}pez} M.~A.,  {L{\'o}pez-S{\'a}nchez} {\'A}.~R.,   {Hopkins} A.~M.,
  2013, \mn@doi [\apj] {10.1088/0004-637X/764/2/178}, \href
  {https://ui.adsabs.h} {764, 178}

\bibitem[\protect\citeauthoryear{{Lilly} et~al.,}{{Lilly}
  et~al.}{2007}]{zcosmos}
{Lilly} S.~J.,  et~al., 2007, \mn@doi [\apjs] {10.1086/516589}, \href
  {https://ui.adsabs.harvard.edu/abs/2007ApJS..172...70L} {172, 70}

\bibitem[\protect\citeauthoryear{{Martins}, {Gonz{\'a}lez Delgado},
  {Leitherer}, {Cervi{\~n}o}  \& {Hauschildt}}{{Martins}
  et~al.}{2005}]{martins2005}
{Martins} L.~P.,  {Gonz{\'a}lez Delgado} R.~M.,  {Leitherer} C.,  {Cervi{\~n}o}
  M.,   {Hauschildt} P.,  2005, \mn@doi [\mnras]
  {10.1111/j.1365-2966.2005.08703.x}, \href
  {https://ui.adsabs.harvard.edu/abs/2005MNRAS.358...49M} {358, 49}

\bibitem[\protect\citeauthoryear{{Mill{\'a}n-Irigoyen}, {Moll{\'a}},
  {Cervi{\~n}o}, {Ascasibar}, {Garc{\'\i}a-Vargas}  \&
  {Coelho}}{{Mill{\'a}n-Irigoyen} et~al.}{2021}]{millan-irigoyen+2021}
{Mill{\'a}n-Irigoyen} I.,  {Moll{\'a}} M.,  {Cervi{\~n}o} M.,  {Ascasibar} Y.,
  {Garc{\'\i}a-Vargas} M.~L.,   {Coelho} P.~R.~T.,  2021, \mn@doi [\mnras]
  {10.1093/mnras/stab1969}, \href
  {https://ui.adsabs.harvard.edu/abs/2021MNRAS.506.4781M} {506, 4781}

\bibitem[\protect\citeauthoryear{{Moreno-Raya}, {L{\'o}pez-S{\'a}nchez},
  {Moll{\'a}}, {Galbany}, {V{\'\i}lchez}  \& {Carnero}}{{Moreno-Raya}
  et~al.}{2016a}]{mr-2016b}
{Moreno-Raya} M.~E.,  {L{\'o}pez-S{\'a}nchez} {\'A}.~R.,  {Moll{\'a}} M.,
  {Galbany} L.,  {V{\'\i}lchez} J.~M.,   {Carnero} A.,  2016a, \mn@doi [\mnras]
  {10.1093/mnras/stw1706}, \href
  {https://ui.adsabs.harvard.edu/abs/2016MNRAS.462.1281M} {462, 1281}

\bibitem[\protect\citeauthoryear{{Moreno-Raya}, {Moll{\'a}},
  {L{\'o}pez-S{\'a}nchez}, {Galbany}, {V{\'\i}lchez}, {Carnero Rosell}  \&
  {Dom{\'\i}nguez}}{{Moreno-Raya} et~al.}{2016b}]{mr-2016a}
{Moreno-Raya} M.~E.,  {Moll{\'a}} M.,  {L{\'o}pez-S{\'a}nchez} {\'A}.~R.,
  {Galbany} L.,  {V{\'\i}lchez} J.~M.,  {Carnero Rosell} A.,   {Dom{\'\i}nguez}
  I.,  2016b, \mn@doi [\apjl] {10.3847/2041-8205/818/1/L19}, \href
  {https://ui.adsabs.harvard.edu/abs/2016ApJ...818L..19M} {818, L19}

\bibitem[\protect\citeauthoryear{{Moreno-Raya}, {Galbany},
  {L{\'o}pez-S{\'a}nchez}, {Moll{\'a}}, {Gonz{\'a}lez-Gait{\'a}n},
  {V{\'\i}lchez}  \& {Carnero}}{{Moreno-Raya}
  et~al.}{2018}]{2018MNRAS.476..307M}
{Moreno-Raya} M.~E.,  {Galbany} L.,  {L{\'o}pez-S{\'a}nchez} {\'A}.~R.,
  {Moll{\'a}} M.,  {Gonz{\'a}lez-Gait{\'a}n} S.,  {V{\'\i}lchez} J.~M.,
  {Carnero} A.,  2018, \mn@doi [\mnras] {10.1093/mnras/sty185}, \href
  {https://ui.adsabs.harvard.edu/abs/2018MNRAS.476..307M} {476, 307}

\bibitem[\protect\citeauthoryear{{Munari}, {Sordo}, {Castelli}  \&
  {Zwitter}}{{Munari} et~al.}{2005}]{Munari2005}
{Munari} U.,  {Sordo} R.,  {Castelli} F.,   {Zwitter} T.,  2005, \mn@doi [\aap]
  {10.1051/0004-6361:20042490}, \href
  {https://ui.adsabs.harvard.edu/abs/2005A&A...442.1127M} {442, 1127}

\bibitem[\protect\citeauthoryear{{Nordin} et~al.,}{{Nordin}
  et~al.}{2011}]{2011ApJ...734...42N}
{Nordin} J.,  et~al., 2011, \mn@doi [\apj] {10.1088/0004-637X/734/1/42}, \href
  {http://adsabs.harvard.edu/abs/2011ApJ...734...42N} {734, 42}

\bibitem[\protect\citeauthoryear{{Olmstead} et~al.,}{{Olmstead}
  et~al.}{2014}]{2014AJ....147...75O}
{Olmstead} M.~D.,  et~al., 2014, \mn@doi [\aj] {10.1088/0004-6256/147/4/75},
  \href {https://ui.adsabs.harvard.edu/abs/2014AJ....147...75O} {147, 75}

\bibitem[\protect\citeauthoryear{{Pan} et~al.,}{{Pan} et~al.}{2014}]{Pan+2014}
{Pan} Y.~C.,  et~al., 2014, \mn@doi [\mnras] {10.1093/mnras/stt2287}, \href
  {https://ui.adsabs.harvard.edu/abs/2014MNRAS.438.1391P} {438, 1391}

\bibitem[\protect\citeauthoryear{{Perlmutter}, {Turner}  \&
  {White}}{{Perlmutter} et~al.}{1999}]{perlmutter1999}
{Perlmutter} S.,  {Turner} M.~S.,   {White} M.,  1999, Physical Review Letters,
  83, 670

\bibitem[\protect\citeauthoryear{{Phillips}}{{Phillips}}{1993}]{phillips1993}
{Phillips} M.~M.,  1993, \mn@doi [\apjl] {10.1086/186970}, \href
  {https://ui.adsabs.harvard.edu/abs/1993ApJ...413L.105P} {413, L105}

\bibitem[\protect\citeauthoryear{{Riess}, {Press}  \& {Kirshner}}{{Riess}
  et~al.}{1996}]{riess1996}
{Riess} A.~G.,  {Press} W.~H.,   {Kirshner} R.~P.,  1996, \mn@doi [\apj]
  {10.1086/178174}, \href
  {https://ui.adsabs.harvard.edu/abs/1996ApJ...473..588R} {473, 588}

\bibitem[\protect\citeauthoryear{{Riess} et~al.,}{{Riess}
  et~al.}{1998}]{riess1998}
{Riess} A.~G.,  et~al., 1998, \mn@doi [\aj] {10.1086/300499}, \href
  {https://ui.adsabs.harvard.edu/abs/1998AJ....116.1009R} {116, 1009}

\bibitem[\protect\citeauthoryear{{Rigault} et~al.,}{{Rigault}
  et~al.}{2020}]{Rigault+2020}
{Rigault} M.,  et~al., 2020, \mn@doi [\aap] {10.1051/0004-6361/201730404},
  \href {https://ui.adsabs.harvard.edu/abs/2020A&A...644A.176R} {644, A176}

\bibitem[\protect\citeauthoryear{{Roman} et~al.,}{{Roman}
  et~al.}{2018}]{roman2018}
{Roman} M.,  et~al., 2018, \mn@doi [\aap] {10.1051/0004-6361/201731425}, \href
  {https://ui.adsabs.harvard.edu/abs/2018A&A...615A..68R} {615, A68}

\bibitem[\protect\citeauthoryear{{Sako} et~al.,}{{Sako}
  et~al.}{2018}]{Sako+2018}
{Sako} M.,  et~al., 2018, \mn@doi [\pasp] {10.1088/1538-3873/aab4e0}, \href
  {https://ui.adsabs.harvard.edu/abs/2018PASP..130f4002S} {130, 064002}

\bibitem[\protect\citeauthoryear{{Schlegel}, {Finkbeiner}  \&
  {Davis}}{{Schlegel} et~al.}{1998}]{MapaExti}
{Schlegel} D.~J.,  {Finkbeiner} D.~P.,   {Davis} M.,  1998, \mn@doi [\apj]
  {10.1086/305772}, \href
  {https://ui.adsabs.harvard.edu/abs/1998ApJ...500..525S} {500, 525}

\bibitem[\protect\citeauthoryear{{Scolnic} et~al.,}{{Scolnic}
  et~al.}{2018a}]{2018ApJ...859..101S}
{Scolnic} D.~M.,  et~al., 2018a, \mn@doi [\apj] {10.3847/1538-4357/aab9bb},
  \href {https://ui.adsabs.harvard.edu/abs/2018ApJ...859..101S} {859, 101}

\bibitem[\protect\citeauthoryear{{Scolnic} et~al.,}{{Scolnic}
  et~al.}{2018b}]{Scolnic+2018}
{Scolnic} D.~M.,  et~al., 2018b, \mn@doi [\apj] {10.3847/1538-4357/aab9bb},
  \href {https://ui.adsabs.harvard.edu/abs/2018ApJ...859..101S} {859, 101}

\bibitem[\protect\citeauthoryear{{Scoville}}{{Scoville}}{2007}]{COSMOS}
{Scoville} N. e.~a.,  2007, \apjs, 172, 1

\bibitem[\protect\citeauthoryear{{Smith} et~al.,}{{Smith}
  et~al.}{2020}]{Smith+2020}
{Smith} M.,  et~al., 2020, \mn@doi [\mnras] {10.1093/mnras/staa946}, \href
  {https://ui.adsabs.harvard.edu/abs/2020MNRAS.494.4426S} {494, 4426}

\bibitem[\protect\citeauthoryear{{Sullivan} et~al.,}{{Sullivan}
  et~al.}{2006}]{2006ApJ...648..868S}
{Sullivan} M.,  et~al., 2006, \mn@doi [\apj] {10.1086/506137}, \href
  {http://adsabs.harvard.edu/abs/2006ApJ...648..868S} {648, 868}

\bibitem[\protect\citeauthoryear{{Sullivan} et~al.,}{{Sullivan}
  et~al.}{2010}]{2010MNRAS.406..782S}
{Sullivan} M.,  et~al., 2010, \mn@doi [\mnras]
  {10.1111/j.1365-2966.2010.16731.x}, \href
  {http://adsabs.harvard.edu/abs/2010MNRAS.406..782S} {406, 782}

\bibitem[\protect\citeauthoryear{{Sullivan} et~al.,}{{Sullivan}
  et~al.}{2011}]{2011ApJ...737..102S}
{Sullivan} M.,  et~al., 2011, \mn@doi [\apj] {10.1088/0004-637X/737/2/102},
  \href {http://adsabs.harvard.edu/abs/2011ApJ...737..102S} {737, 102}

\bibitem[\protect\citeauthoryear{{Suzuki} et~al.,}{{Suzuki}
  et~al.}{2012}]{suzuki}
{Suzuki} N.,  et~al., 2012, \mn@doi [\apj] {10.1088/0004-637X/746/1/85}, \href
  {https://ui.adsabs.harvard.edu/abs/2012ApJ...746...85S} {746, 85}

\bibitem[\protect\citeauthoryear{{Timmes}, {Brown}  \& {Truran}}{{Timmes}
  et~al.}{2003}]{timmes2003}
{Timmes} F.~X.,  {Brown} E.~F.,   {Truran} J.~W.,  2003, \mn@doi [\apjl]
  {10.1086/376721}, \href
  {https://ui.adsabs.harvard.edu/abs/2003ApJ...590L..83T} {590, L83}

\bibitem[\protect\citeauthoryear{{Tody}}{{Tody}}{1993}]{iraf}
{Tody} D.,  1993, in {Hanisch} R.~J.,  {Brissenden} R.~J.~V.,   {Barnes} J.,
  eds,  Astronomical Society of the Pacific Conference Series Vol. 52,
  Astronomical Data Analysis Software and Systems II. p.~173

\bibitem[\protect\citeauthoryear{{Tremonti} et~al.,}{{Tremonti}
  et~al.}{2004}]{tremonti2004}
{Tremonti} C.~A.,  et~al., 2004, \mn@doi [\apj] {10.1086/423264}, \href
  {http://cdsads.u-strasbg.fr/abs/2004ApJ...613..898T} {613, 898}

\bibitem[\protect\citeauthoryear{{Tripp}}{{Tripp}}{1998}]{tripp1998}
{Tripp} R.,  1998, \aap, \href
  {https://ui.adsabs.harvard.edu/abs/1998A&A...331..815T} {331, 815}

\bibitem[\protect\citeauthoryear{{Trump} et~al.,}{{Trump}
  et~al.}{2009}]{magellan}
{Trump} J.~R.,  et~al., 2009, \mn@doi [\apj] {10.1088/0004-637X/696/2/1195},
  \href {https://ui.adsabs.harvard.edu/abs/2009ApJ...696.1195T} {696, 1195}

\bibitem[\protect\citeauthoryear{{Uddin}, {Mould}, {Lidman}, {Ruhlmann-Kleider}
   \& {Zhang}}{{Uddin} et~al.}{2017a}]{Uddin2017}
{Uddin} S.~A.,  {Mould} J.,  {Lidman} C.,  {Ruhlmann-Kleider} V.,   {Zhang}
  B.~R.,  2017a, \mn@doi [\apj] {10.3847/1538-4357/aa8df7}, \href
  {https://ui.adsabs.harvard.edu/abs/2017ApJ...848...56U} {848, 56}

\bibitem[\protect\citeauthoryear{{Uddin}, {Mould}  \& {Wang}}{{Uddin}
  et~al.}{2017b}]{Uddin2018}
{Uddin} S.~A.,  {Mould} J.,   {Wang} L.,  2017b, \mn@doi [\apj]
  {10.3847/1538-4357/aa93e9}, \href
  {https://ui.adsabs.harvard.edu/abs/2017ApJ...850..135U} {850, 135}

\bibitem[\protect\citeauthoryear{{Uddin} et~al.,}{{Uddin}
  et~al.}{2020}]{Uddin2020}
{Uddin} S.~A.,  et~al., 2020, \mn@doi [\apj] {10.3847/1538-4357/abafb7}, \href
  {https://ui.adsabs.harvard.edu/abs/2020ApJ...901..143U} {901, 143}

\bibitem[\protect\citeauthoryear{{Vazdekis}, {S{\'a}nchez-Bl{\'a}zquez},
  {Falc{\'o}n-Barroso}, {Cenarro}, {Beasley}, {Cardiel}, {Gorgas}  \&
  {Peletier}}{{Vazdekis} et~al.}{2010}]{vazdekis2010}
{Vazdekis} A.,  {S{\'a}nchez-Bl{\'a}zquez} P.,  {Falc{\'o}n-Barroso} J.,
  {Cenarro} A.~J.,  {Beasley} M.~A.,  {Cardiel} N.,  {Gorgas} J.,   {Peletier}
  R.~F.,  2010, \mn@doi [\mnras] {10.1111/j.1365-2966.2010.16407.x}, \href
  {https://ui.adsabs.harvard.edu/abs/2010MNRAS.404.1639V} {404, 1639}

\bibitem[\protect\citeauthoryear{{Wolf} et~al.,}{{Wolf}
  et~al.}{2016}]{Wolf+2016}
{Wolf} R.~C.,  et~al., 2016, \mn@doi [\apj] {10.3847/0004-637X/821/2/115},
  \href {https://ui.adsabs.harvard.edu/abs/2016ApJ...821..115W} {821, 115}

\makeatother
\end{thebibliography}

\bsp	
\label{lastpage}
\end{document}